\documentclass[preprint,nofootinbib,aps,superscriptaddress,eqsecnum]{revtex4-2} 
\usepackage[colorlinks,citecolor=red]{hyperref}
\usepackage[usenames]{color}
\usepackage{xcolor}
\usepackage{float}
\pdfoutput=1
\usepackage{tabularx}
\usepackage{graphicx}
\usepackage{amsmath}
\usepackage{amsfonts}
\usepackage{amssymb}
\usepackage[justification=centering]{caption} 
\usepackage{caption}
\usepackage{subcaption}
\usepackage{color}
\allowdisplaybreaks
\def\bea{\begin{eqnarray}}
	\def\eea{\end{eqnarray}}
\def\be{\begin{equation}}
	\def\ee{\end{equation}}

\begin{document}
	\title{A $\Gamma_{3}$ modular symmetric approach for two-zero textures in left-right symmetric model}
	\author{Ankita Kakoti}
	\email{kakotiankita97@gmail.com}
	\affiliation{Department of Physics, Sibsagar University, Sibsagar 785665, India}
	
	\author{Happy Borgohain}
	\email{haps.tezu@gmail.com}
	\affiliation{Department of Physics, Silapathar College, Silapathar, 787059, India}

	\begin{abstract}
The observed pattern for neutrino masses and mixing provides compelling evidence for Beyond Standard Model physics which further motivates the search for predictive frameworks that can simultaneously address flavor structure and its phenomenological consequences. This work particularly investigates the realization of all possible seven two-zero neutrino mass textures within the generic left-right symmetric model with $A_{4}$ modular symmetry. By considering modular weights 4,8 and 10, we systematically construct all the possible classes of 2-0 textures without the introduction of any flavon fields which enhances the predictive power of the framework. In this work, we also identify the texture classes capable of simultaneously accommodating current neutrino data, reproducing the observed baryon asymmetry and also yielding experimentally testable results for the effective Majorana neutrino mass for new physics contributions of neutrinoless double beta decay.

	\end{abstract}
	\maketitle
	\newpage

		\section{\label{lrsm11}\textbf{Introduction}}

	Development in the area of particle physics and cosmology has taken the forefront in today's era of theoretical research. Although the Standard Model (SM)\cite{PhysRevLett.19.1264} aligns remarkably well with numerous experimental results, it does not account for several essential phenomena, such as the origin of neutrino masses and flavor mixing, the baryon asymmetry of the Universe, and the characteristics of dark matter and dark energy. Specifically, the observation of neutrino oscillations\cite{PhysRevLett.81.1562} has confirmed that neutrinos have small yet non-zero masses and experience flavor mixing, representing one of the most compelling pieces of evidence for theories beyond the celebrated Standard Model (BSM). In addition, SM also could not explain the intrinsic nature of neutrinos (whether Dirac or Majorana) and also mass ordering of neutrinos (whether normal or inverted). These are some open questions which strongly motivate extensions of SM which can successfully blend neutrino physics and cosmology.\\
	It is already well known that one of the most compelling mechanisms for generating light neutrino masses is the seesaw mechanism, where heavy right-handed (RH) neutrinos carrying Majorana masses are taken into account. An important consequence of taking these heavy RH states into picture is that these heavy states can dynamically generate the observed matter-antimatter asymmetry of the universe through leptogenesis. As such, an ultraviolet complete theory where the ingredients required for seesaw mechanism are naturally present will be studied in this work. The theory is generally named as left-right symmetric theory and the associated neutrino mass model is called left-right symmetric model (LRSM), which is based on the gauge group $SU(3)_{C}\otimes SU(2)_{R}\otimes SU(2)_{L} \otimes U(1)_{B-L}$\cite{Pati:1974yy,Mohapatra:1974gc,Mohapatra:1979ia,Senjanovic:1978ev}.\\
    Although the LRSM successfully explains the origin of neutrino masses, it does not determine the flavor structure of the Yukawa sector. Flavor symmetries are therefore introduced to constrain the fermion mass matrices and enhance the predictive power of the model. In this regard, modular flavor symmetry has emerged as an attractive alternative to conventional discrete flavor symmetries. By enhancing Yukawa couplings to modular forms, modular-invariant theories eliminate the demand for flavon fields and drastically reduce the number of free parameters, leading to highly predictive fermion mass models. The modular symmetric approach has recently gained utmost importance and  implementation within the LRSM thereby providing an unified description of fermion masses and mixing.
The increasing precision of neutrino oscillation experiments, including Super-Kamiokande\cite{Scholberg:1999ar}, T2K\cite{T2K:2011qtm}, NOvA\cite{Habig:2012uva}, KamLAND\cite{KamLAND:2004mhv}, and JUNO\cite{JUNO:2025gmd}, along with global assessments like NuFIT, offers demanding tests of neutrino mass theories. Future investigations, notably DUNE\cite{PhysRevLett.123.131803} and Hyper-Kamiokande\cite{Yokoyama:2017mnt}, will further increase the precision of neutrino oscillation parameters, imposing stringent limits on various theoretical frameworks.  
The neutrino mass matrix structures provides viable insight into the underlying flavor symmetry. In particular, texture-zero mass matrices garnered considerable attention because they provide predicted relations among the neutrino masses and mixing parameters. Within the framework of modular symmetric LRSM, zero textures could arise naturally from symmetry considerations and considerably enhance the predictive power of the model.\\
In this work, we aim to re-investigate neutrino mass textures in a modular symmetric Left-Right Symmetric Model utilizing the latest neutrino oscillation data and cosmological constraints. We analyze the phenomenologically viable textures, identify the permissible parameter space, and explore their repercussions for neutrino masses, mixing parameters and CP violation. Our results particularly demonstrate that modular symmetry provides a simple and predictive framework for understanding neutrino flavor within the LRSM. Also to determine whether the obtained texture-zero mass matrices align with the ongoing BSM phenomenology, we have taken into consideration the study of leptogenesis\cite{Davidson:2008bu,Buchmuller:2005eh} and neutrinoless double beta decay $(0\nu\beta\beta)$\cite{Jones:2021cga,Vergados:2012xy,Cirigliano:2022oqy}.\\
Talking about texture-zeros in the neutrino mass matrix, the number of combinations that can be obtained for a $3 \times 3$ matrix is determined using ${}^nC_m$, where $n$ is the number of independent elements in the neutrino mass matrix and $m$ represents the number of zeros. For a symmetric matrix, we have six independent elements and as such for two-zero textures, we can have ${}^6C_2=15$ combinations of the mass matrix. But, according to the latest particle physics and cosmology data only seven of them falls within the allowed regime, which are depicted below.
\begin{equation*}
    A_{1}=\begin{pmatrix}
        0 & 0 & \times\\
        0 & \times & \times\\
        \times& \times & \times
    \end{pmatrix} \hspace{1cm}
     A_{2}=\begin{pmatrix}
        0 & \times & 0\\
        \times & \times & \times\\
        0& \times & \times
    \end{pmatrix}
\end{equation*}
\begin{equation*}
    B_{1}=\begin{pmatrix}
        \times & \times & 0\\
        \times & 0 & \times\\
       0& \times & \times
    \end{pmatrix}\hspace{0.5cm}
    B_{2}=\begin{pmatrix}
        \times & 0 & \times\\
        0 & \times & \times\\
        \times& \times & 0
    \end{pmatrix}\hspace{0.5cm}
    B_{3}=\begin{pmatrix}
        \times & 0 & \times\\
        0 & 0 & \times\\
        \times& \times & \times
    \end{pmatrix}\hspace{0.5cm}
    B_{4}=\begin{pmatrix}
        \times & \times & 0\\
        \times & \times & \times\\
        0& \times & 0
    \end{pmatrix}
\end{equation*}
\begin{equation*}
C=\begin{pmatrix}
        \times & \times & \times\\
        \times & 0 & \times\\
        \times& \times & 0
    \end{pmatrix}
\end{equation*}
where, $\times$ denote the non-zero entries in the mass matrix.\\
The paper has been organized in the manner where in section \ref{lrsm12} we present a brief discussion the model framework, with subsection 
\ref{lrsm12b} showing the realization of LRSM with $A_{4}$ modular symmetry, section \ref{lrsm13} being the core of the manuscript provides detailed modular symmetric realization of two-zero textures in LRSM, in section \ref{lrsm14} we present the phenomenological analysis and results for resonant leptogenesis and neutrinoless double beta decay and section \ref{lrsm15} presents the discussion and conclusion of the work.

			\section{\label{lrsm12}\textbf{The Model Framework}}
\subsubsection{\label{lrsm12a}\textbf{Left-Right Symmetric Model}}
Consisting of one Higgs bidoublet $\phi (1,2,2,0)$ and two scalar triplets given by $\Delta_L(1,3,1,2)$ and $\Delta_R(1,1,3,2)$, the scalar sector of the model takes part in spontaneous symmetry breaking in two steps. Firstly, the LRSM gauge group is broken down to SM gauge group by VEV of $\Delta_R$, hence giving mass to heavy RH neutrinos and SM gauge group is finally broken down to $U(1)_{em}$ gauge group by VEV of $\phi$ and tiny VEV of $\Delta_L$\cite{Grimus:1993fx,BhupalDev:2018xya,Corrigan:2015kfu}.\\
The super-potential for the model incorporating the scalar sector is given by,
\begin{equation}
	\label{E:1}
	\mathcal{W} = {L_{iL}^{T}}Y_{ij}^l \phi{L_{jR}^{c}}+ f_{L,ij}{L_{L,i}}^Ti\sigma_2\Delta_L L_{L,j}+f_{R,ij}{L_{R,i}^{C}}^{T}i\sigma_2\Delta_R L_{R,j}^{C}
	\end{equation}
	\begin{equation}
		\supset M_{D}\nu_{L}\nu_{R}+M_{L}\nu_{L}\nu_{L}+M_{R}\nu_{R}\nu_{R}
	\end{equation}
where, $L_L$ and $L_R$ are the left-handed and right-handed lepton fields. $Y^l$ being the Yukawa coupling corresponding to leptons. $f_L$ and $f_R$ are the Majorana Yukawa couplings and are equal because of the discrete left-right symmetry. The family indices $i,j$ runs from $1$ to $3$ representing the three generations of the fermions. \\ 	The resultant light neutrino mass of LRSM is expressed as a sum of the type-I and type-II seesaw mass terms, given as,
\begin{equation}
	\label{E:3}
	M_{\nu} = M_{\nu}^I + M_{\nu}^{II}
\end{equation}
where,
\begin{equation}
	\label{E:4}
	M_{\nu}^I = M_{D}M_{R}^{-1}M_{D}^T
\end{equation}
is the type-I seesaw mass, and type-II seesaw mass is given by
\begin{equation}
	\label{E:5}
	M_{\nu}^{II} = M_{L}
\end{equation}
$M_D$ is the Dirac mass matrix and $M_R$ is the right-handed Majorana mass matrix, where, $M_R = \sqrt{2}v_{R}f_{R}$ and $M_L = \sqrt{2}v_{L}f_{L}$. $v_R$ and $v_{L}$ are the respective VEVs of $\Delta_R$ and $\Delta_L$. The magnitudes of the VEVs follows the relation, $|v_L|^2 < |k^{2} +k'^{2}| < |v_R|^2$.  \\
The light neutrino mass obtained can be expressed in terms of a matrix given as\cite{Borgohain:2017akh},
\begin{equation}
	M_{\nu}=\begin{pmatrix}
		M_{L} & M_{D}\\
		M_{D}^{T} & M_{R}
	\end{pmatrix}
\end{equation}
This matrix is a $6 \times 6$ matrix which can be diagonalized by a unitary matrix as follows,
\begin{equation}
	\label{E:22}
	\nu^{T}M_{\nu}\nu = \begin{pmatrix}
		\hat M_{\nu} & 0\\
		0 & \hat M_{R}
	\end{pmatrix}
\end{equation}
where, $\nu$ represents the diagonalizing matrix of the full neutrino mass matrix, $M_{\nu}$,$\hat{M_{\nu}} = diag(m_1,m_2,m_3)$, with $m_i$ being the light neutrino masses and $\hat{M_{R}} = diag(M_1,M_2,M_3)$, with $M_i$ being the heavy right-handed neutrino masses.\\
The diagonalizing matrix can be represented as,\\
\begin{equation}
	\label{E:23}
	\nu = \begin{pmatrix}
		U & S\\
		T & V
	\end{pmatrix} \approx \begin{pmatrix}
		1-\frac{1}{2}RR^\dagger & R\\
		-R^\dagger & 1-\frac{1}{2}R^\dagger R
	\end{pmatrix} \begin{pmatrix}
		V_{\nu} & 0\\
		0 & V_R
	\end{pmatrix}
\end{equation}
where, $R$ describes the left-right mixing and is given by,\\
\begin{equation}
	\label{E:24}
	R = M_{D}M_{R}^{-1} + O(M_{D}^3(M_{R}^{-1})).
\end{equation}
The matrices $U,V,S$ and $T$ are as follows,
\begin{equation}
	\label{E:25}
	U = [1-\frac{1}{2}M_{D}M_{R}^{-1}(M_D M_{R}^{-1})^\dagger]V_{\nu}
\end{equation}
\begin{equation}
	\label{E:26}
	V = [1-\frac{1}{2}(M_{D}M_{R}^{-1})^{\dagger} M_D M_{R}^{-1}]V_{\nu}
\end{equation}
\begin{equation}
	\label{E:27}
	S = M_D M_{R}^{-1} v_{R} f_{R}
\end{equation}
\begin{equation}
	\label{E:28}
	T = -(M_D M_{R}^{-1})^\dagger V_{\nu}
\end{equation}
\subsubsection{\label{lrsm12b}\textbf{LRSM with $A_{4}$  modular symmetry}}
Modular symmetry has gained utmost importance in the area of model building for the fact that, the use of modular symmetry constrains the use of extra fields for realization of a particular model\cite{Feruglio:2017spp,Ding:2023htn,Kobayashi:2024hkk,CentellesChulia:2023osj}.  When using modular symmetry, the Yukawa couplings are expressed in terms of modular Yukawa forms $Y$. In the current work, we have illustrated the implementation of $\Gamma_{3}$ modular group in the context of LRSM. $\Gamma_{3}$ modular group is isomorphic to non-abelian discrete symmetry group $A_{4}$ and as such, the particle content within the model will be assigned respective charges under the LRSM gauge group and each of the particle will have a corresponding modular weight. \\
After incorporating $A_{4}$ modular symmetry, the superpotential of LRSM will be expressed in terms of modular Yukawa forms as shown in equation 
\begin{equation}
	\label{e1}
	\mathcal{W}=Y_{LR}L_{L}^{T}\phi L_{R}^{c}+Y_{L}L_{L}^{T}i\sigma_2\Delta_L L_{L}+Y_{R}L_{R}^{{c}^{T}} i\sigma_2\Delta_R L_{R}^{c}
\end{equation}
where, $Y_{LR}$ represents the modular Yukawa form for Dirac mass term and similarly $Y_{L}$ and $Y_{R}$ are respectively couplings for left-handed and right-handed Majorana mass terms.\\
The number of modular forms will depend upon the weight of the group under consideration. The number of modular forms required for the construction of a model under modular symmetry is given in table \ref{t1}.
\begin{table}[H]
	\begin{center}
		\begin{tabular}{|c|c|c|}
			\hline
			N & No. of modular forms & $\Gamma(N)$ \\
			\hline
			2 & k + 1 & $S_3$ \\
			\hline
			3 & 2k + 1 & $A_4$ \\
			\hline
			4 & 4k + 1 & $S_4$ \\
			\hline
			5 & 10k + 1 & $A_5$ \\
			\hline 
			6 & 12k &  \\
			\hline
			7 & 28k - 2 & \\
			\hline
		\end{tabular}
		\caption{\label{t1}No. of modular forms corresponding to modular weight 2k.}
	\end{center}
\end{table}
The charge assignments for the particle content of the model with incorporation of modular symmetry is given in table \ref{t2}.
\begin{table}[H]
	\begin{center}
		\begin{tabular}{|c|c|c|c|c|c|c|}
			\hline
			Gauge group & $L_L$ & $L_R^{c}$ & $\phi$ & $\Delta_L$ & $\Delta_R$ \\
			\hline
			$SU(3)_C$ & 1 & 1 & 1 & 1 & 1\\
			\hline
			$SU(2)_L$ & 2 & 1 & 2 & 3 & 1 \\
			\hline
			$SU(2)_R$ & 1 & 2 & 2 & 1 & 3 \\
			\hline
			$U(1)_{B-L}$ & -1 & -1 & 0 & 2 & 2 \\
			\hline 
		\end{tabular}
		\caption{\label{t2}Charge assignments for the particle content of the model.}
	\end{center}
\end{table}

Our work mainly revolves around the study of assigning different modular weights to the particle content of the model which consequently gives rise to two-zero textures in the resulting light neutrino mass matrix. The succeeding section elaborately describes the origin of two-zero textures in the neutrino mass matrix as a result of varying modular weights.
\section{\label{lrsm13}\textbf{Two-zero textures in modular $A_{4}$ LRSM}}
\subsection{For $k_{{Y}_{max}}$=4}
\begin{figure}[H]
	\centering
	\includegraphics[scale=0.25]{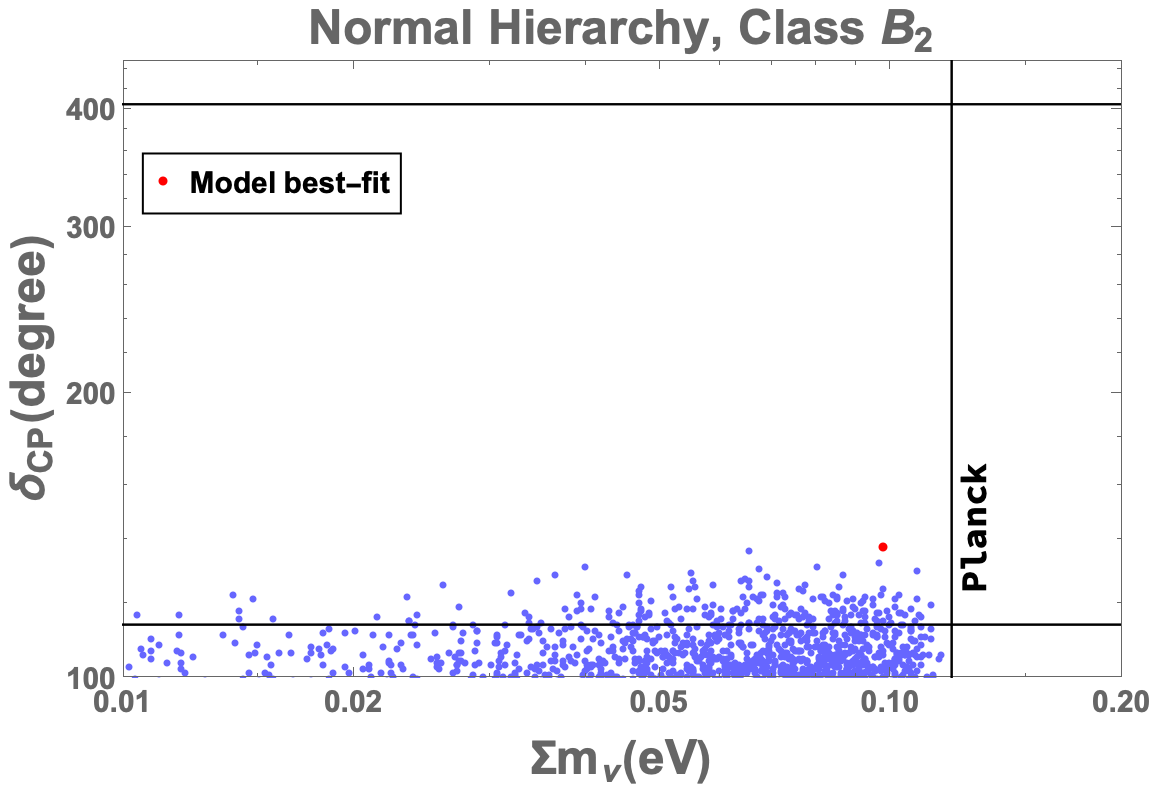}
	\includegraphics[scale=0.25]{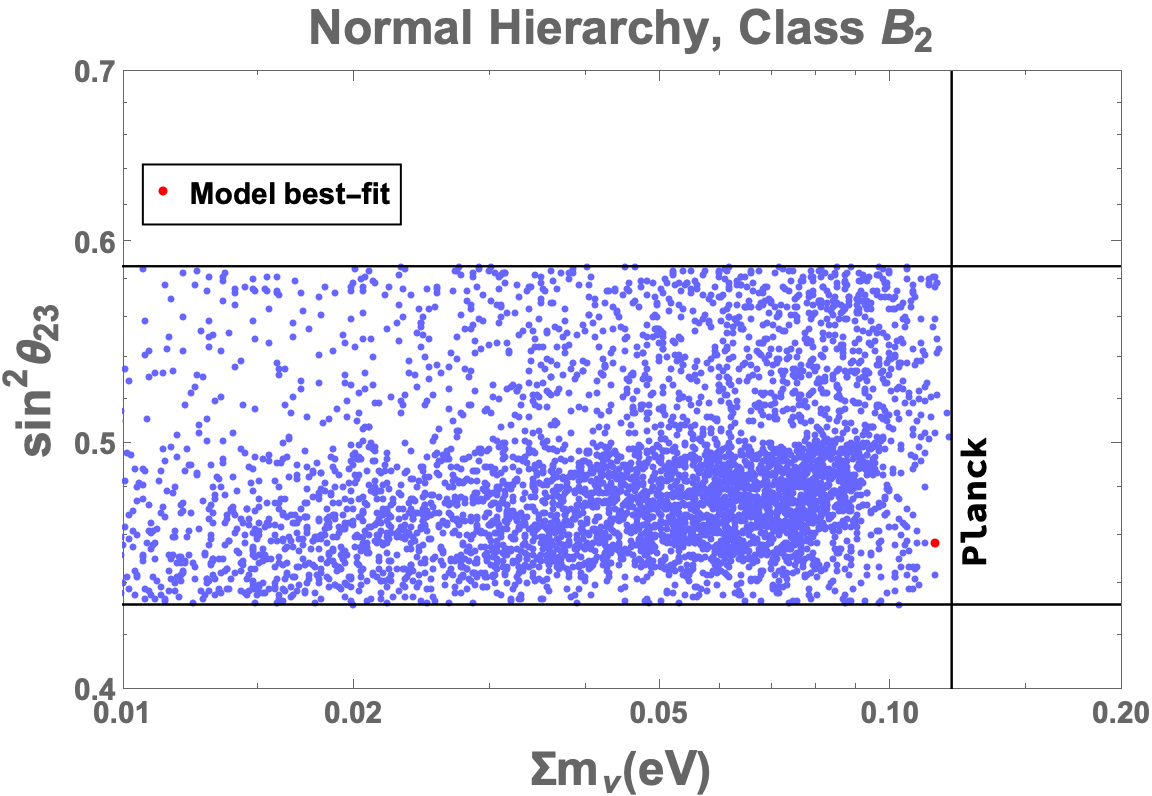}
		\includegraphics[scale=0.25]{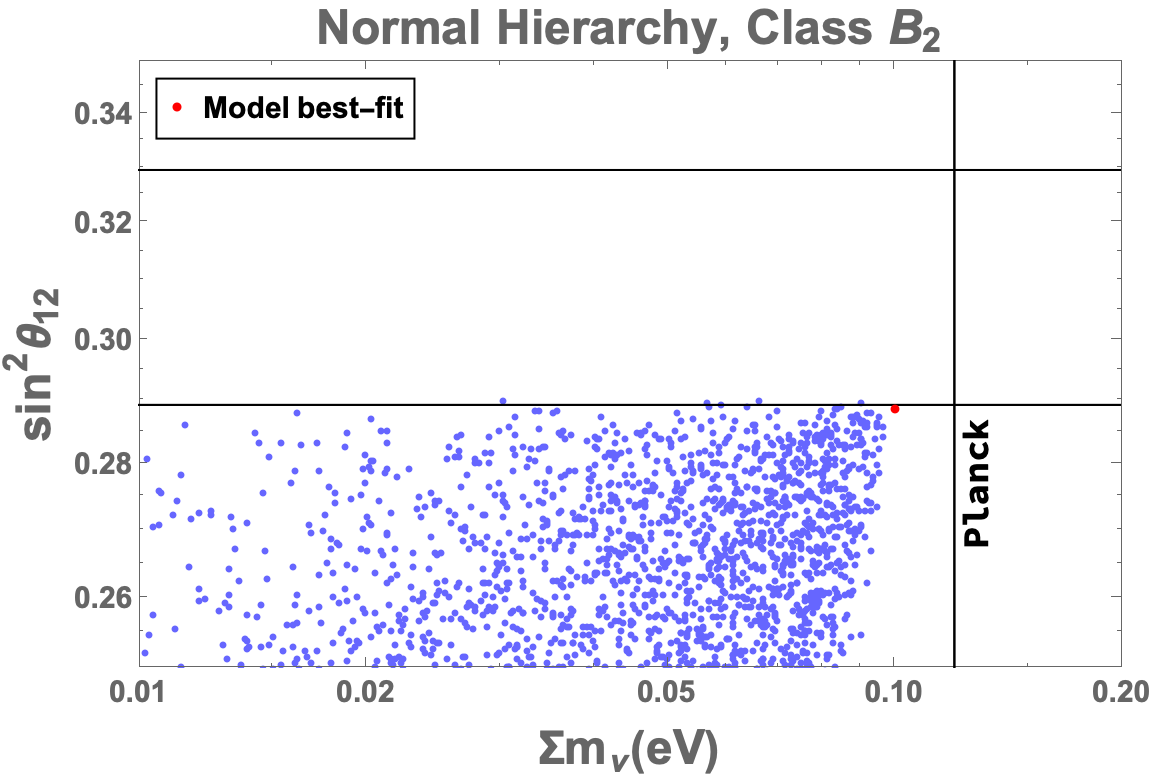}
	\includegraphics[scale=0.25]{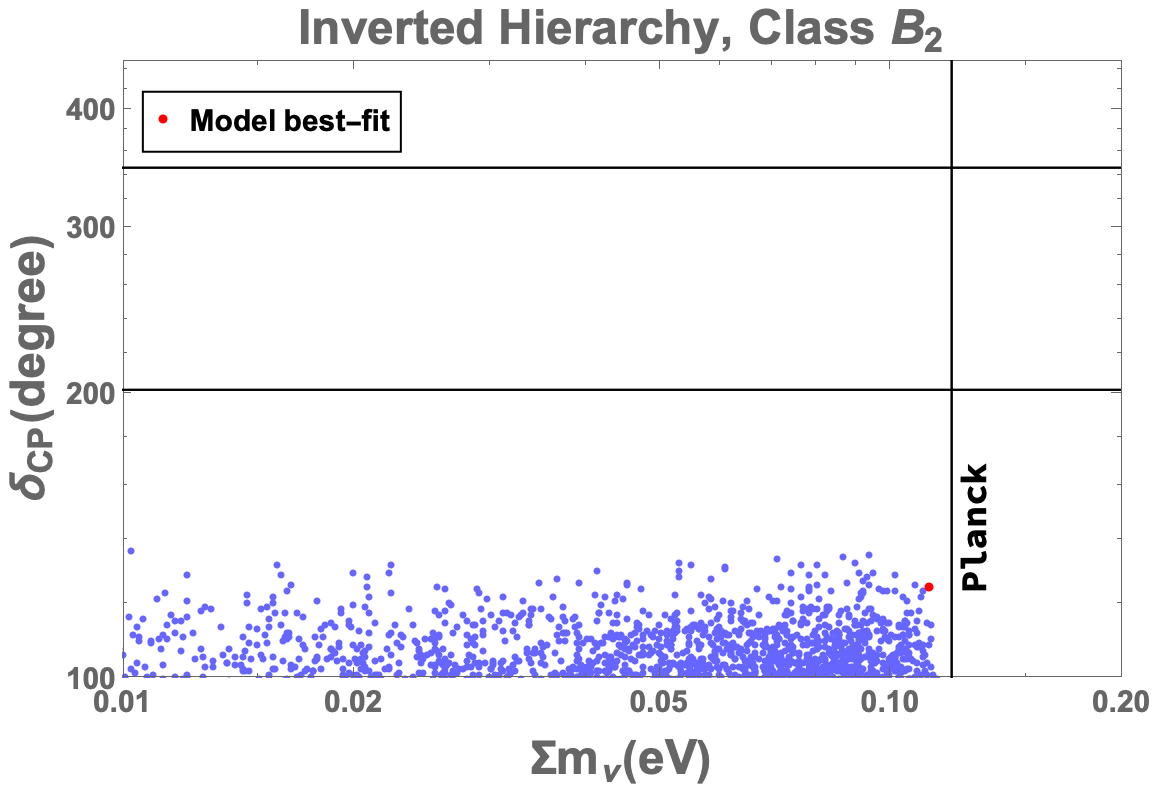}
		\includegraphics[scale=0.25]{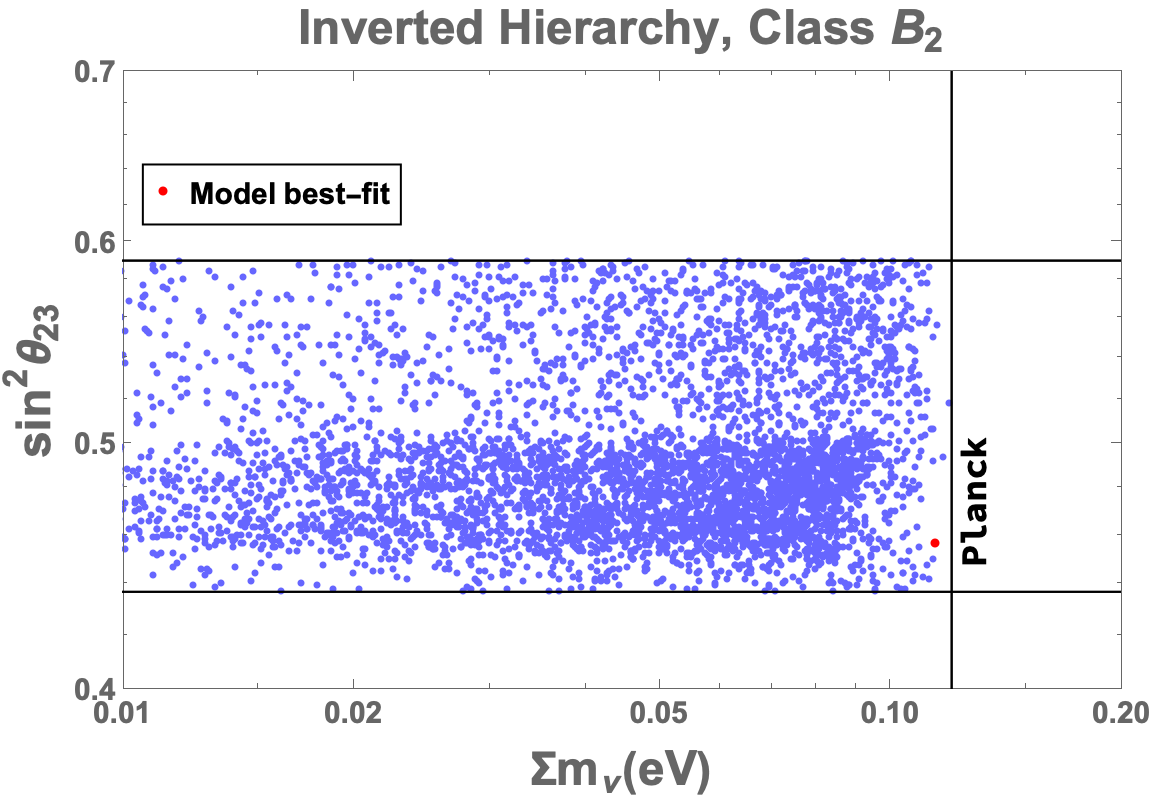}
	\includegraphics[scale=0.25]{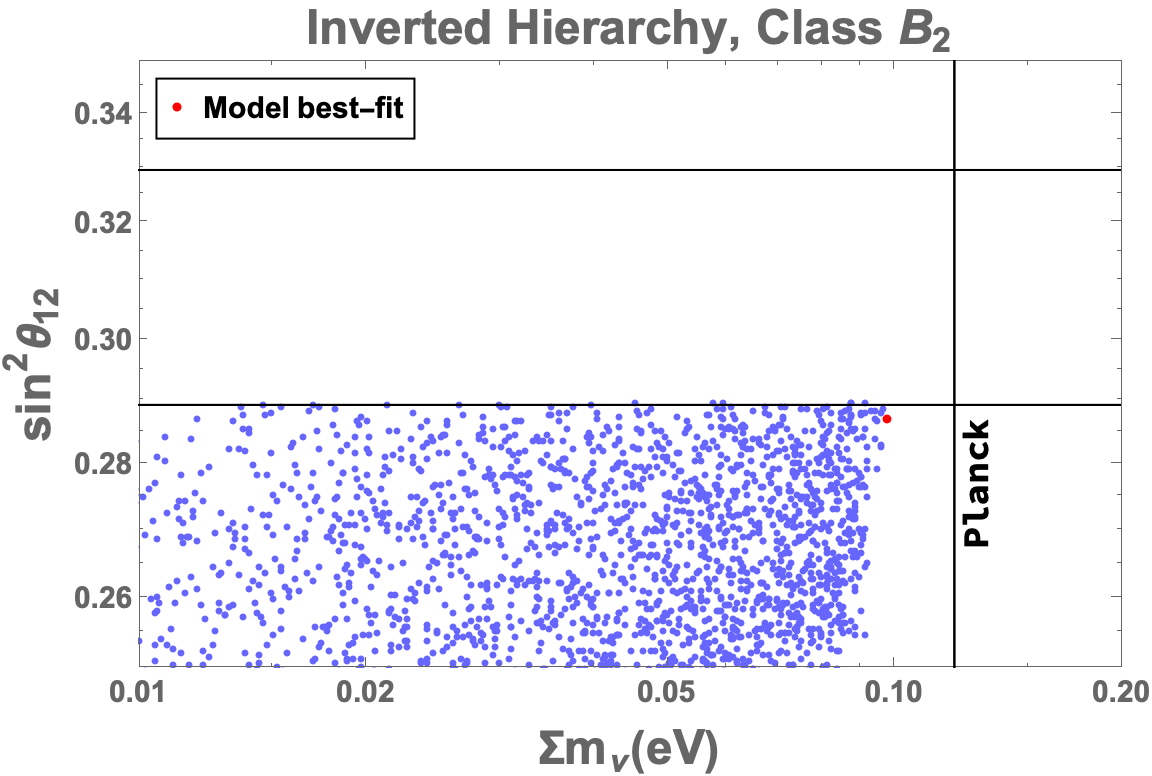}
		\caption {\label{f1} Neutrino oscillation parameters with sum of neutrino masses, where the black horizontal lines depict the 3$\sigma$ range\cite{Esteban:2024eli} and the vertical line represents the Planck bound on $\Sigma m_{\nu}$\cite{Planck:2018vyg}. The red dot represents the model best-fit value.}
	\end{figure}

For $k_{Y}=4$, the modular Yukawa forms have  two singlets $1$, $1'$ and one triplet $3$, which are expressed in terms of $(Y_{1},Y_{2},Y_{3})$ as\cite{Zhang:2019ngf},
\begin{equation}
	\label{ee1}
	Y_{(1)}^{4} = Y_{1}^{2} + 2 Y_{2} Y_{3};
	Y_{(1')}^{4} = Y_{3}^{2} + 2 Y_{1} Y_{2};
	Y_{(3)}^{4} = \begin{pmatrix}
		Y_{1}^{2}-Y_{2}Y_{3}\\
		Y_{3}^{2}-Y_{1}Y_{2}\\
		Y_{2}^{2}-Y_{1}Y_{3}
	\end{pmatrix}
\end{equation}
\begin{table}[H]
	\begin{center}
		\begin{tabular}{|c|c|c|c|c|c|c|c|c|c|}
			\hline
			Gauge group & $L_{L_1}$ & $L_{L_2}$ & $L_{L_3}$ & $L_{R_1}^{c}$ & $L_{R_2}^{c}$ & $L_{R_3}^{c}$ & $\phi$ & $\Delta_L$ & $\Delta_R$ \\
			\hline
			$A_{4}$ & 1 & $1^{\prime}$ & $1^{\prime\prime}$ & 1 & $1^{\prime\prime}$ & $1^{\prime}$ & 1 &1 & 1\\
			\hline
			$k$ & -2 & -2 & -2 & -2 & -2 & -2 & 0 & 0 & 0 \\
			\hline
		\end{tabular}
		\caption{\label{t3}Charge assignments for the particle content of the model.}
	\end{center}
\end{table}
The superpotential associated with the left-right coupling (Dirac mass term) is given by,
\begin{equation}
	\begin{split}
	\label{e19}
	\mathcal{W_{D}}=L_{L_1}^{T}\phi L_{R_1}^{c}Y_{1}^{4}+L_{L_1}^{T}\phi L_{R_2}^{c}Y_{1'}^{4}+L_{L_2}^{T}\phi L_{R_2}^{c}Y_{1}^{4}\\ +L_{L_2}^{T}\phi L_{R_3}^{c}Y_{1'}^{4}+L_{L_3}^{T}\phi L_{R_1}^{c}Y_{1'}^{4}+L_{L_3}^{T}\phi L_{R_3}^{c}Y_{1}^{4}
\end{split}
\end{equation}
From equation \eqref{e19}, the Dirac mass matrix can be given as,
\begin{equation}
	\label{e20}
	M_{D}=v\begin{pmatrix}
	Y_{1}^{4} & Y_{1'}^{4} & 0\\
	0 & Y_{1}^{4} & Y_{1'}^{4}\\
	Y_{1'}^{4} & 0 & Y_{1}^{4}
	\end{pmatrix}
\end{equation}
For right-handed neutrino Majorana mass, the superpotential is given as,
\begin{equation}
	\begin{split}	
	\label{e21}
	\mathcal{W_{R}}=L_{R_1}^{c{T}} i\sigma_2 \Delta_{R}L_{R_1}^{c}Y_{1}^{4}+L_{R_1}^{c{T}} i\sigma_2 \Delta_{R} L_{R_2}^{c}Y_{1'}^{4}+L_{R_2}^{c{T}} i\sigma_2 \Delta_{R} L_{R_1}^{c}Y_{1'}^{4}\\+L_{R_2}^{c{T}} i\sigma_2 \Delta_{R} L_{R_3}^{c}Y_{1}^{4}+L_{R_3}^{c{T}} i\sigma_2 \Delta_{R} L_{R_2}^{c}Y_{1}^{4}+L_{R_3}^{c{T}} i\sigma_2  \Delta_{R}L_{R_3}^{c}Y_{1'}^{4}
\end{split}
\end{equation}
The right-handed Majorana mass matrix is given as,
\begin{equation}
	\label{e22}
	M_{R}=v_{R}\begin{pmatrix}
		Y_{1}^{4} & Y_{1'}^{4} & 0\\
		Y_{1'}^{4} &0 & Y_{1}^{4}\\
		0 & Y_{1}^{4} & Y_{1'}^{4}
	\end{pmatrix}
\end{equation}
The left-handed Majorana mass matrix is given as,
\begin{equation}
	\begin{split}
	\label{e23}
	\mathcal{W_{L}}=L_{L_1}^{c{T}} i\sigma_2\Delta_{L} L_{L_1}^{c}Y_{1}^{4}+L_{L_1}^{c{T}} i\sigma_2 \Delta_{L}L_{L_3}^{c}Y_{1}^{4}+L_{L_2}^{c{T}} i\sigma_2\Delta_{L} L_{L_2}^{c}Y_{1'}^{4}\\+L_{L_2}^{c{T}} i\sigma_2 \Delta_{L}L_{L_3}^{c}Y_{1}^{4}+L_{L_3}^{c{T}} i\sigma_2 \Delta_{L}L_{L_1}^{c}Y_{1'}^{4}+L_{L_3}^{c{T}} i\sigma_2 \Delta_{L}L_{L_2}^{c}Y_{1}^{4}
\end{split}
\end{equation}
\begin{equation}
	\label{e24}
	M_{L}=v_{L}\begin{pmatrix}
		Y_{1}^{4} & 0 &  Y_{1'}^{4}\\
		 0 &Y_{1'}^{4}& Y_{1}^{4}\\
		Y_{1'}^{4} & Y_{1}^{4} & 0
	\end{pmatrix}
\end{equation}
The resulting light neutrino mass given as a summation of type-I and type-II seesaw masses is given as,
\begin{equation}
	\label{e25}
	\mathcal{M_{\nu}}= \frac{v^2+v_{L}v_{R}}{v_{R}}\begin{pmatrix}
		Y_{1}^{2}+2 Y_{2}Y_{3} & 0 & 2 Y_{1}Y_{2}+Y_{3}^{2}\\
		0 &2 Y_{1}Y_{2}+Y_{3}^{2} & Y_{1}^{2}+2 Y_{2}Y_{3} \\
	2 Y_{1}Y_{2}+Y_{3}^{2} & Y_{1}^{2}+2 Y_{2}Y_{3} & 0	
			\end{pmatrix}
\end{equation}
Equation \eqref{e25} represents \textbf{Class $B_2$} of two-zero neutrino mass texture.
To check the viability of the texture obtained, we have obtained the data points corresponding to neutrino oscillation parameters and checked whether they lie in the Nufit range. For Class $B_{2}$, the plots are depicted in figure \ref{f1}.
For $k_{Y_{max}}=4$, we can now interchange the $A_{4}$ charge and weight assignments for fields $L_{L_2} \leftrightarrow L_{L_3}$, and $L_{R_2} \leftrightarrow L_{R_3}$, which as we will see will lead to Class $B_{1}$ of two-zero texture. The superpotential in this case is given as,
\begin{equation}
	\begin{split}
		\label{e26}
		\mathcal{W_{D}}=L_{L_1}^{T}\phi L_{R_1}^{c}Y_{1}^{4}+L_{L_1}^{T}\phi L_{R_3}^{c}Y_{1'}^{4}+L_{L_2}^{T}\phi L_{R_1}^{c}Y_{1'}^{4}\\ +L_{L_2}^{T}\phi L_{R_2}^{c}Y_{1}^{4}+L_{L_3}^{T}\phi L_{R_2}^{c}Y_{1'}^{4}+L_{L_3}^{T}\phi L_{R_3}^{c}Y_{1}^{4}
	\end{split}
\end{equation}
From equation \eqref{e26}, the Dirac mass matrix can be written as,
\begin{equation}
	\label{e27}
	M_{D}=v\begin{pmatrix}
		Y_{1}^{4} & 0 & Y_{1'}^{4}\\
		Y_{1'}^{4} & Y_{1}^{4} & 0\\
		 0 & Y_{1'}^{4} & Y_{1}^{4}
	\end{pmatrix}
\end{equation}
The superpotential corresponding to right-handed neutrino mass matrix is given by equation \eqref{e28},
\begin{equation}
	\begin{split}	
		\label{e28}
		\mathcal{W_{R}}=L_{R_1}^{c{T}} i\sigma_2 \Delta_{R}L_{R_1}^{c}Y_{1}^{4}+L_{R_1}^{c{T}} i\sigma_2 \Delta_{R} L_{R_3}^{c}Y_{1'}^{4}+L_{R_2}^{c{T}} i\sigma_2 \Delta_{R} L_{R_2}^{c}Y_{1}^{4}\\+L_{R_2}^{c{T}} i\sigma_2 \Delta_{R} L_{R_3}^{c}Y_{1}^{4}+L_{R_3}^{c{T}} i\sigma_2 \Delta_{R} L_{R_1}^{c}Y_{1'}^{4}+L_{R_3}^{c{T}} i\sigma_2  \Delta_{R}L_{R_2}^{c}Y_{1}^{4}
	\end{split}
\end{equation}
\begin{equation}
	\label{e29}
	M_{R}=v_{R}\begin{pmatrix}
		Y_{1}^{4} & 0 & Y_{1'}^{4}\\
		 0 & Y_{1'}^{4} & Y_{1}^{4}\\
		Y_{1'}^{4} & Y_{1}^{4} & 0
	\end{pmatrix}
\end{equation}
For left-handed neutrinos, the following is the superpotential and corresponding mass matrix,
\begin{equation}
	\begin{split}
		\label{e30}
		\mathcal{W_{L}}=L_{L_1}^{{T}} i\sigma_2\Delta_{L} L_{L_1}Y_{1}^{4}+L_{L_1}^{{T}} i\sigma_2 \Delta_{L}L_{L_2}Y_{1'}^{4}+L_{L_2}^{{T}} i\sigma_2\Delta_{L} L_{L_1}Y_{1'}^{4}\\+L_{L_2}^{{T}} i\sigma_2 \Delta_{L}L_{L_3}Y_{1}^{4}+L_{L_3}^{{T}} i\sigma_2 \Delta_{L}L_{L_2}Y_{1}^{4}+L_{L_3}^{{T}} i\sigma_2 \Delta_{L}L_{L_3}Y_{1'}^{4}
	\end{split}
\end{equation}
\begin{equation}
	\label{e31}
	M_{L}=v_{L}\begin{pmatrix}
		Y_{1}^{4} & Y_{1'}^{4} & 0 \\
		Y_{1'}^{4} & 0  & Y_{1}^{4}\\
		0 & Y_{1}^{4} & Y_{1'}^{4}
	\end{pmatrix}
\end{equation}
The resulting light neutrino mass matrox is given as,
\begin{equation}
	\label{e32}
	\mathcal{M_{\nu}}= \frac{v^2+v_{L}v_{R}}{v_{R}}\begin{pmatrix}
		Y_{1}^{2}+2 Y_{2}Y_{3} & 2 Y_{1}Y_{2}+Y_{3}^{2} & 0 \\
		2 Y_{1}Y_{2}+Y_{3}^{2} & 0 & Y_{1}^{2}+2 Y_{2}Y_{3} \\
		0 & Y_{1}^{2}+2 Y_{2}Y_{3} & 2 Y_{1}Y_{2}+Y_{3}^{2}
	\end{pmatrix}
\end{equation}
\begin{figure}[H]
	\centering

	\includegraphics[scale=0.3]{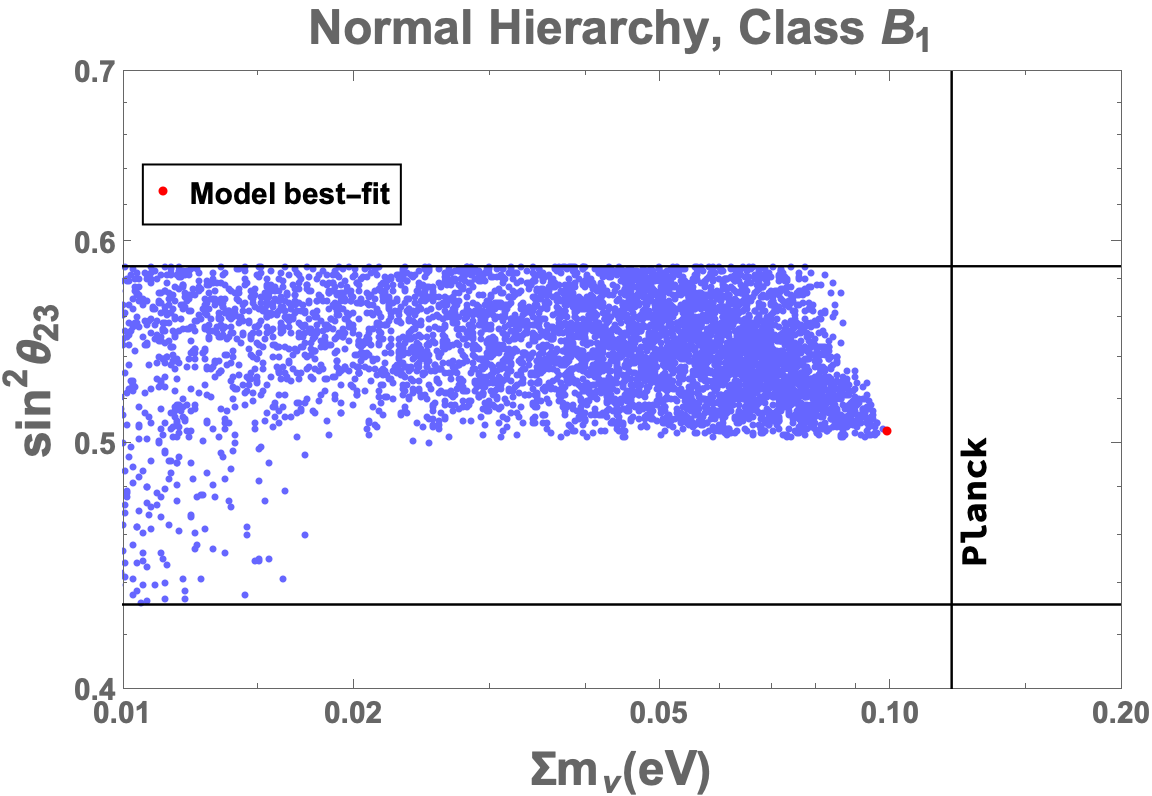}
	\includegraphics[scale=0.3]{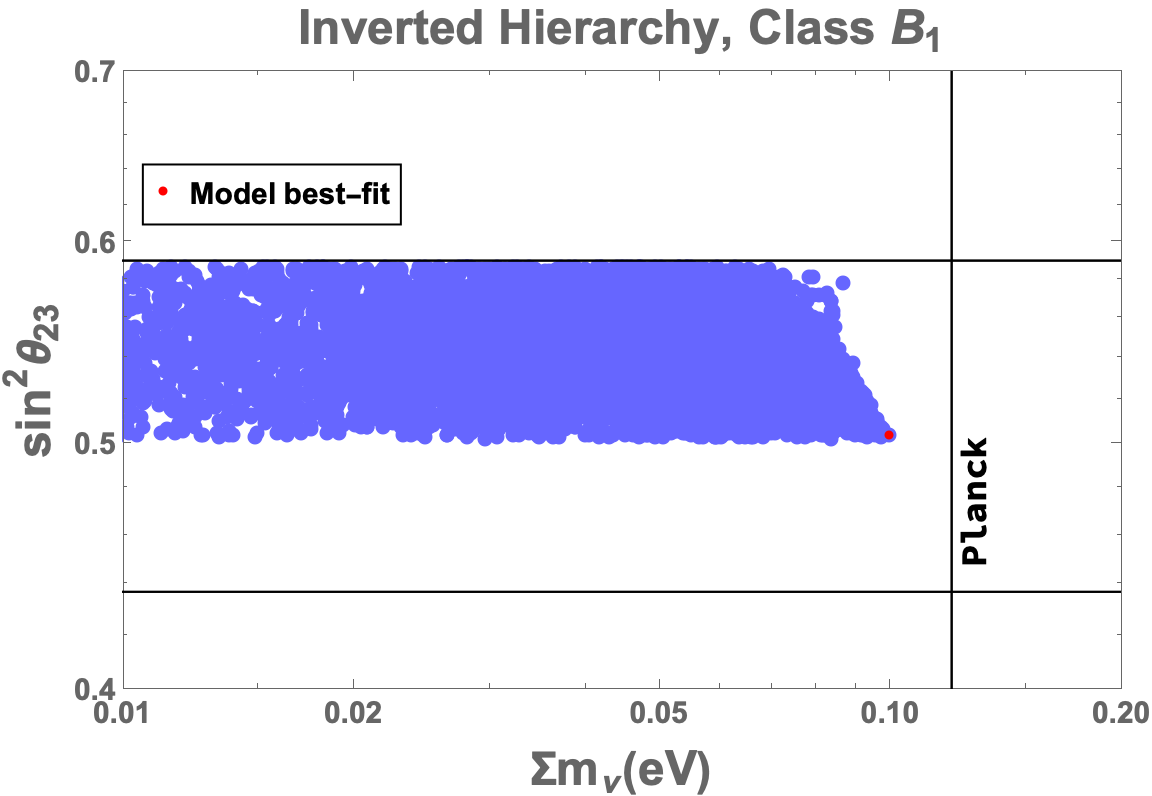}
	\includegraphics[scale=0.3]{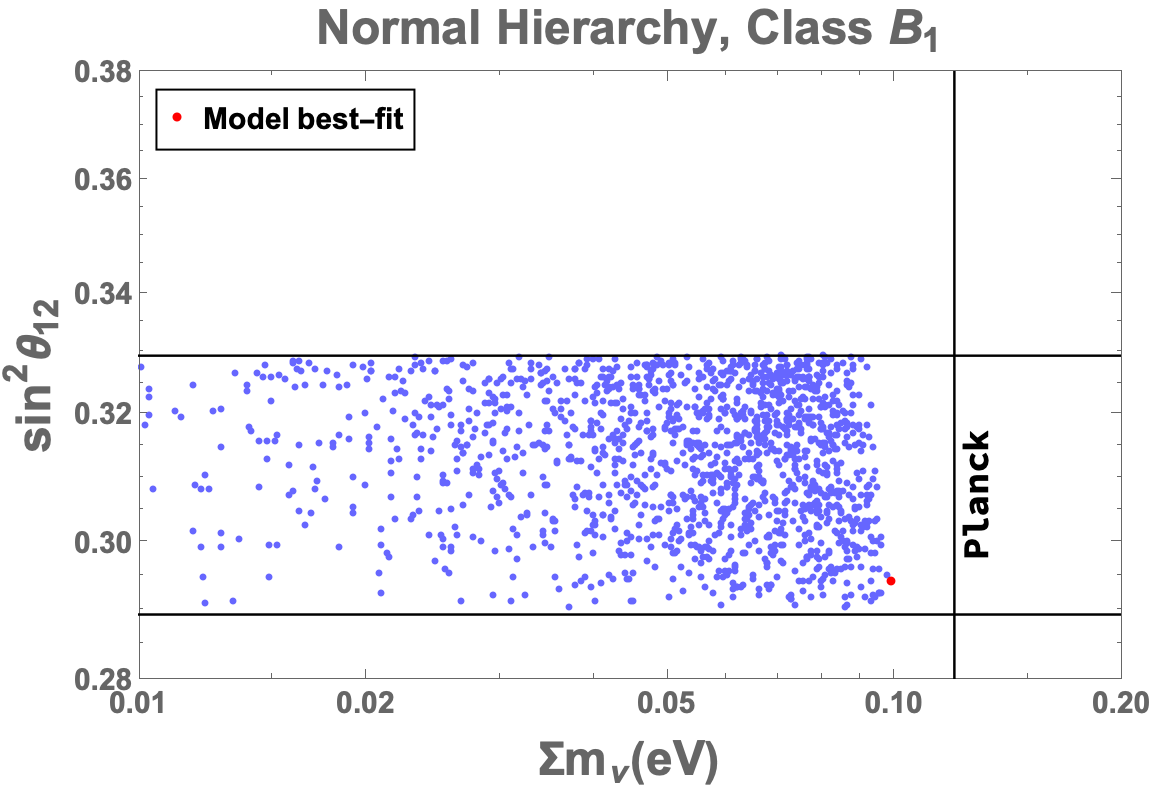}
	\includegraphics[scale=0.3]{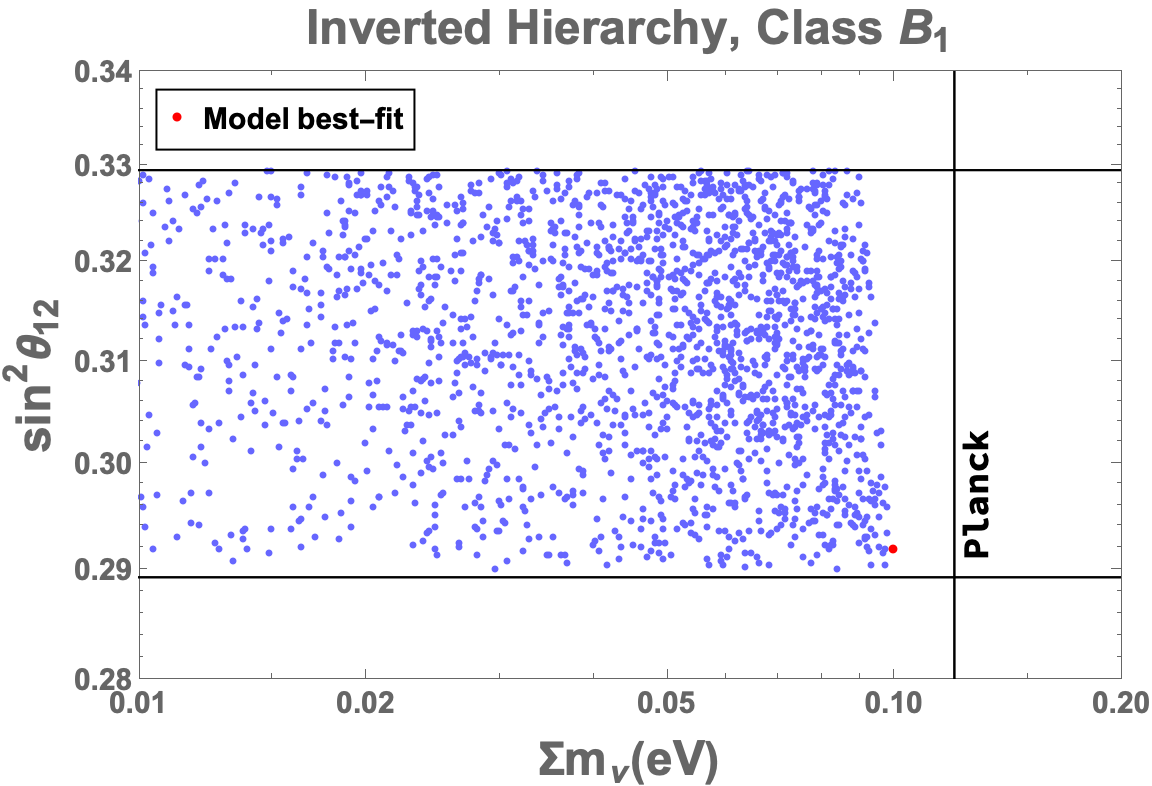}
	\caption{\label{f2} Neutrino oscillation parameters with sum of neutrino masses, where the black horizontal lines depict the 3$\sigma$ range\cite{Esteban:2024eli} and the vertical line represents the Planck bound on $\Sigma m_{\nu}$\cite{Planck:2018vyg}. The red dot represents the model best-fit value.}
\end{figure}
For Class $B_{1}$, no data points satisfying the 3$\sigma$ range for $\delta_{CP}$ was obatined for both normal and inverted mass orderings. As such figure \ref{f2} shows only the variation of atmospheric mixing angle and solar mixing angle with sum of neutrino masses.
\subsection{For $k_{Y_{max}}=8$}
For $k_{Y}=8$, there will be nine number of modular forms, three singlets $1$,$1'$,$1''$ and two triplets $3_{1}$, $3_{2}$ under $A_{4}$, represented as,
\begin{equation*}
	\label{}
	Y_{(1)}^{8} = (Y_{1}^{2}+2Y_{2}Y_{3})^{2};
	Y_{(1')}^{8} = (Y_{1}^{2}+2Y_{2}Y_{3})(Y_{3}^{2}+2Y_{1}Y_{2});
	Y_{(1'')}^{8} = (Y_{3}^{2}+2Y_{1}Y_{3})^{2}
\end{equation*}
\begin{equation}
	\label{ee2}
		Y_{(3_{1})}^{8} = (Y_{1}^{2}+2Y_{2}Y_{3})\begin{pmatrix}
		Y_{1}^{2}-Y_{2}Y_{3}\\
		Y_{3}^{2}-Y_{1}Y_{2}\\
		Y_{2}^{2}-Y_{1}Y_{3}
	\end{pmatrix};
		Y_{(3_{2})}^{8} = (Y_{3}^{2}+2Y_{1}Y_{2})\begin{pmatrix}
		Y_{2}^{2}-Y_{1}Y_{3}\\
		Y_{1}^{2}-Y_{2}Y_{3}\\
		Y_{3}^{2}-Y_{1}Y_{2}
	\end{pmatrix}
	\end{equation}
	For $k_{Y}=6$, there are two triplets and one singlet, and in this case we will be using only the singlet under weight 6, which is represented as,
	\begin{equation}
		Y_{(1)}^{6}=Y_{1}^{3}+Y_{1}^{3}+Y_{1}^{3}-3Y_{1}Y_{2}Y_{3}
	\end{equation}
	\begin{table}[H]
		\begin{center}
			\begin{tabular}{|c|c|c|c|c|c|c|c|c|c|}
				\hline
				Gauge group & $L_{L_1}$ & $L_{L_2}$ & $L_{L_3}$ & $L_{R_1}^{c}$ & $L_{R_2}^{c}$ & $L_{R_3}^{c}$ & $\phi$ & $\Delta_L$ & $\Delta_R$ \\
				\hline
				$A_{4}$ & 1 & $1^{\prime}$ & $1^{\prime\prime}$ & 1 & $1^{\prime\prime}$ & $1^{\prime}$ & 1 &1 & 1\\
				\hline
				$k$ & -4 & -4 & -2 & 0 & -2 & -2 & 0 & 0 & 0 \\
				\hline
			\end{tabular}
			\caption{\label{t4}Charge assignments for the particle content of the model.}
		\end{center}
	\end{table}
	The superpotential corresponding to Dirac mass term as well as right-handed and left-handed Majorana terms is given by,
	\begin{equation}
		\begin{split}
		\label{e33}
		\mathcal{W}_{B_4}=L_{L_1}^{T}\phi L_{R_1}^{c}Y_{1}^{4}+L_{L_2}^{T}\phi L_{R_2}^{c}Y_{1}^{6}+L_{L_3}^{T}\phi L_{R_3}^{c}Y_{1}^{4}+L_{R_1}^{c{T}} i\sigma_2 \Delta_{R}L_{R_1}^{c}Y_{1}^{0}\\+(L_{R_2}^{c{T}} i\sigma_2 \Delta_{R} L_{R_3}^{c}Y_{1}^{4}+L_{R_3}^{c{T}} i\sigma_2 \Delta_{R} L_{R_2}^{c}Y_{1}^{4})+L_{R_3}^{c{T}} i\sigma_2 \Delta_{R} L_{R_3}^{c}Y_{1'}^{4}\\+L_{L_1}^{{T}} i\sigma_2\Delta_{L} L_{L_1}Y_{1}^{8}+(L_{L_1}^{{T}} i\sigma_2 \Delta_{L}L_{L_2}Y_{1^{\prime\prime}}^{8}+L_{L_2}^{{T}} i\sigma_2\Delta_{L} L_{L_1}Y_{1^{\prime\prime}}^{8})\\+L_{L_2}^{{T}} i\sigma_2 \Delta_{L}L_{L_2}Y_{1'}^{8}+(L_{L_2}^{{T}} i\sigma_2 \Delta_{L}L_{L_3}Y_{1}^{6}+L_{L_3}^{{T}} i\sigma_2 \Delta_{L}L_{L_2}Y_{1}^{6})
		\end{split}
	\end{equation}
\begin{equation}
	M_D =
	\begin{pmatrix}
		Y_{1}^{4} & 0 & 0 \\
		0 & Y_{1}^{6} & 0\\
		0 & 0 & Y_{1}^{4}
	\end{pmatrix},
	\qquad
	M_R =
	\begin{pmatrix}
		1 & 0 & 0\\
		0 &	0 & Y_{1}^{4}\\
		0 &  Y_{1}^{4} &  Y_{1'}^{4}
	\end{pmatrix},
	\qquad
		M_L =
	\begin{pmatrix}
		Y_{1}^{8} & Y_{1^{\prime\prime}}^{8}  & 0\\
		Y_{1^{\prime\prime}}^{8} & Y_{1'}^{8} & Y_{1}^{6}\\
		0 &  Y_{1}^{6} &  0
	\end{pmatrix}
\end{equation}
$Y_{1}^{0}$ is a constant, which appears as unity in the $11$ element of right-handed Majorana mass matrix.
The resulting light neutrino mass matrix is given as,
\begin{equation}
	\resizebox{\textwidth}{!}{$
	\label{e34}
	M_{\nu}=\begin{pmatrix}
		\frac{v^{2}+v_{L}v_{R}}{v_{R}}(Y_{1}^{2}+2 Y_{2}Y_{3})^{2} & v_{L}(2 Y_{1}Y_{2}+Y_{3})^{2} & 0\\
		v_{L}(2 Y_{1}Y_{2}+Y_{3})^{2} & \frac{(2 Y_{1}Y_{2}+Y_{3}^{2})\Bigg(v_{L}(Y_{1}^{2}+2 Y_{2}Y_{3})^{3}-\frac{v^{2}(Y_{1}^{3}+Y_{2}^{3}+Y_{3}^{3}-3Y_{1}Y_{2}Y_{3})^{2}}{v_{R}}\Bigg)}{(Y_{1}^{2}+2 Y_{2}Y_{3})^{2}} & \frac{(v^{2}+v_{L}v_{R})(Y_{1}^{3}+Y_{2}^{3}+Y_{3}^{3}-3Y_{1}Y_{2}Y_{3})}{v_{R}}\\
		0 & \frac{(v^{2}+v_{L}v_{R})(Y_{1}^{3}+Y_{2}^{3}+Y_{3}^{3}-3Y_{1}Y_{2}Y_{3})}{v_{R}} &0 
	\end{pmatrix}
	$}
\end{equation}
From equation \eqref{e34}, it is seen that the charge and modular weights assigned above results in Class $B_{4}$ of two-zero texture.
\begin{figure}[H]
	\centering
	\includegraphics[scale=0.3]{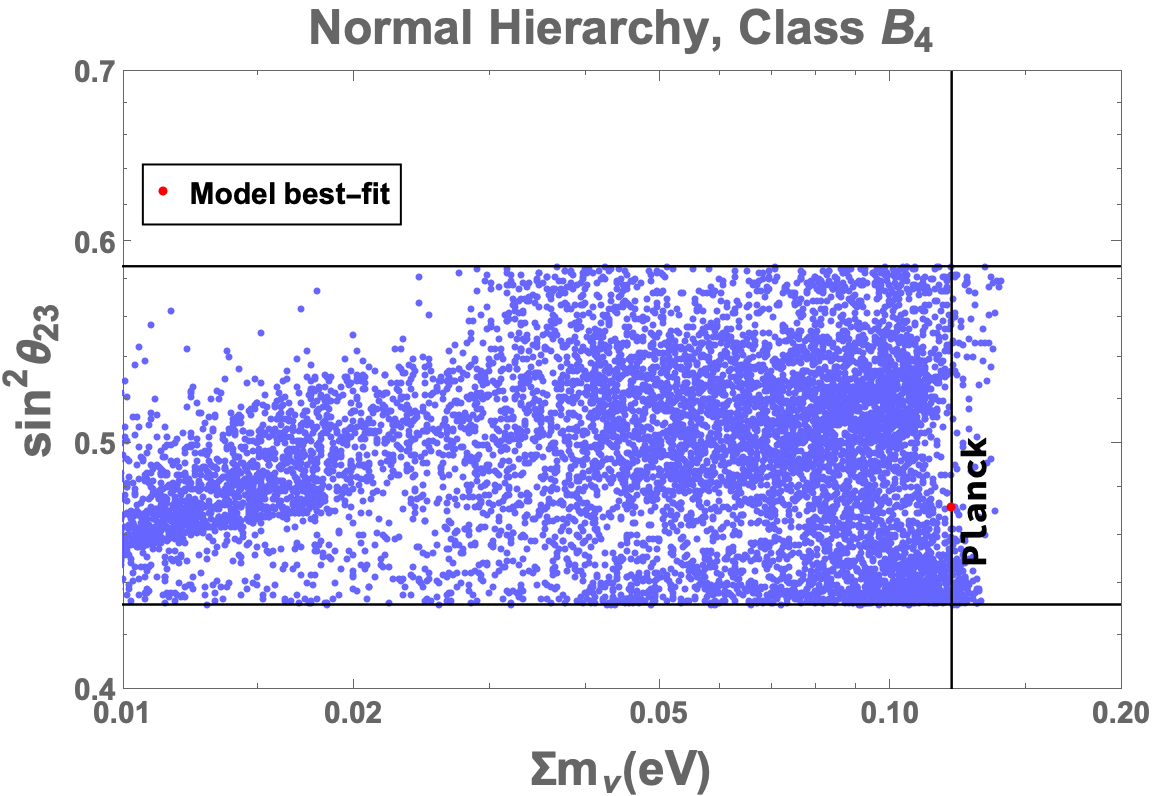}
	\includegraphics[scale=0.3]{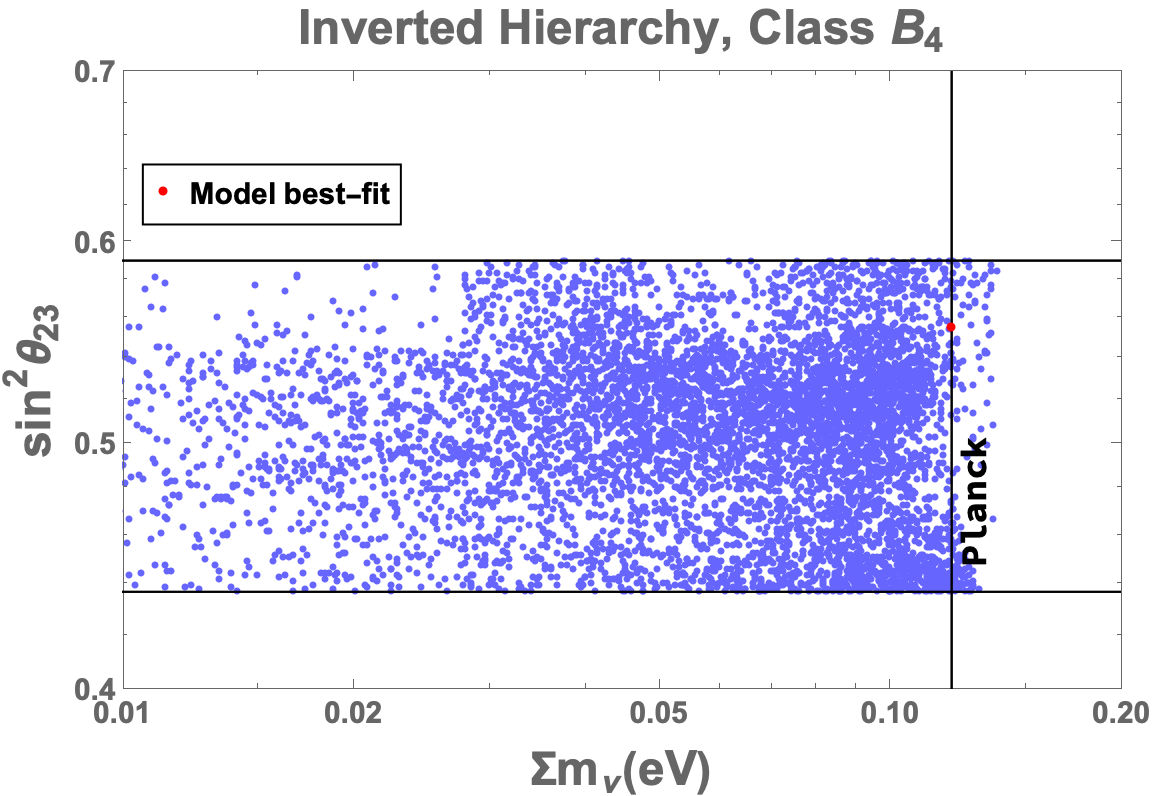}
	\includegraphics[scale=0.3]{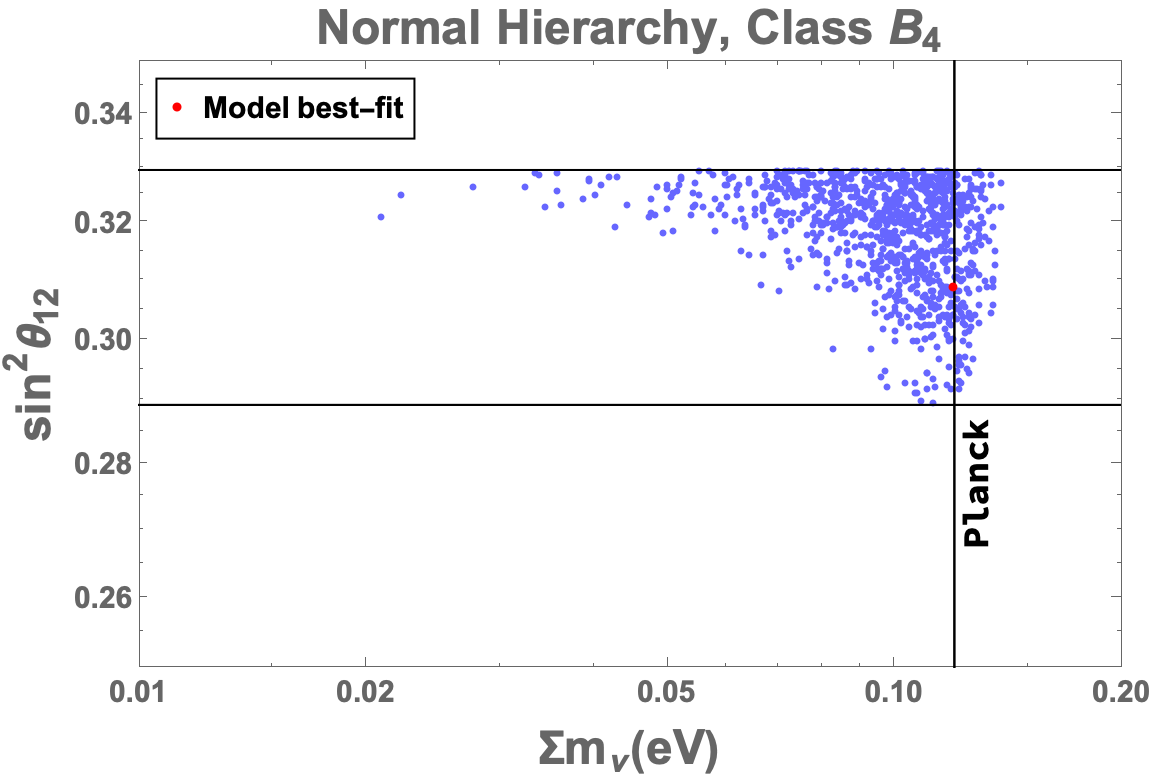}
	\includegraphics[scale=0.3]{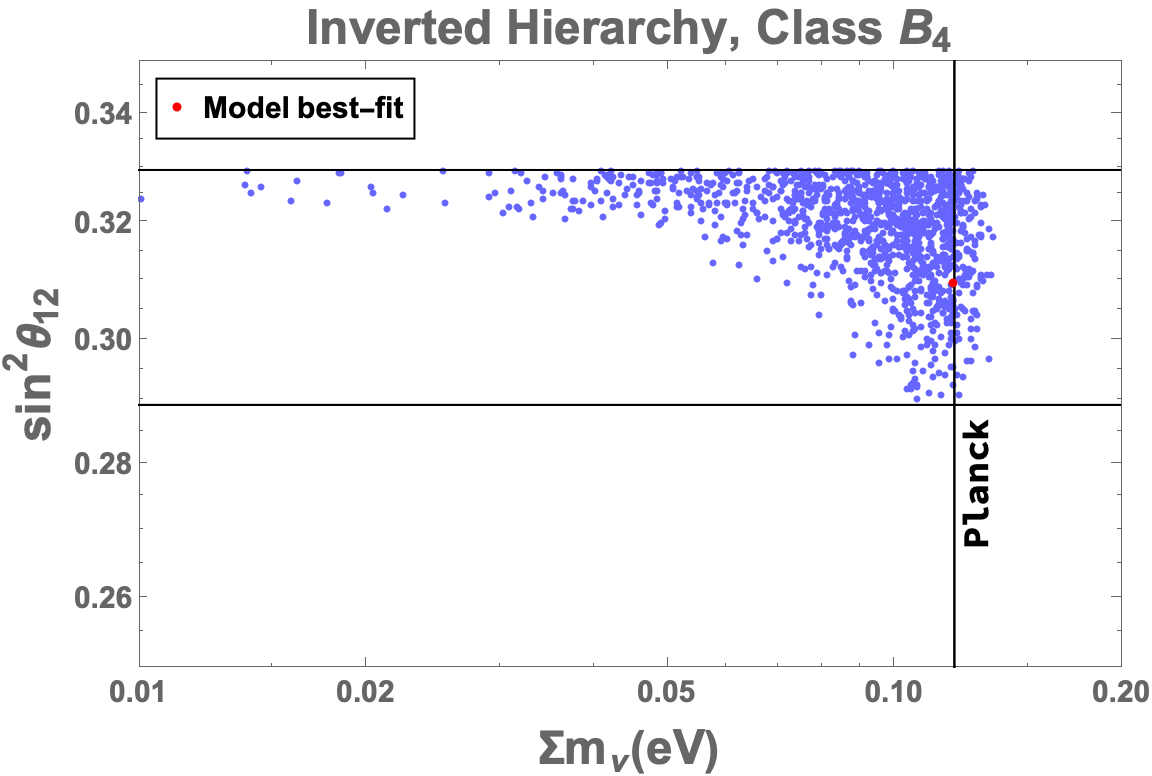}
	\caption {\label{f3} Neutrino oscillation parameters with sum of neutrino masses, where the black horizontal lines depict the 3$\sigma$ range\cite{Esteban:2024eli} and the vertical line represents the Planck bound on $\Sigma m_{\nu}$\cite{Planck:2018vyg}. The red dot represents the model best-fit value.}
\end{figure}
Interchanging the charge and weight assignments of the particle fields as $L_{L_2}\leftrightarrow L_{L_3}$ and $L_{R_2}\leftrightarrow L_{R_3}$, which is shown in table \ref{t5}.
\begin{table}[H]
	\begin{center}
		\begin{tabular}{|c|c|c|c|c|c|c|c|c|c|}
			\hline
			Gauge group & $L_{L_1}$ & $L_{L_2}$ & $L_{L_3}$ & $L_{R_1}^{c}$ & $L_{R_2}^{c}$ & $L_{R_3}^{c}$ & $\phi$ & $\Delta_L$ & $\Delta_R$ \\
			\hline
			$A_{4}$ & 1 & $1^{\prime\prime}$ & $1^{\prime}$ & 1 & $1^{\prime}$ & $1^{\prime\prime}$ & 1 &1 & 1\\
			\hline
			$k$ & -4 & -2 & -4 & 0 & -2 & -2 & 0 & 0 & 0 \\
			\hline
		\end{tabular}
		\caption{\label{t5}Charge assignments for the particle content of the model.}
	\end{center}
\end{table}
The superpotential for the above assignments can be given as,
	\begin{equation}
\begin{split}
		\label{e35}
		\mathcal{W}_{B_3}=L_{L_1}^{T}\phi L_{R_1}^{c}Y_{1}^{4}+L_{L_2}^{T}\phi L_{R_2}^{c}Y_{1}^{4}+L_{L_3}^{T}\phi L_{R_3}^{c}Y_{1}^{6}+L_{R_1}^{c{T}} i\sigma_2 \Delta_{R}L_{R_1}^{c}Y_{1}^{0}\\+(L_{R_2}^{c{T}} i\sigma_2 \Delta_{R} L_{R_3}^{c}Y_{1}^{4}+L_{R_3}^{c{T}} i\sigma_2 \Delta_{R} L_{R_2}^{c}Y_{1}^{4})+L_{R_2}^{c{T}} i\sigma_2 \Delta_{R} L_{R_2}^{c}Y_{1'}^{4}\\+L_{L_1}^{{T}} i\sigma_2\Delta_{L} L_{L_1}Y_{1}^{8}+(L_{L_1}^{{T}} i\sigma_2 \Delta_{L}L_{L_3}Y_{1^{\prime\prime}}^{8}+L_{L_3}^{{T}} i\sigma_2\Delta_{L} L_{L_1}Y_{1^{\prime\prime}}^{8})\\+L_{L_3}^{{T}} i\sigma_2 \Delta_{L}L_{L_3}Y_{1'}^{8}+(L_{L_2}^{{T}} i\sigma_2 \Delta_{L}L_{L_3}Y_{1}^{6}+L_{L_3}^{{T}} i\sigma_2 \Delta_{L}L_{L_2}Y_{1}^{6})
	\end{split}
\end{equation}
\begin{equation}
	M_D =
	\begin{pmatrix}
		Y_{1}^{4} & 0 & 0 \\
		0 & Y_{1}^{4} & 0\\
		0 & 0 & Y_{1}^{6}
	\end{pmatrix},
	\qquad
	M_R =
	\begin{pmatrix}
		1 & 0 & 0\\
		0 &	Y_{1'}^{4} & Y_{1}^{4}\\
		0 &  Y_{1}^{4} &  0
	\end{pmatrix},
	\qquad
	M_L =
	\begin{pmatrix}
		Y_{1}^{8} & 0  & Y_{1^{\prime\prime}}^{8}\\
		0 & 0 & Y_{1}^{6}\\
		Y_{1^{\prime\prime}}^{8} &  Y_{1}^{6} &  Y_{1'}^{8}
	\end{pmatrix}
\end{equation}
The resulting mass matrix is given as,
\begin{equation}
\resizebox{\textwidth}{!}{$
	\label{e36}
	M_{\nu}=\begin{pmatrix}
		\frac{v^{2}+v_{L}v_{R}}{v_{R}}(Y_{1}^{2}+2 Y_{2}Y_{3})^{2} & 0 & v_{L}(2 Y_{1}Y_{2}+Y_{3})^{2}\\
		0 & 0 & \frac{(Y_{1}^{3}+Y_{2}^{3}+Y_{3}^{3}-3Y_{1}Y_{2}Y_{3})(v_{L}v_{R}(Y_{1}^{2}+2 Y_{2}Y_{3})+v^{2}(2 Y_{1}Y_{2}+Y_{3}))}{v_{R}(Y_{1}^{2}+2 Y_{2}Y_{3})}\\
		v_{L}(2 Y_{1}Y_{2}+Y_{3})^{2} & \frac{(Y_{1}^{3}+Y_{2}^{3}+Y_{3}^{3}-3Y_{1}Y_{2}Y_{3})(v_{L}v_{R}(Y_{1}^{2}+2 Y_{2}Y_{3})+v^{2}(2 Y_{1}Y_{2}+Y_{3}))}{v_{R}(Y_{1}^{2}+2 Y_{2}Y_{3})} & \frac{(2 Y_{1}Y_{2}+Y_{3}^{2})\Bigg(v_{L}(Y_{1}^{2}+2 Y_{2}Y_{3})^{3}-\frac{v^{2}(Y_{1}^{3}+Y_{2}^{3}+Y_{3}^{3}-3Y_{1}Y_{2}Y_{3})^{2}}{v_{R}}\Bigg)}{(Y_{1}^{2}+2 Y_{2}Y_{3})^{2}}
	\end{pmatrix}
	$}
\end{equation}
Equation \eqref{e36} represents Class $B_{3}$ of two-zero texture of neutrino mass matrix.
\begin{figure}[H]
	\centering
	\includegraphics[scale=0.3]{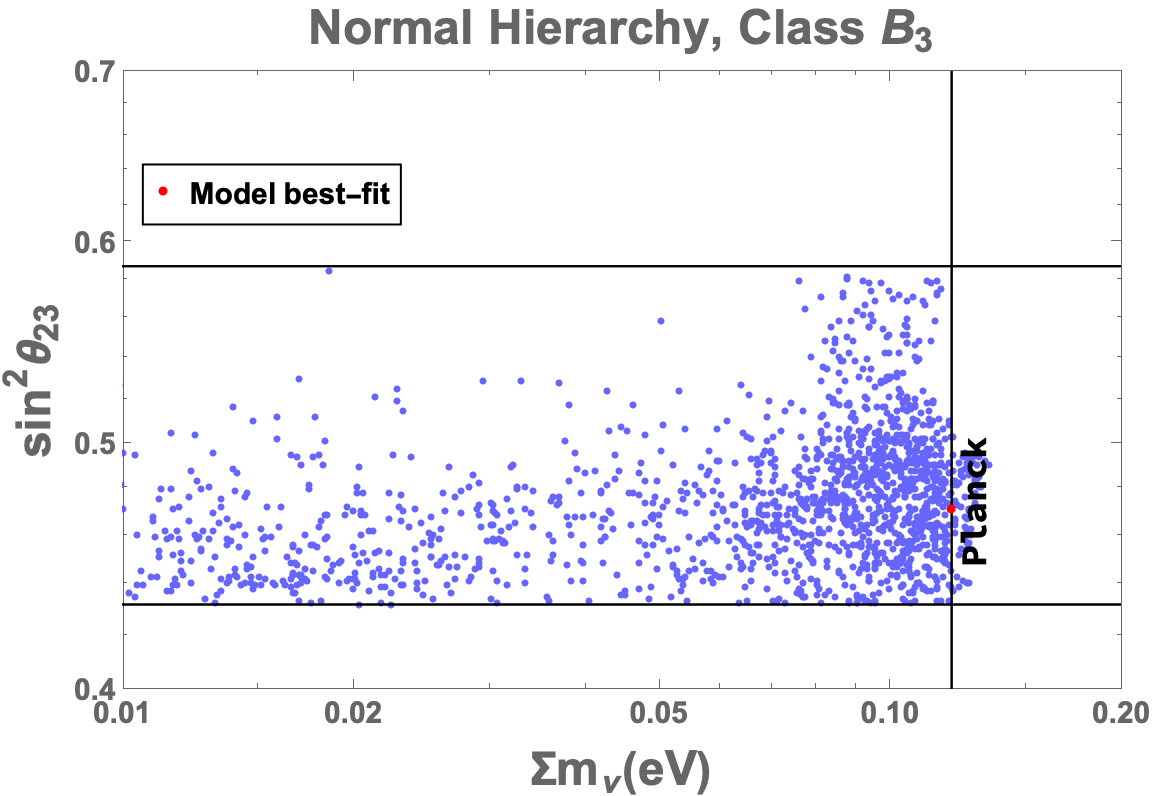}
	\includegraphics[scale=0.3]{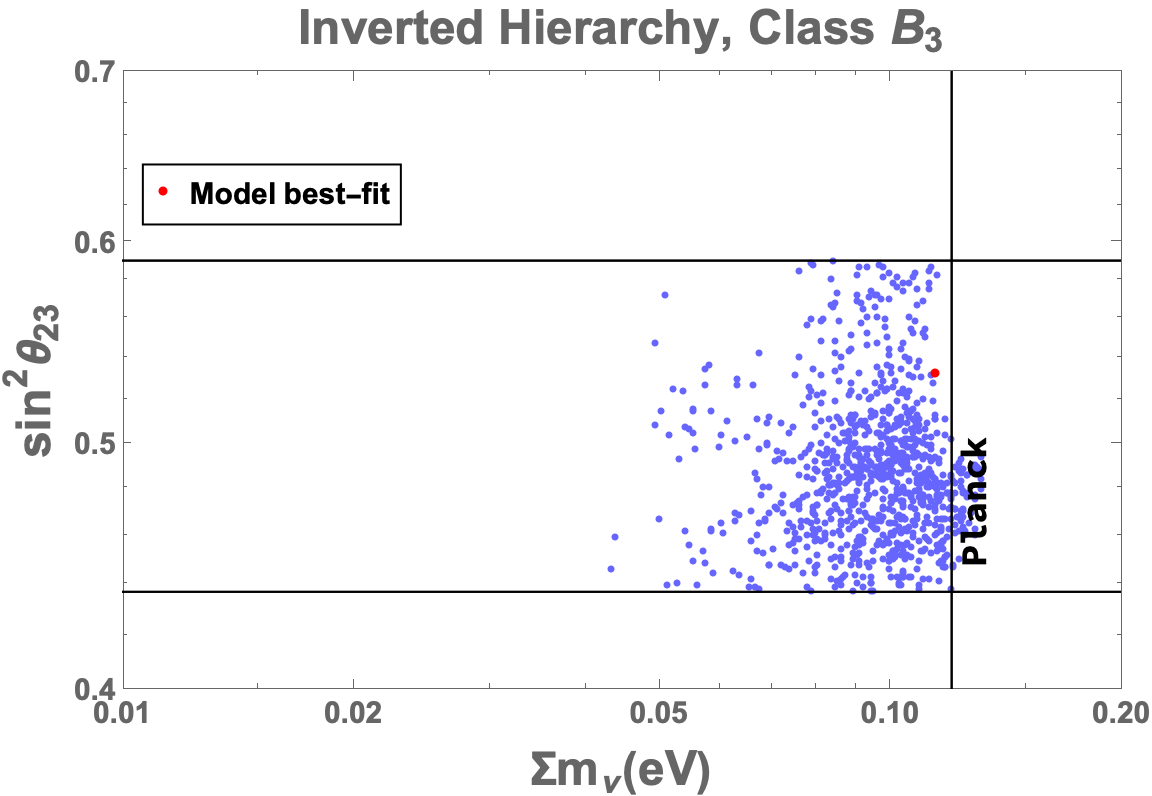}
	\includegraphics[scale=0.3]{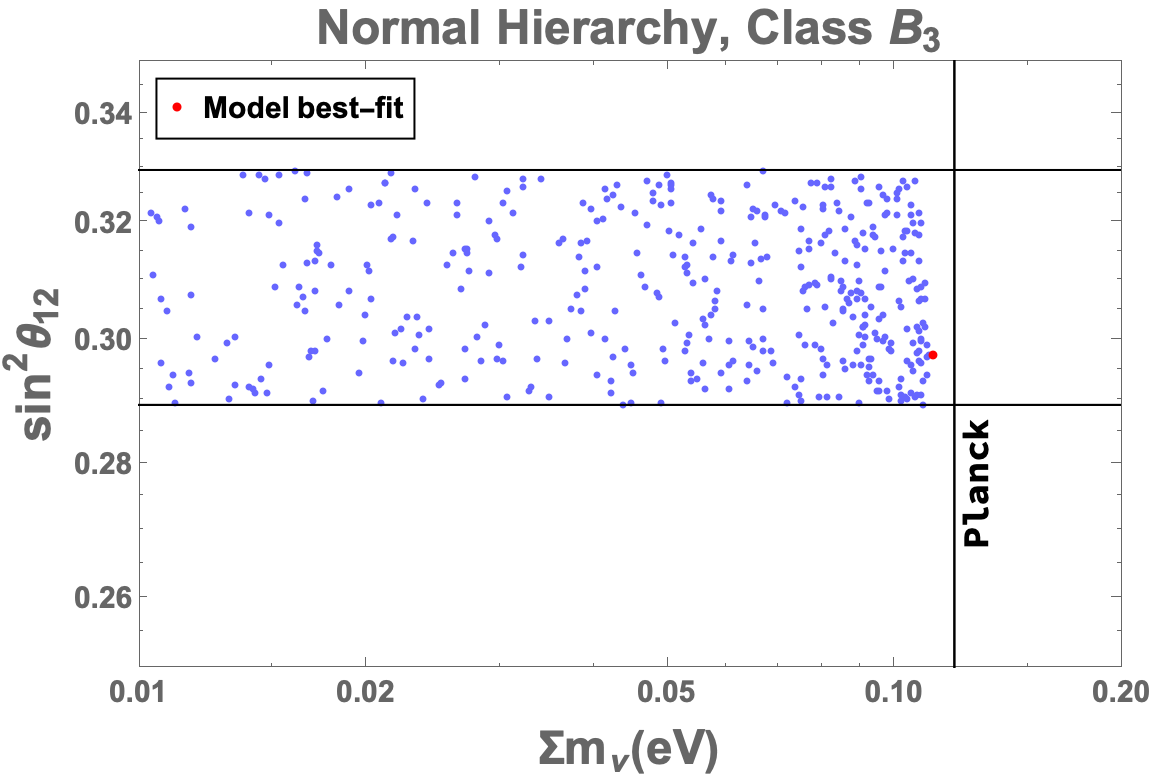}
	\includegraphics[scale=0.3]{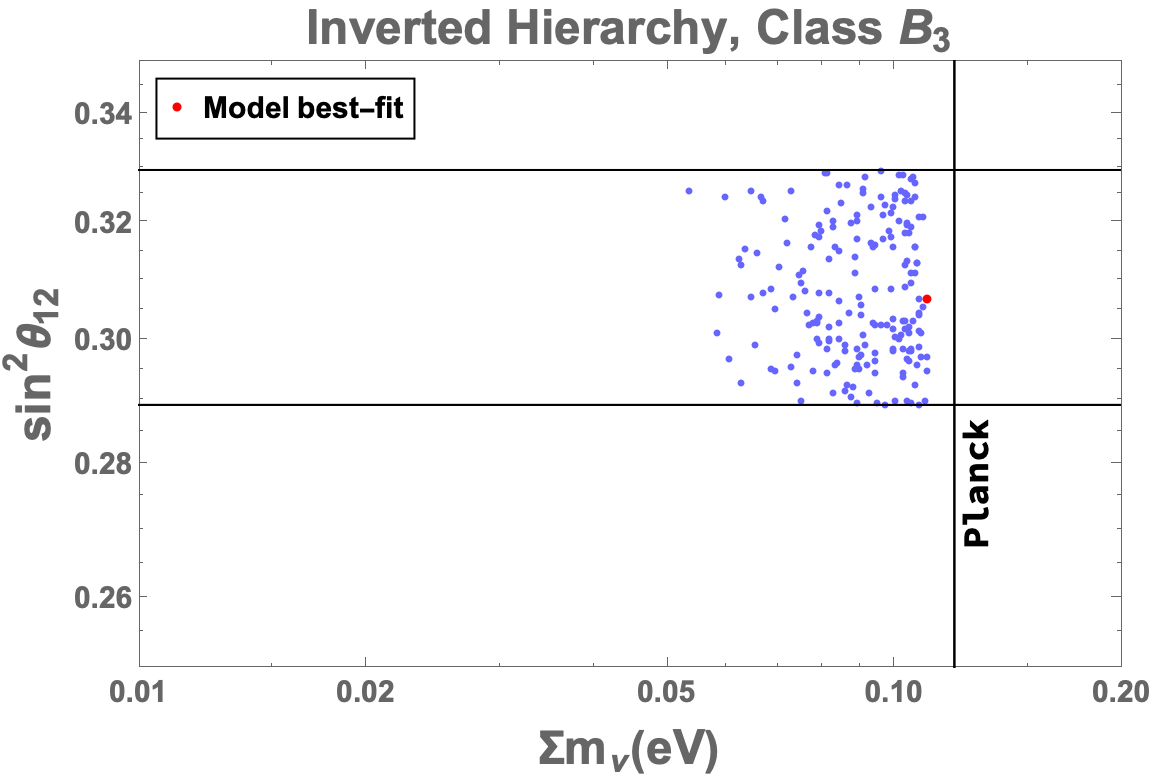}
	\caption {\label{f4} Neutrino oscillation parameters with sum of neutrino masses, where the black horizontal lines depict the 3$\sigma$ range\cite{Esteban:2024eli} and the vertical line represents the Planck bound on $\Sigma m_{\nu}$\cite{Planck:2018vyg}. The red dot represents the model best-fit value.}
\end{figure}
Again interchanging the charge and weight assignments of the particle fields as $L_{L_3}\leftrightarrow L_{L_1}$ and $L_{R_3}\leftrightarrow L_{R_1}$, which is shown in table \ref{t5}.
\begin{table}[H]
	\begin{center}
		\begin{tabular}{|c|c|c|c|c|c|c|c|c|c|}
			\hline
			Gauge group & $L_{L_1}$ & $L_{L_2}$ & $L_{L_3}$ & $L_{R_1}^{c}$ & $L_{R_2}^{c}$ & $L_{R_3}^{c}$ & $\phi$ & $\Delta_L$ & $\Delta_R$ \\
			\hline
			$A_{4}$ & $1^{\prime\prime}$ & $1^{\prime}$ & 1 & $1^{\prime}$ & $1^{\prime\prime}$ & 1 & 1 &1 & 1\\
			\hline
			$k$ & -2 & -4 & -4 & -2 & -2 & 0 & 0 & 0 & 0 \\
			\hline
		\end{tabular}
		\caption{\label{t6}Charge assignments for the particle content of the model.}
	\end{center}
\end{table}
The superpotential for the above assignments can be given as,
\begin{equation}
	\begin{split}
		\label{e37}
		\mathcal{W}_{A_2}=L_{L_1}^{T}\phi L_{R_1}^{c}Y_{1}^{4}+L_{L_2}^{T}\phi L_{R_2}^{c}Y_{1}^{6}+L_{L_3}^{T}\phi L_{R_3}^{c}Y_{1}^{4}+L_{R_1}^{c{T}} i\sigma_2 \Delta_{R}L_{R_1}^{c}Y_{1'}^{4}\\+(L_{R_1}^{c{T}} i\sigma_2 \Delta_{R} L_{R_2}^{c}Y_{1}^{4}+L_{R_2}^{c{T}} i\sigma_2 \Delta_{R} L_{R_1}^{c}Y_{1}^{4})+(L_{R_2}^{c{T}} i\sigma_2 \Delta_{R}L_{R_3}^{c}Y_{1^{\prime\prime}}^{8}+\\L_{R_3}^{c{T}} i\sigma_2 \Delta_{R}L_{L_2}^{c}Y_{1^{\prime\prime}}^{8})+ L_{R_2}^{c{T}} i\sigma_2 \Delta_{2}L_{R_2}^{c}Y_{1'}^{4}+L_{R_3}^{c{T}} i\sigma_2 \Delta_{R} L_{R_3}^{c}Y_{1}^{0}\\+(L_{L_1}^{{T}} i\sigma_2\Delta_{L} L_{L_2}Y_{1}^{6}+L_{L_2}^{{T}} i\sigma_2 \Delta_{L}L_{L_1}Y_{1}^{6})+L_{L_2}^{{T}} i\sigma_2\Delta_{L} L_{L_2}Y_{1}^{8})\\+(L_{L_2}^{{T}} i\sigma_2 \Delta_{L}L_{L_3}Y_{1^{\prime\prime}}^{8}+L_{L_3}^{{T}} i\sigma_2 \Delta_{L}L_{L_2}Y_{1^{\prime\prime}}^{8})+L_{L_3}^{{T}} i\sigma_2 \Delta_{L}L_{L_3}Y_{1}^{8}
	\end{split}
\end{equation}
\begin{equation}
	M_D =
	\begin{pmatrix}
		Y_{1}^{4} & 0 & 0 \\
		0 & Y_{1}^{6} & 0\\
		0 & 0 & Y_{1}^{4}
	\end{pmatrix},
	\qquad
	M_R =
	\begin{pmatrix}
		Y_{1'}^{4} & Y_{1}^{4} & 0\\
		Y_{1}^{4} &	Y_{1'}^{4} & Y_{1^{\prime\prime}}^{8}\\
		0 &  Y_{1^{\prime\prime}}^{8} &  1
	\end{pmatrix},
	\qquad
	M_L =
	\begin{pmatrix}
		0 & Y_{1}^{6}  & 0\\
		Y_{1}^{6} & Y_{1'}^{8} & Y_{1^{\prime\prime}}^{8}\\
		0 &   Y_{1^{\prime\prime}}^{8} &  Y_{1}^{8}
	\end{pmatrix}
\end{equation}
The resulting mass matrix is given as,
\begin{equation}
	\resizebox{\textwidth}{!}{$
		\label{e38}
		M_{\nu}=\begin{pmatrix}
			0&  \frac{(v^{2}+v_{L}v_{R})(Y_{1}^{3}+Y_{2}^{3}+Y_{3}^{3}-3Y_{1}Y_{2}Y_{3})}{v_{R}} & 0\\
			\frac{(v^{2}+v_{L}v_{R})(Y_{1}^{3}+Y_{2}^{3}+Y_{3}^{3}-3Y_{1}Y_{2}Y_{3})}{v_{R}} & \frac{(2 Y_{1}Y_{2}+Y_{3}^{2})\Bigg(v_{L}(Y_{1}^{2}+2 Y_{2}Y_{3})^{3}-\frac{v^{2}(Y_{1}^{3}+Y_{2}^{3}+Y_{3}^{3}-3Y_{1}Y_{2}Y_{3})^{2}}{v_{R}}\Bigg)}{(Y_{1}^{2}+2 Y_{2}Y_{3})^{2}} & v_{L}(2 Y_{1}Y_{2}+Y_{3})^{2}\\
			0 & v_{L}(2 Y_{1}Y_{2}+Y_{3})^{2}& (Y_{1}^{2}+2Y_{2}Y_{3})\Bigg(v_{L}+\frac{v^{2}(Y_{1}^{2}+2 Y_{2}Y_{3})}{v_{R}}\Bigg)
		\end{pmatrix}
		$}
\end{equation}
Equation \eqref{e38} represents Class $A_{2}$ of two-zero texture of neutrino mass matrix.
\begin{figure}[H]
	\centering
	\includegraphics[scale=0.3]{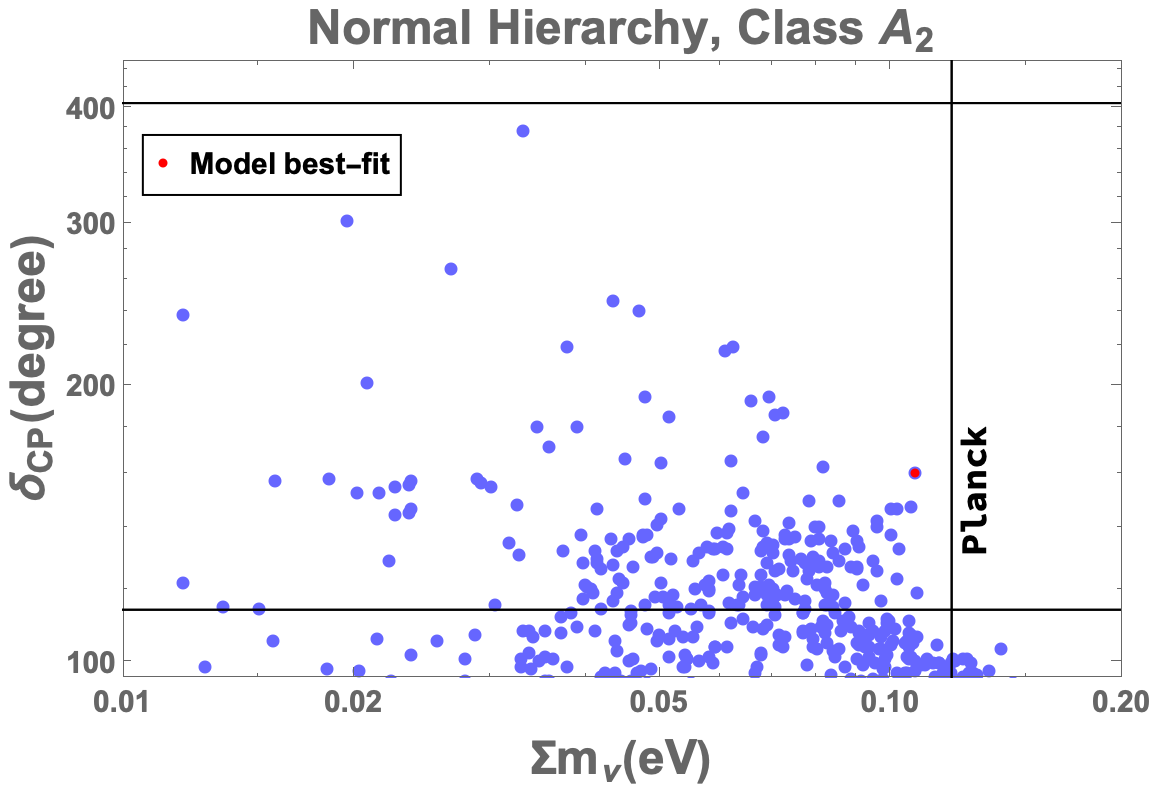}
	\includegraphics[scale=0.3]{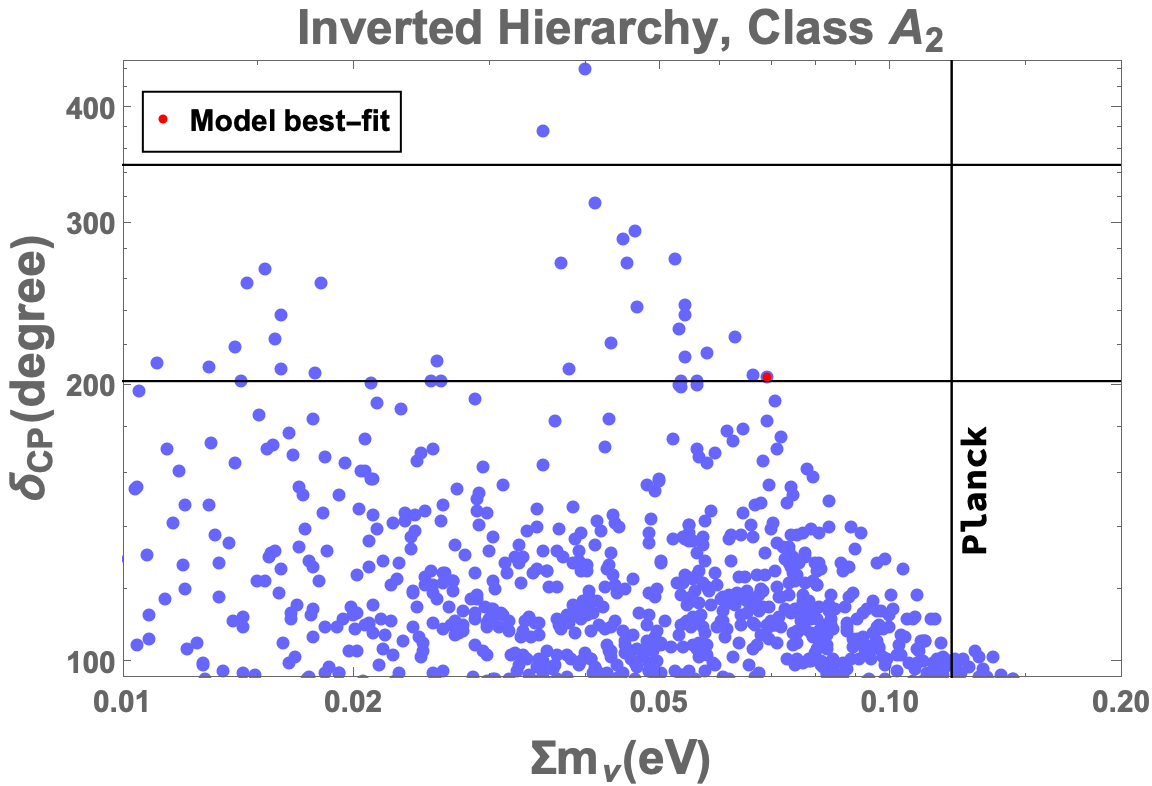}
	\includegraphics[scale=0.3]{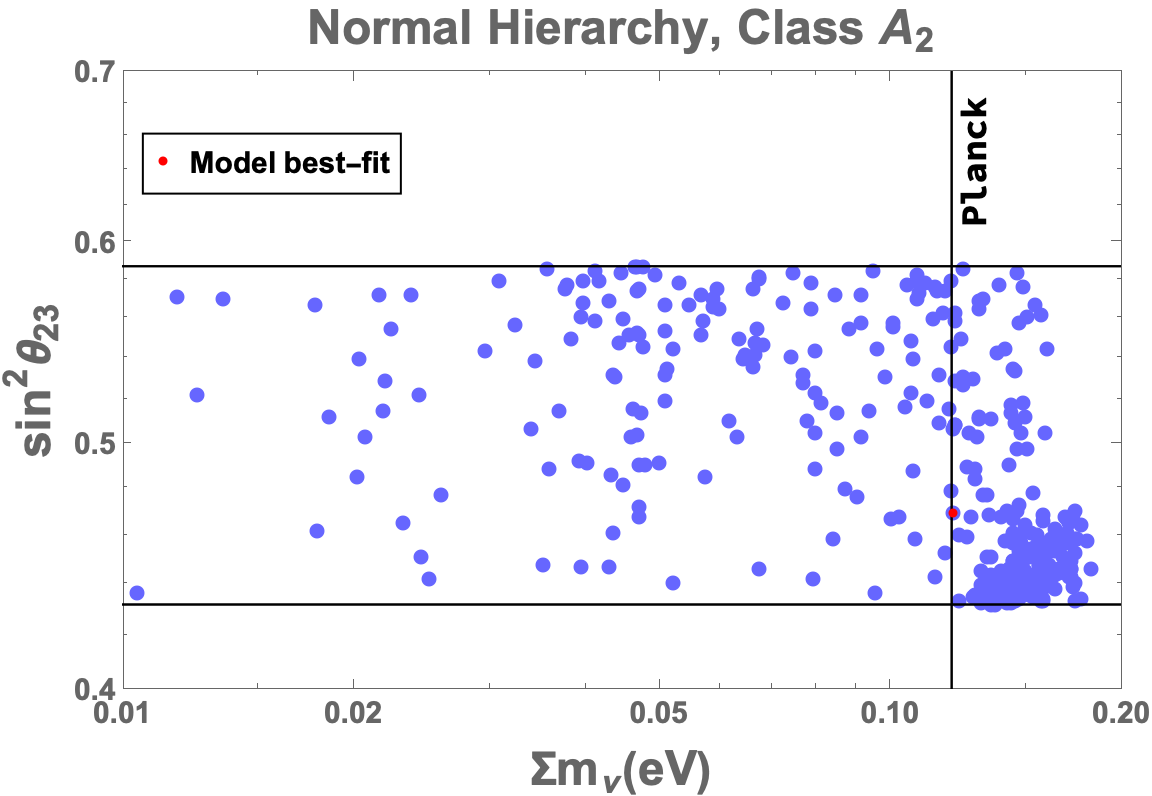}
	\includegraphics[scale=0.3]{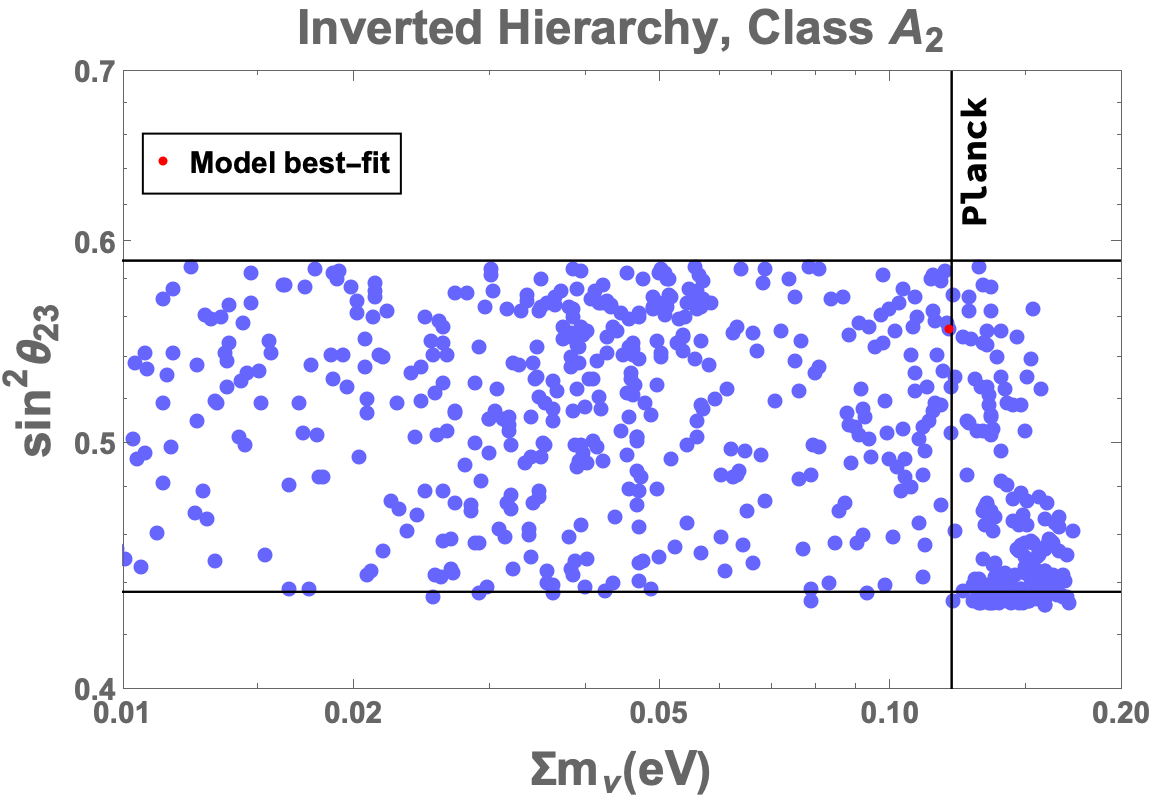}
	\caption {\label{f5} Neutrino oscillation parameters with sum of neutrino masses, where the black horizontal lines depict the 3$\sigma$ range\cite{Esteban:2024eli} and the vertical line represents the Planck bound on $\Sigma m_{\nu}$\cite{Planck:2018vyg}. The red dot represents the model best-fit value.}
\end{figure}
Finally interchanging the assignments of the fields using the criteria $L_{L_1}\leftrightarrow L_{L_2}\leftrightarrow L_{L_3}\leftrightarrow L_{L_1}$ and $L_{R_1}\leftrightarrow L_{R_2}\leftrightarrow L_{R_3}\leftrightarrow L_{R_1}$, which is represented in table \ref{t7}
\begin{table}[H]
	\begin{center}
		\begin{tabular}{|c|c|c|c|c|c|c|c|c|c|}
			\hline
			Gauge group & $L_{L_1}$ & $L_{L_2}$ & $L_{L_3}$ & $L_{R_1}^{c}$ & $L_{R_2}^{c}$ & $L_{R_3}^{c}$ & $\phi$ & $\Delta_L$ & $\Delta_R$ \\
			\hline
			$A_{4}$ & $1^{\prime\prime}$ & 1 & $1^{\prime}$ & $1^{\prime}$ & 1 & $1^{\prime\prime}$ & 1 &1 & 1\\
			\hline
			$k$ & -2 & -4 & -4 & -2 & 0 & -2 & 0 & 0 & 0 \\
			\hline
		\end{tabular}
		\caption{\label{t7}Charge assignments for the particle content of the model.}
	\end{center}
\end{table}
The superpotential for the above assignments can be given as,
\begin{equation}
	\begin{split}
		\label{e39}
		\mathcal{W}_{A_1}=L_{L_1}^{T}\phi L_{R_1}^{c}Y_{1}^{4}+L_{L_2}^{T}\phi L_{R_2}^{c}Y_{1}^{6}+L_{L_3}^{T}\phi L_{R_3}^{c}Y_{1}^{4}+L_{R_1}^{c{T}} i\sigma_2 \Delta_{R}L_{R_1}^{c}Y_{1'}^{4}\\+(L_{R_1}^{c{T}} i\sigma_2 \Delta_{R} L_{R_2}^{c}Y_{1}^{4}+L_{R_2}^{c{T}} i\sigma_2 \Delta_{R} L_{R_1}^{c}Y_{1}^{4})+L_{R_3}^{c{T}} i\sigma_2 \Delta_{R} L_{R_3}^{c}Y_{1}^{0}\\+(L_{L_1}^{{T}} i\sigma_2\Delta_{L} L_{L_2}Y_{1}^{6}+L_{L_2}^{{T}} i\sigma_2 \Delta_{L}L_{L_1}Y_{1}^{6})+L_{L_2}^{{T}} i\sigma_2\Delta_{L} L_{L_2}Y_{1}^{8}\\+(L_{L_2}^{{T}} i\sigma_2 \Delta_{L}L_{L_3}Y_{1^{\prime\prime}}^{8}+L_{L_3}^{{T}} i\sigma_2 \Delta_{L}L_{L_2}Y_{1^{\prime\prime}}^{8})+L_{L_3}^{{T}} i\sigma_2 \Delta_{L}L_{L_3}Y_{1}^{8}
	\end{split}
\end{equation}
\begin{equation}
	M_D =
	\begin{pmatrix}
		Y_{1}^{4} & 0 & 0 \\
		0 & Y_{1}^{4} & 0\\
		0 & 0 & Y_{1}^{6}
	\end{pmatrix},
	\qquad
	M_R =
	\begin{pmatrix}
		Y_{1'}^{4} & 0 & Y_{1}^{4}\\
		0 &	1 & 0\\
		Y_{1}^{4} &  0 &  0
	\end{pmatrix},
	\qquad
	M_L =
	\begin{pmatrix}
		0 & 0  & Y_{1}^{6}\\
		0 & Y_{1}^{8} & Y_{1^{\prime\prime}}^{8}\\
		Y_{1}^{6}\ &   Y_{1^{\prime\prime}}^{8} &  Y_{1'}^{8}
	\end{pmatrix}
\end{equation}
The resulting mass matrix is given as,
\begin{equation}
	\resizebox{\textwidth}{!}{$
		\label{e40}
		M_{\nu}=\begin{pmatrix}
			0& 0&  \frac{(v^{2}+v_{L}v_{R})(Y_{1}^{3}+Y_{2}^{3}+Y_{3}^{3}-3Y_{1}Y_{2}Y_{3})}{v_{R}} \\
			0 & \frac{(v^{2}+v_{L}v_{R})(Y_{1}^{2}+2Y_{2}Y_{3})^{2}}{v_{R}}& v_{L}(2 Y_{1}Y_{2}+Y_{3})^{2}\\
			 \frac{(v^{2}+v_{L}v_{R})(Y_{1}^{3}+Y_{2}^{3}+Y_{3}^{3}-3Y_{1}Y_{2}Y_{3})}{v_{R}}  & v_{L}(2 Y_{1}Y_{2}+Y_{3})^{2}& \frac{(2 Y_{1}Y_{2}+Y_{3}^{2})\Bigg(v_{L}(Y_{1}^{2}+2 Y_{2}Y_{3})^{3}-\frac{v^{2}(Y_{1}^{3}+Y_{2}^{3}+Y_{3}^{3}-3Y_{1}Y_{2}Y_{3})^{2}}{v_{R}}\Bigg)}{(Y_{1}^{2}+2 Y_{2}Y_{3})^{2}} 
		\end{pmatrix}
		$}
\end{equation}
Equation \eqref{e40} represents Class $A_{1}$ of two-zero texture of neutrino mass matrix.
\begin{figure}[H]
	\centering
	\includegraphics[scale=0.25]{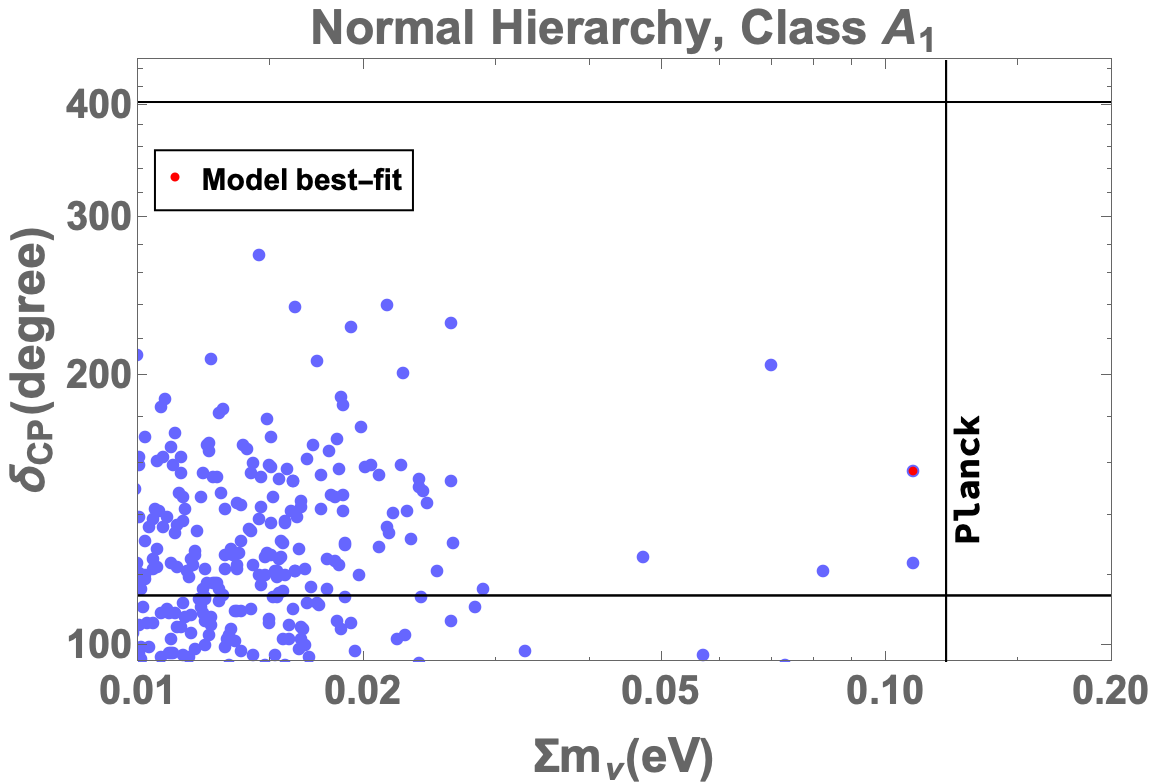}
	\includegraphics[scale=0.25]{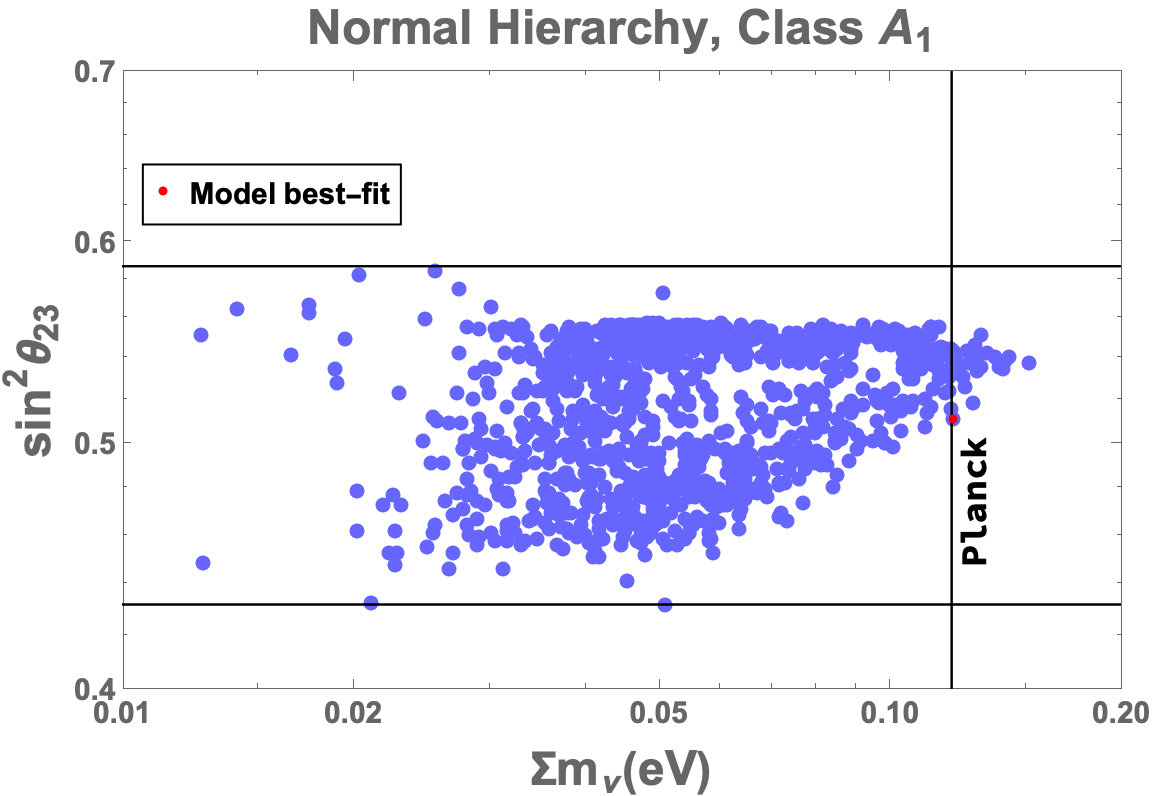}
	\includegraphics[scale=0.25]{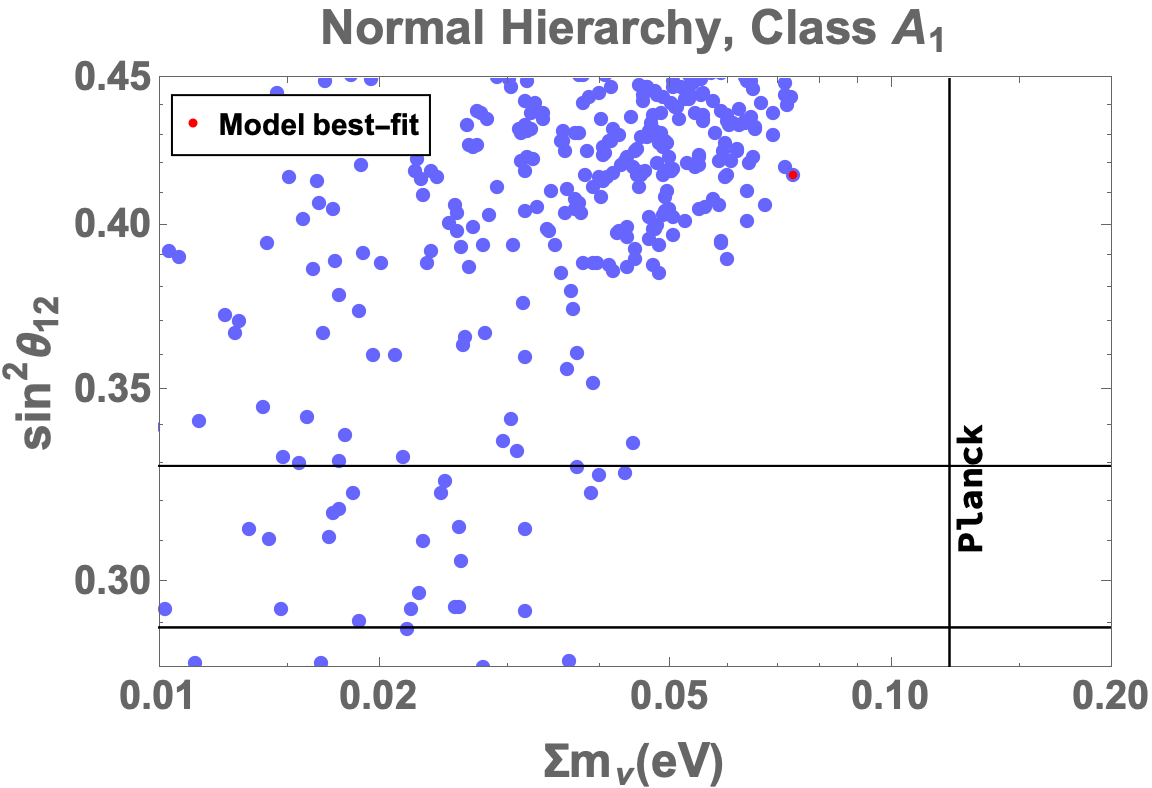}
		\includegraphics[scale=0.25]{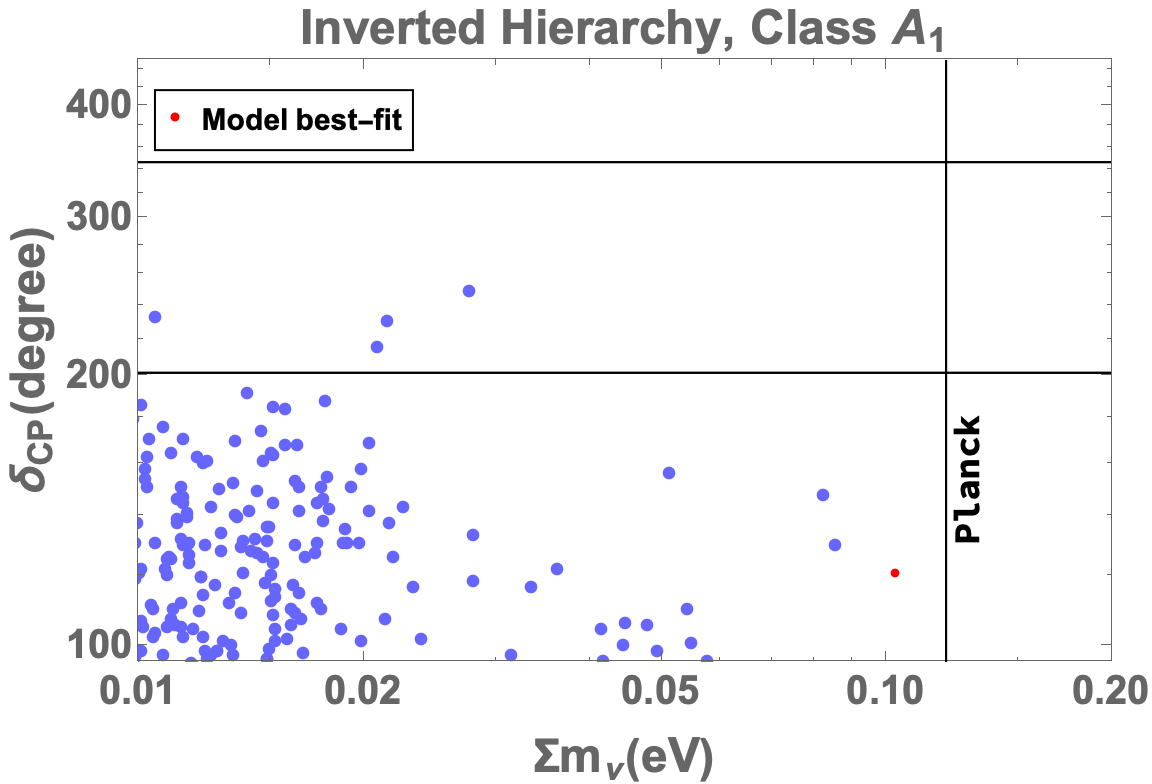}
			\includegraphics[scale=0.25]{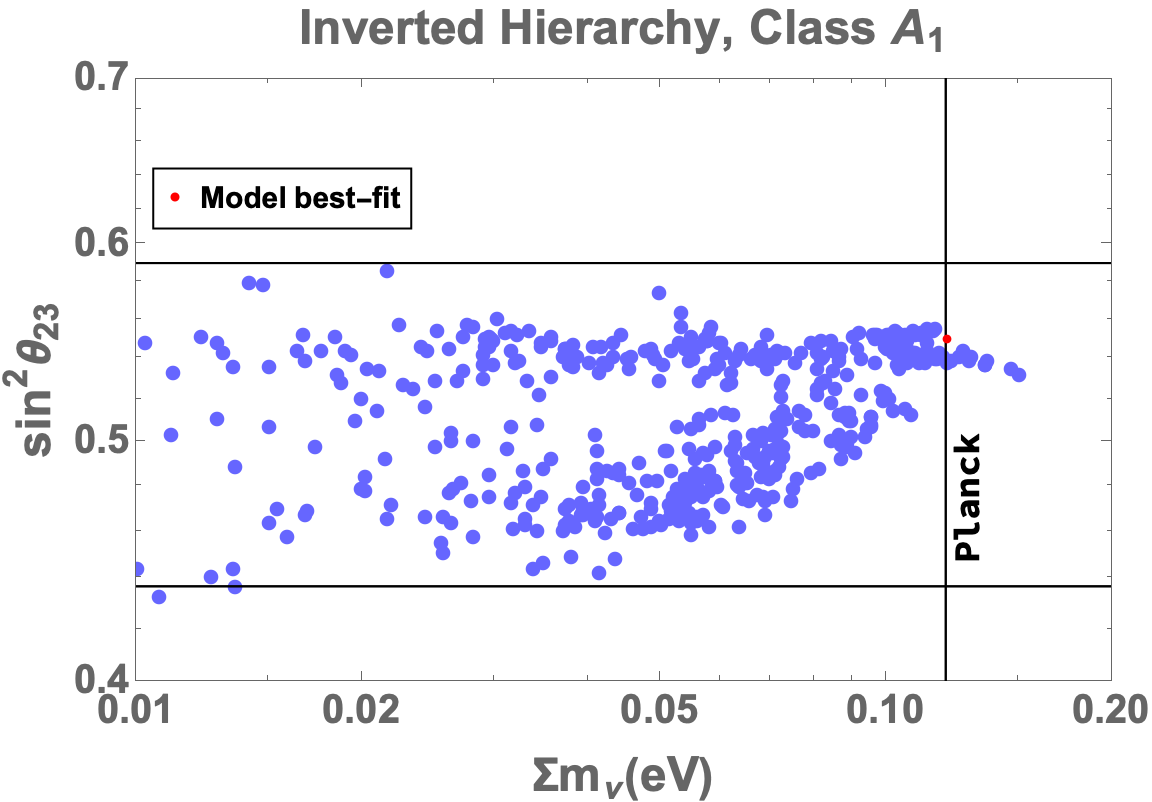}
		\includegraphics[scale=0.25]{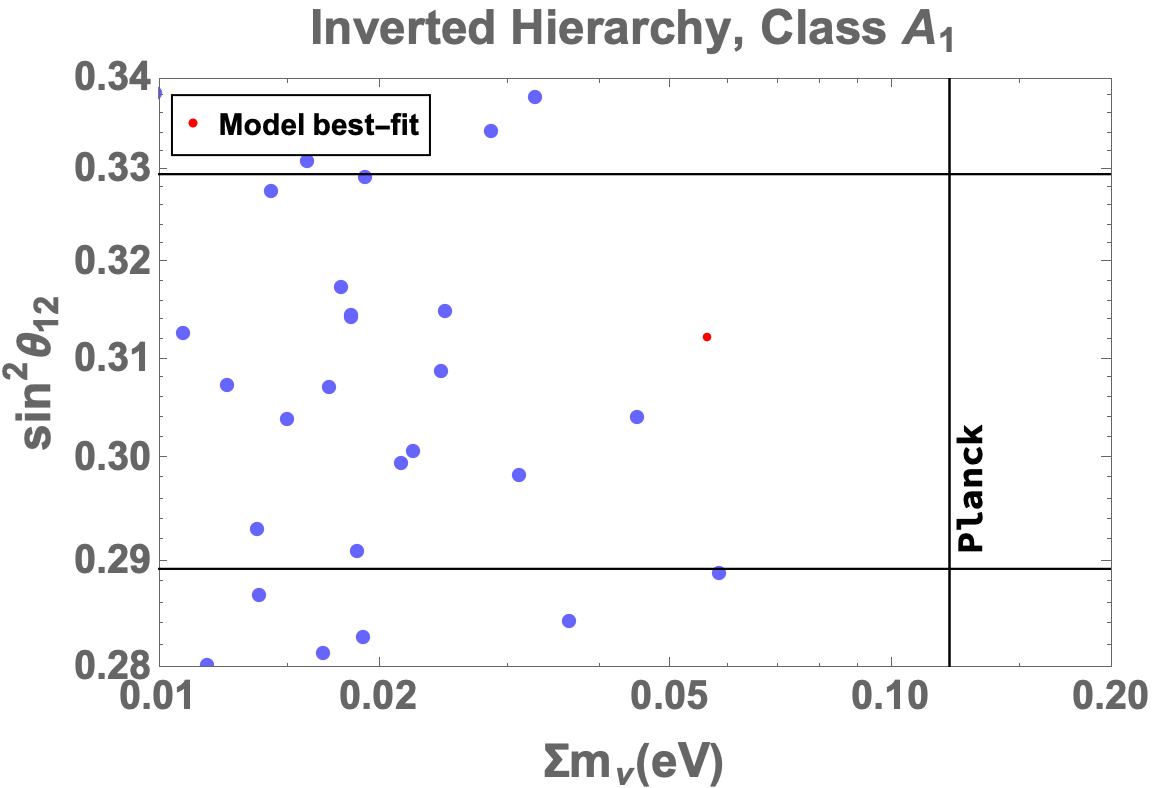}
		\caption {\label{f6} Neutrino oscillation parameters with sum of neutrino masses, where the black horizontal lines depict the 3$\sigma$ range\cite{Esteban:2024eli} and the vertical line represents the Planck bound on $\Sigma m_{\nu}$\cite{Planck:2018vyg}. The red dot represents the model best-fit value.}
\end{figure}
\subsection{For $k_{Y_{max}}=10$}
For modular weight 10, there are nine number of forms, three singlets $1$,$1'$ and three triplets $3_{1}$, $3_{2}$ and $3_{3}$ under $A_{4}$, represented as,
\begin{equation*}
Y_{(1)}^{10} = (Y_{1}^{2}+2Y_{2}Y_{3})(Y_{1}^{3}+Y_{2}^{3}+Y_{3}^{2}-3Y_{1}Y_{2}Y_{3});
		Y_{(1')}^{10} = (Y_{3}^{2}+2Y_{1}Y_{2})(Y_{1}^{3}+Y_{2}^{3}+Y_{3}^{2}-3Y_{1}Y_{2}Y_{3})
\end{equation*}
\begin{equation*}
	Y_{(3_{1})}^{10} = (Y_{1}^{2}+2Y_{2}Y_{3})^{2}\begin{pmatrix}
		Y_{1}\\
		Y_{2}\\
		Y_{3}
	\end{pmatrix}
\end{equation*}
\begin{equation}
		\label{e41}
	Y_{(3_{2})}^{10} = (Y_{3}^{2}+2Y_{1}Y_{2})^{2}\begin{pmatrix}
		Y_{2}\\
		Y_{3}\\
		Y_{1}
		\end{pmatrix};
			Y_{(3_{3})}^{10} = (Y_{1}^{2}+2Y_{2}Y_{3})(Y_{3}^{2}+2Y_{1}Y_{2})\begin{pmatrix}
			Y_{3}\\
			Y_{1}\\
			Y_{2}
		\end{pmatrix}
\end{equation}
The charge and modular weight assignments for $k_{Y_{max}=10}$ is given by,
\begin{table}[H]
	\begin{center}
		\begin{tabular}{|c|c|c|c|c|c|c|c|c|c|}
			\hline
			Gauge group & $L_{L_1}$ & $L_{L_2}$ & $L_{L_3}$ & $L_{R_1}^{c}$ & $L_{R_2}^{c}$ & $L_{R_3}^{c}$ & $\phi$ & $\Delta_L$ & $\Delta_R$ \\
			\hline
			$A_{4}$ & $1$ & $1^{\prime}$ & $1^{\prime\prime}$ & 1 & $1^{\prime\prime}$ & $1^{\prime}$ & 1 &1 & 1\\
			\hline
			$k$ & -5 & -3 & -3 & -5 & -3 & -3 & 0 & 0 & 0 \\
			\hline
		\end{tabular}
		\caption{\label{t8}Charge assignments for the particle content of the model.}
	\end{center}
\end{table}
The superpotential in this case is given by equation \eqref{e42}

	\begin{equation}
		\resizebox{\linewidth}{!}{$
		\begin{split}
		\label{e42}
		\mathcal{W}_{C}=L_{L_1}^{T}\phi L_{R_1}^{c}Y_{1}^{10}+(L_{L_1}^{T}\phi L_{R_2}^{c}Y_{1'}^{8}+L_{L_2}^{T}\phi L_{R_1}^{c}Y_{1'}^{8})+(L_{L_1}^{T}\phi L_{R_3}^{c}Y_{1^{\prime\prime}}^{8}+L_{L_3}^{T}\phi L_{R_1}^{c}Y_{1^{\prime\prime}}^{8})\\+L_{L_2}^{T}\phi L_{R_2}^{c}Y_{1}^{6}+L_{L_3}^{T}\phi L_{R_3}^{c}Y_{1}^{6}+L_{R_1}^{c{T}} i\sigma_2 \Delta_{R}L_{R_1}^{c}Y_{1}^{10}+(L_{R_1}^{c{T}} i\sigma_2 \Delta_{R} L_{R_2}^{c}Y_{1'}^{8}+L_{R_2}^{c{T}} i\sigma_2 \Delta_{R} L_{R_1}^{c}Y_{1'}^{8})\\+(L_{R_1}^{c{T}} i\sigma_2 \Delta_{R} L_{R_3}^{c}Y_{1^{\prime\prime}}^{8}+L_{R_3}^{c{T}} i\sigma_2 \Delta_{R} L_{R_1}^{c}Y_{1^{\prime\prime}}^{8})+(L_{R_2}^{c{T}} i\sigma_2 \Delta_{R} L_{R_3}^{c}Y_{1}^{6}+L_{R_2}^{c{T}} i\sigma_2 \Delta_{R} L_{R_3}^{c}Y_{1}^{6})+L_{L_1}^{c{T}} i\sigma_2\Delta_{L} L_{L_1}^{c}Y_{1}^{10}\\+(L_{L_1}^{{T}} i\sigma_2 \Delta_{L}L_{L_2}Y_{1^{\prime\prime}}^{8}+L_{L_2}^{{T}} i\sigma_2 \Delta_{L}L_{L_1}Y_{1^{\prime\prime}}^{8})+(L_{L_1}^{{T}} i\sigma_2 \Delta_{L}L_{L_3}Y_{1^{\prime}}^{8}+L_{L_3}^{{T}} i\sigma_2 \Delta_{L}L_{L_1}Y_{1^{\prime}}^{8})+(L_{L_2}^{{T}} i\sigma_2 \Delta_{L}L_{L_3}Y_{1}^{6}+L_{L_3}^{{T}} i\sigma_2 \Delta_{L}L_{L_2}Y_{1}^{6})
\end{split}
$}
\end{equation}
 \begin{equation}
 	M_D =
 	\begin{pmatrix}
 		Y_{1}^{10} & Y_{1'}^{8} & Y_{1^{\prime\prime}}^{8} \\
 		Y_{1'}^{8} & Y_{1}^{6} & 0\\
 		Y_{1^{\prime\prime}}^{8} & 0 & Y_{1}^{6}
 	\end{pmatrix},
 	\qquad
 	M_R =
 	\begin{pmatrix}
 		Y_{1}^{10} &Y_{1^{\prime}}^{8}  & Y_{1^{\prime\prime}}^{8}\\
 		Y_{1^{\prime}}^{8} & 0 & Y_{1}^{6}\\
 		Y_{1^{\prime\prime}}^{8} &  Y_{1}^{6} &  0
 	\end{pmatrix},
 	\qquad
 	M_L =
 \begin{pmatrix}
 	Y_{1}^{10} & Y_{1^{\prime\prime}}^{8}  & Y_{1^{\prime}}^{8}\\
 	Y_{1^{\prime\prime}}^{8} & 0 & Y_{1}^{6}\\
 	Y_{1^{\prime}}^{8} &  Y_{1}^{6} &  0
 \end{pmatrix}
 \end{equation}
 The resulting light neutrino mass matrix is given as,
 \begin{equation}
 	\resizebox{\textwidth}{!}{$
 		\label{e43}
 		M_{\nu}=\begin{pmatrix}
 			\frac{(v^{2}+v_{L}v_{R})(Y_{1}+Y_{2}+Y_{3})(Y_{1}^{2}+2Y_{2}Y_{3})(Y_{1}^{3}+Y_{2}^{3}+Y_{3}^{3}-3Y_{1}Y_{2}Y_{3})}{v_{R}} & \frac{(v^{2}+v_{L}v_{R})(2Y_{1}Y_{2}+Y_{3})^{2}}{v_{R}}&  \frac{(v^{2}+v_{L}v_{R})(Y_{1}^{2}+2Y_{2}Y_{3})(2Y_{1}Y_{2}+Y_{3}^{2})}{v_{R}} \\
 			\frac{(v^{2}+v_{L}v_{R})(2Y_{1}Y_{2}+Y_{3})^{2}}{v_{R}} & 0 & \frac{(v^{2}+v_{L}v_{R})(Y_{1}^{3}+Y_{2}^{3}+Y_{3}^{3}-3Y_{1}Y_{2}Y_{3})}{v_{R}}\\
 			\frac{(v^{2}+v_{L}v_{R})(Y_{1}^{2}+2Y_{2}Y_{3})(2Y_{1}Y_{2}+Y_{3}^{2})}{v_{R}}  & \frac{(v^{2}+v_{L}v_{R})(Y_{1}^{3}+Y_{2}^{3}+Y_{3}^{3}-3Y_{1}Y_{2}Y_{3})}{v_{R}} & 0
 		\end{pmatrix}
 		$}
 \end{equation}
 Equation \eqref{e43} represents Class C of two-zero neutrino mass texture.
 \begin{figure}[H]
 	\centering
 	\includegraphics[scale=0.25]{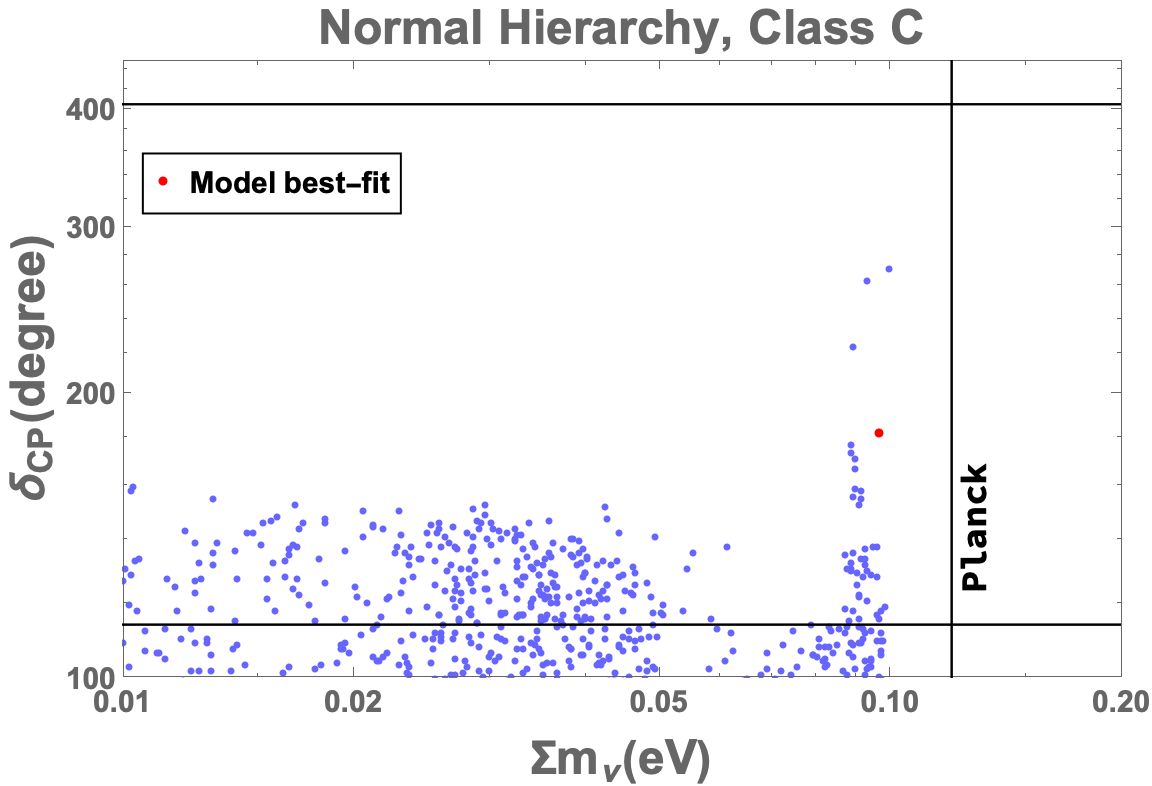}
 	\includegraphics[scale=0.25]{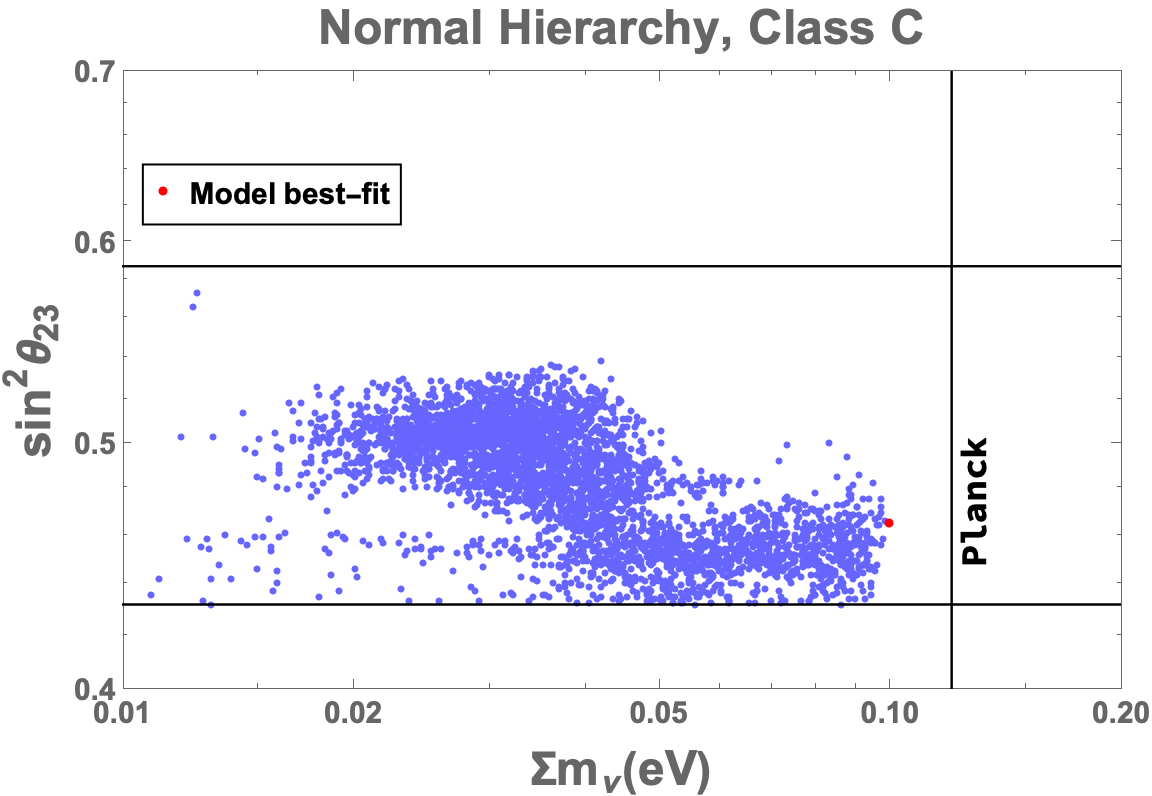}
 	\includegraphics[scale=0.25]{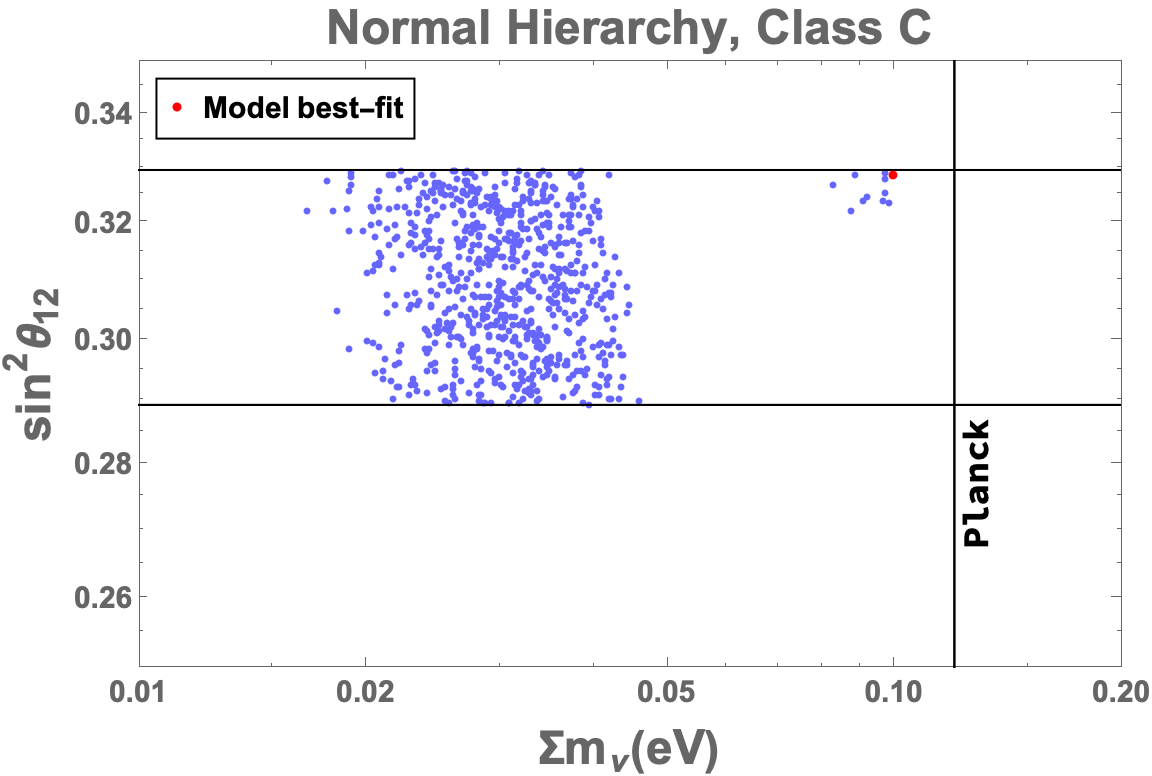}
 	\includegraphics[scale=0.25]{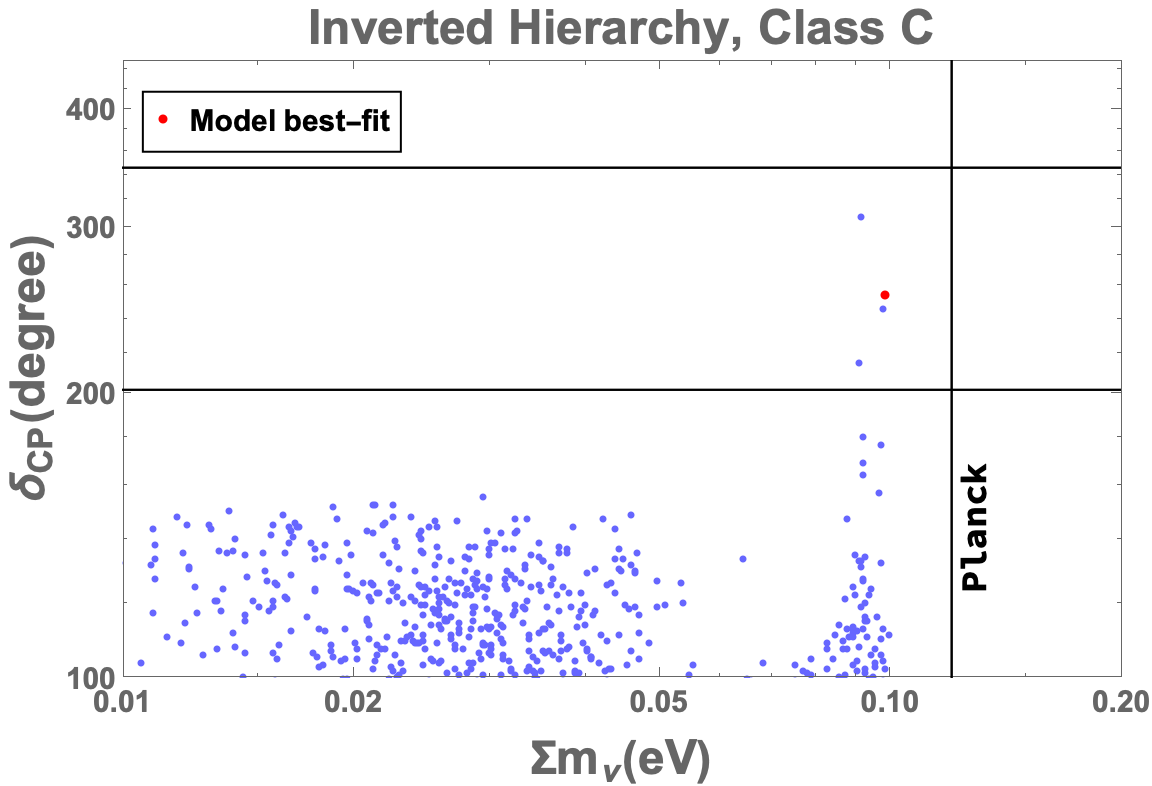}
 		\includegraphics[scale=0.25]{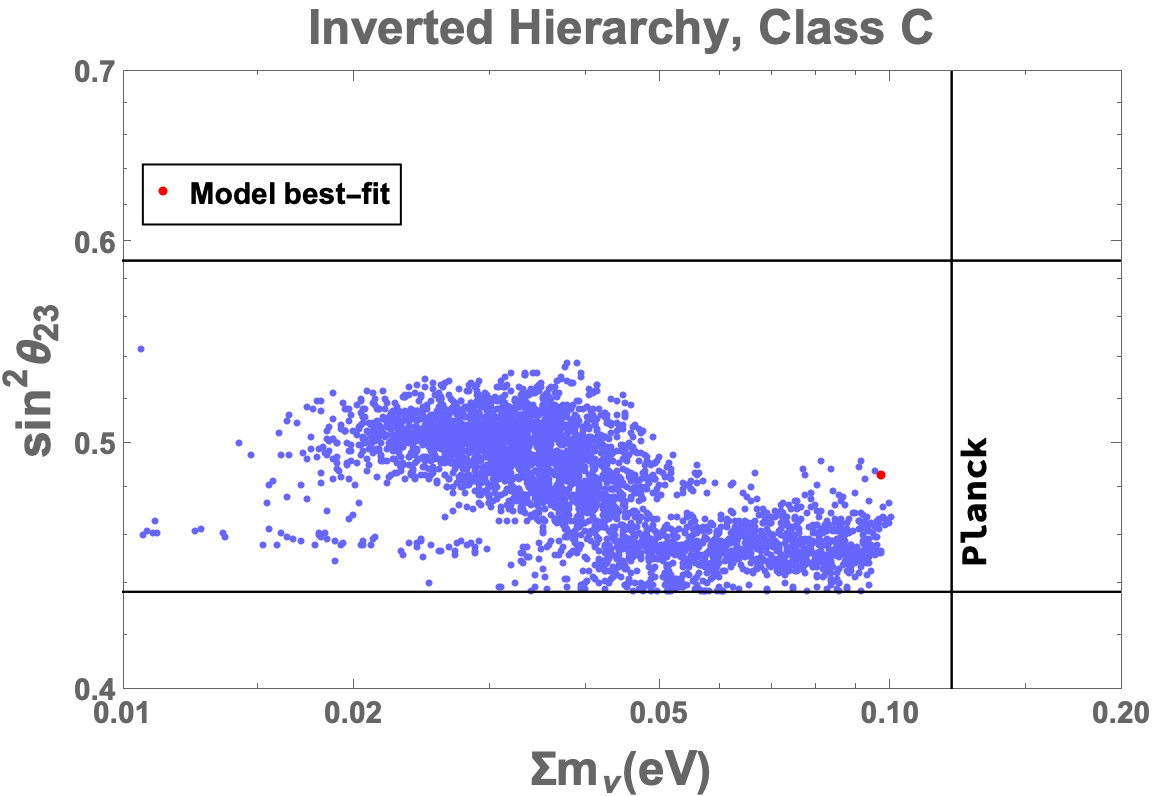}
 	\includegraphics[scale=0.25]{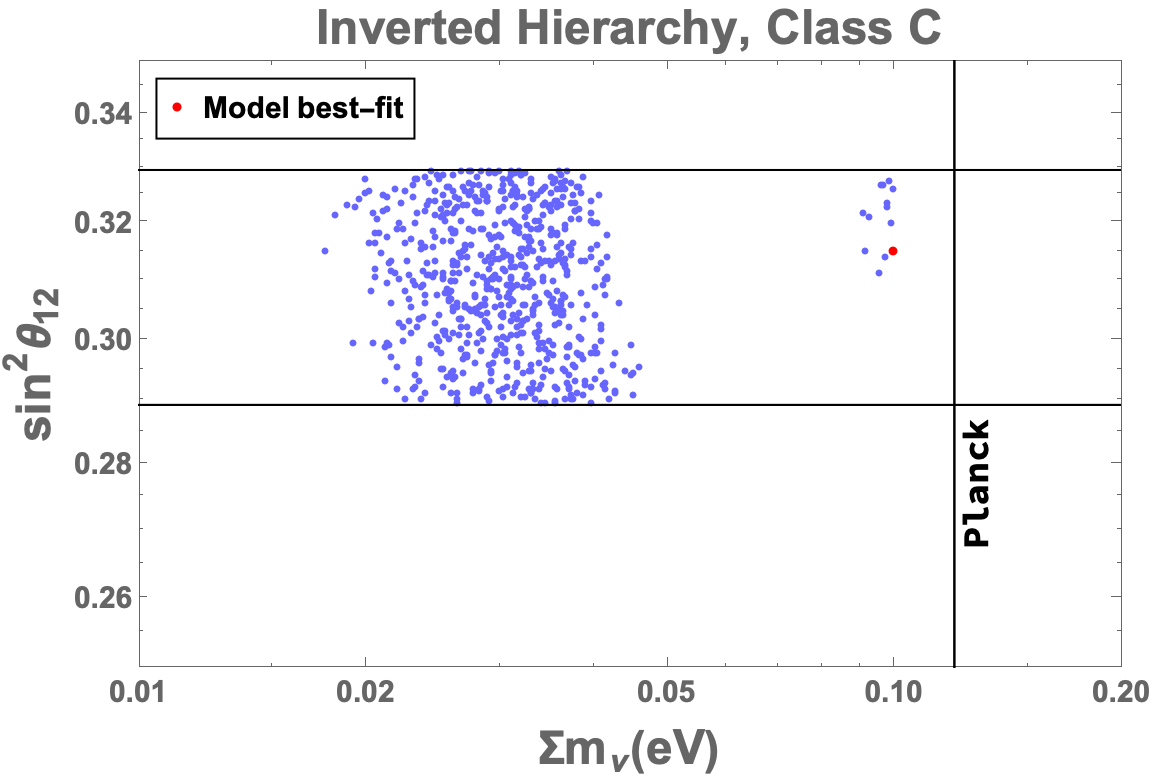}
 	\caption {\label{f7} Neutrino oscillation parameters with sum of neutrino masses, where the black horizontal lines depict the 3$\sigma$ range\cite{Esteban:2024eli} and the vertical line represents the Planck bound on $\Sigma m_{\nu}$\cite{Planck:2018vyg}. The red dot represents the model best-fit value.}
 \end{figure}

 \section{\label{lrsm14}Phenomenological Analysis and Results}
 \subsection{Resonant leptogenesis}
 
An adequate amount of lepton asymmetry can be generated by the phenomenon of resonant leptogenesis (RL) in TeV scale seesaw models \cite{Asaka:2018hyk,Blanchet:2009bu,Flanz:1996fb,Dev:2015cxa,Pilaftsis:2003gt}. For RL, two of the right-handed Majorana neutrinos need to be degenerate which has been found to be so in the present work. The presence of RH neutrinos and scalar triplets in the framework of LRSM suggests that their decays can give rise to lepton asymmetry. The net lepton asymmetry will therefore be generated due to the two seesaw terms. The rate at which the RH neutrinos decay is governed by the Yukawa couplings, and is hence given by \cite{Borgohain:2017akh},
 \begin{equation}
 	\label{E:8}
 	\Gamma_{i}=(Y_{\nu}^{\dagger}Y_{\nu})_{ii} \frac{M_i}{8\pi}
 \end{equation}
 As already mentioned previously that one important condition for RL is that the mass difference of the two heavy RH neutrinos must be comparable to their decay width, i.e., $M_{i}-M_{j}=\Gamma$. In such a case, the CP asymmetry may become very large. The CP violating asymmetry is thus given by \cite{Xing:2015fdg},
 \begin{equation}
 	\label{E:9}
 	\epsilon_{i}=\frac{Im[(Y_{\nu}^{\dagger}Y_{\nu})_{ij}^2]}{(Y_{\nu}^{\dagger}Y_{\nu})_{11}(Y_{\nu}^{\dagger}Y_{\nu})_{22}}.\frac{(M_{i}^{2}-M_{j}^{2})M_{i}\Gamma_{j}}{(M_{i}^{2}-M_{j}^{2})^{2}+M_{i}^{2}\Gamma_{j}^{2}}
 \end{equation}
 The variables $i,j$ run over 1 and 2 and $i\neq j$.\\
 The CP asymmetries $\epsilon_{1}$ and $\epsilon_{2}$ can give rise to a net lepton asymmetry, provided the expansion rate of the universe is larger than $\Gamma_{1}$ and $\Gamma_{2}$. This can further be converted into baryon asymmetry of the universe by $B+L$ violating sphaleron processes\cite{Hong:2023zrf,Kolb:1990vq}.\\
  The CP violating asymmetries $\epsilon_{1}$ and $\epsilon_{2}$ can give rise to net lepton number asymmetry, provided the expansion rate of the universe is larger than $\Gamma_{1}$ and $\Gamma_{2}$.The net baryon asymmetry is then calculated using the following relation \cite{Buchmuller:2004tu},
 \begin{equation}
 	\label{E:16}
 	\eta_B \approx -0.96 \times 10^{-2}\sum_{i}(k_{i}\epsilon_{i})
 \end{equation}
 $k_i$ being the efficiency factors measuring the washout effects. Some parameters are needed to defined as,
 \begin{equation}
 	\label{E:17}
 	K_i \equiv \frac{\Gamma_{i}}{H}
 \end{equation}
 Equation \eqref{E:17} is defined at temperature $T=M_{i}$. The Hubble's constant is given by, $H\equiv \frac{1.66\sqrt{g_{*}}T^{2}}{M_{Planck}}$, where, $g_{*}=107$ and $M_{Planck} = 1.2 \times 10^{19}$ GeV is the Planck mass. decay width is estimated using \eqref{E:8}. The efficiency factors $k_i$ can be calculated using the formula \cite{Blanchet:2008pw},
 \begin{equation}
 	\label{E:18}
 	k_1 \equiv k_2 \equiv \frac{1}{2}(\Sigma_{i} K_{i})^{-1.2}
 \end{equation}
 Equation \eqref{E:18} holds valid for two nearly degenerate heavy Majorana masses and in most general cases $5 \leq K_{i} \leq 100$. 
  \begin{figure}[H]
 	\centering
 		\includegraphics[scale=0.25]{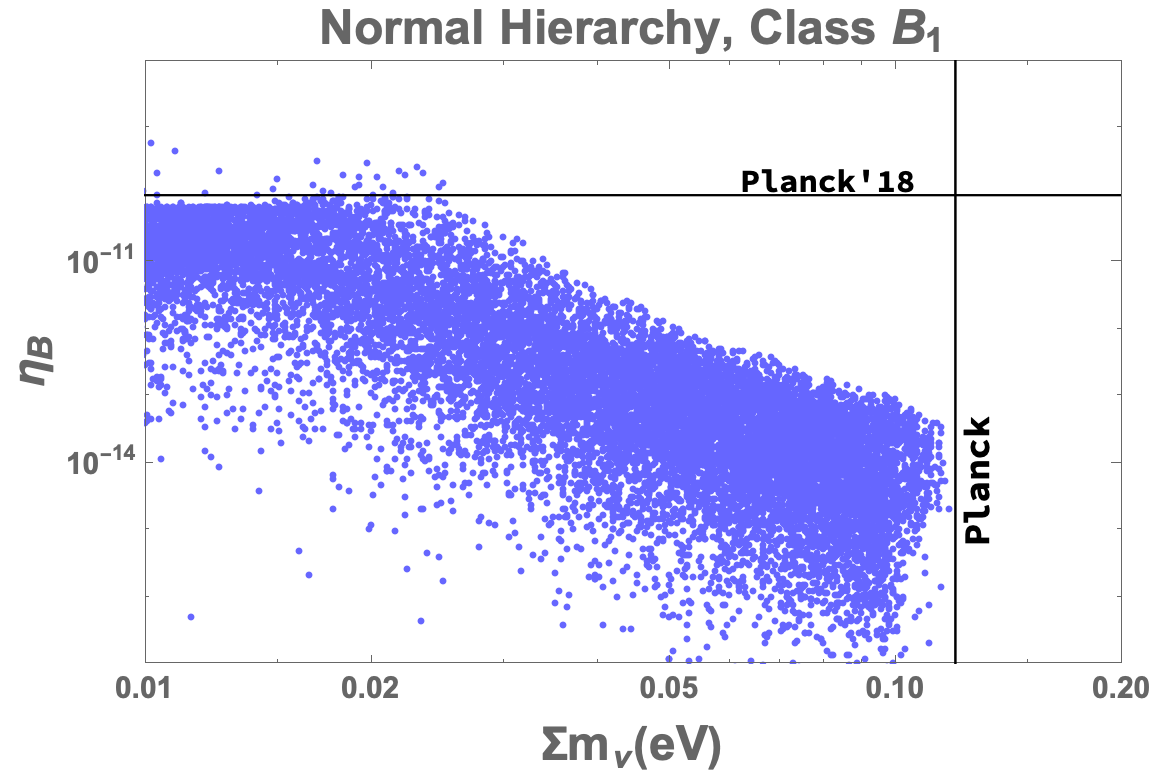}
 	\includegraphics[scale=0.25]{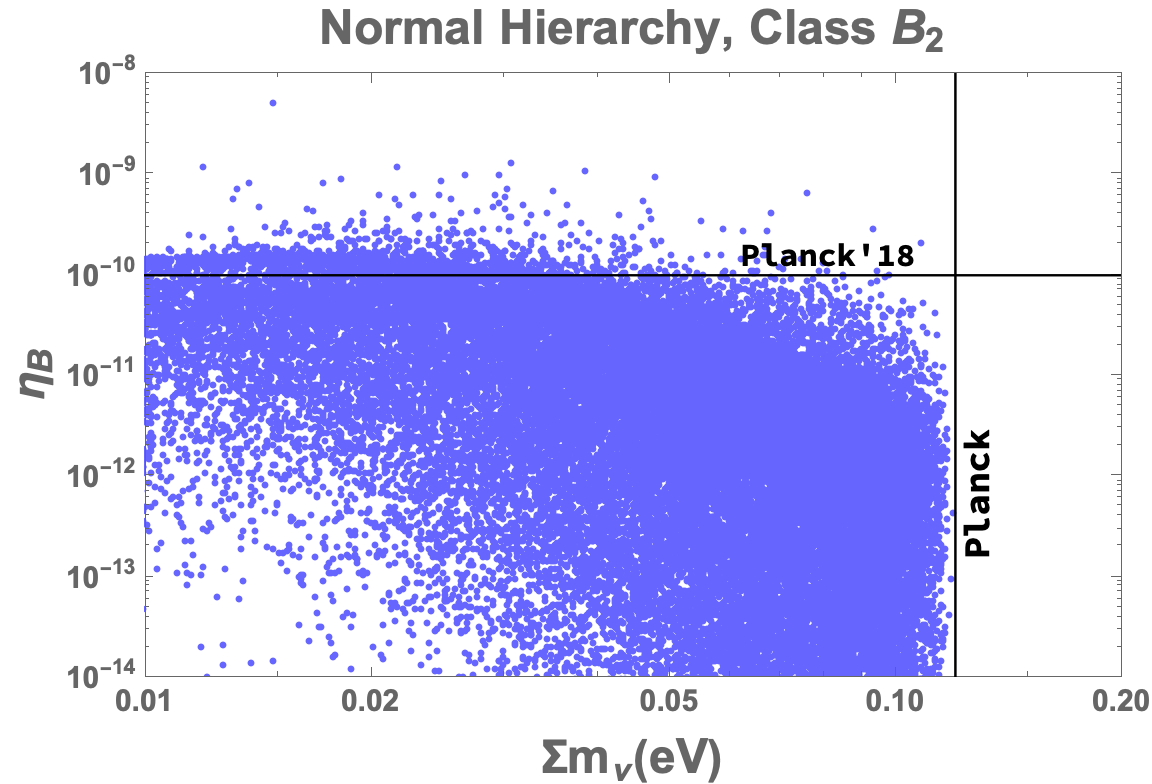}
 		\includegraphics[scale=0.25]{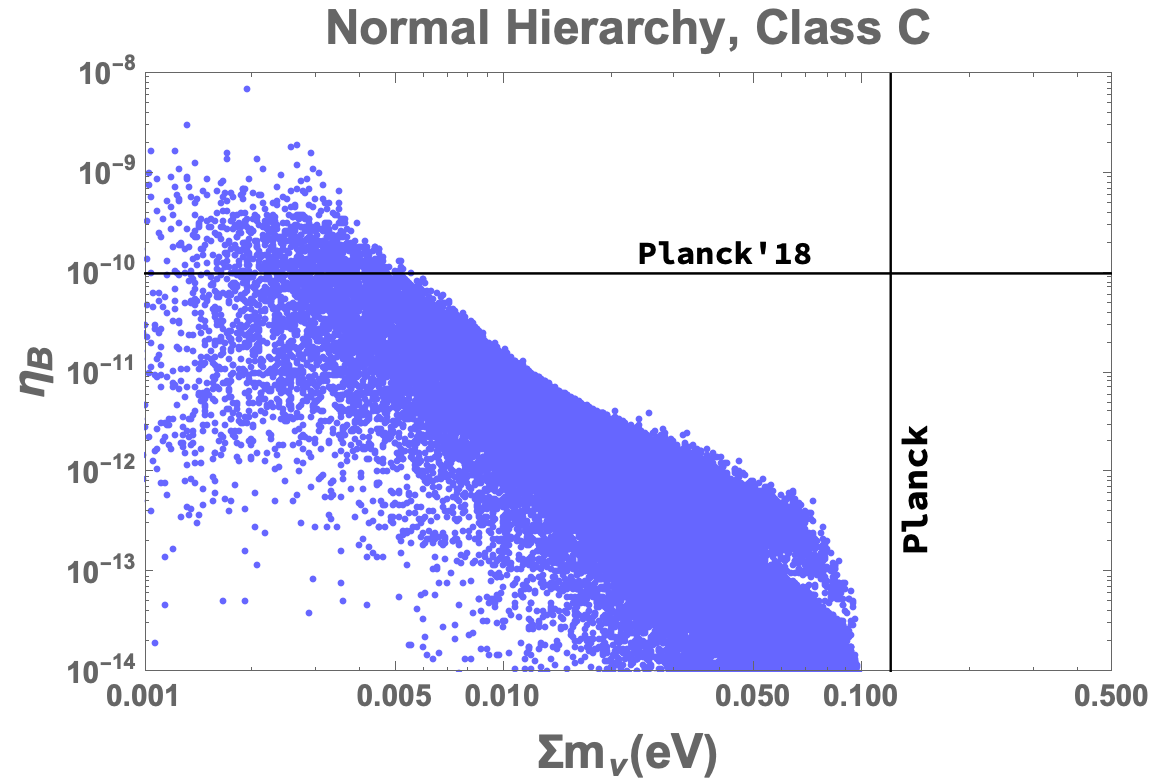}
 	\includegraphics[scale=0.25]{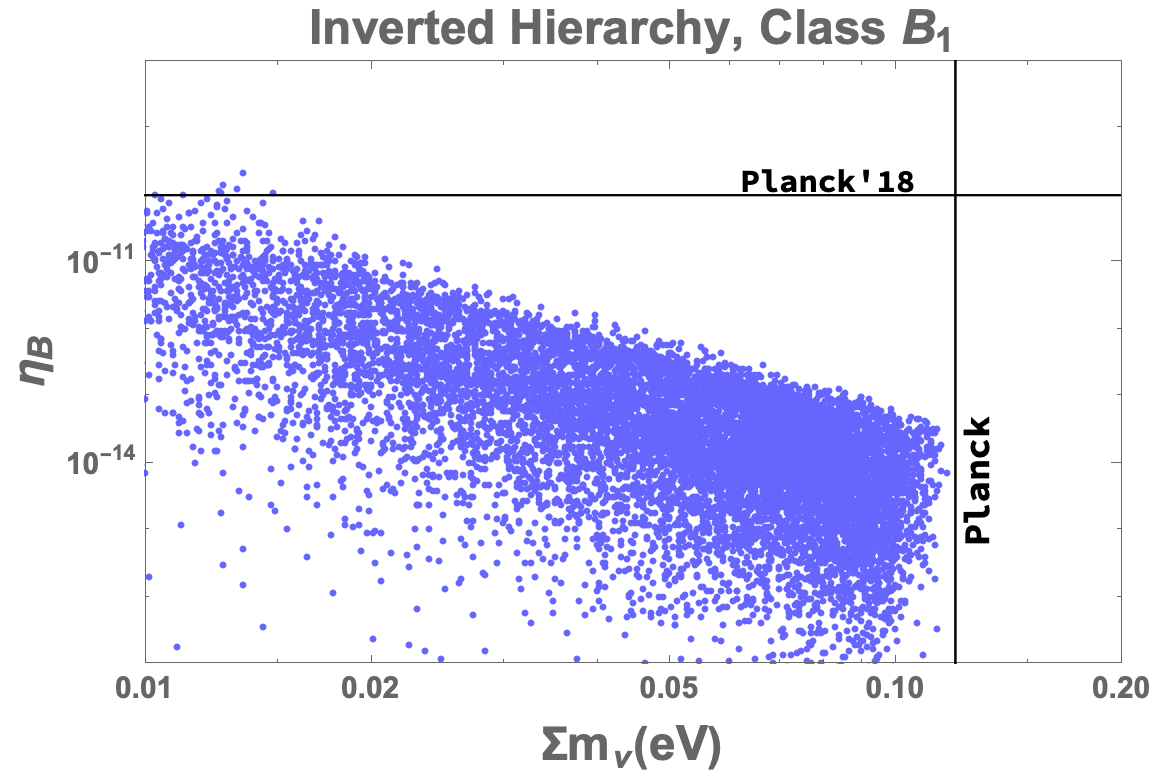}
 	\includegraphics[scale=0.25]{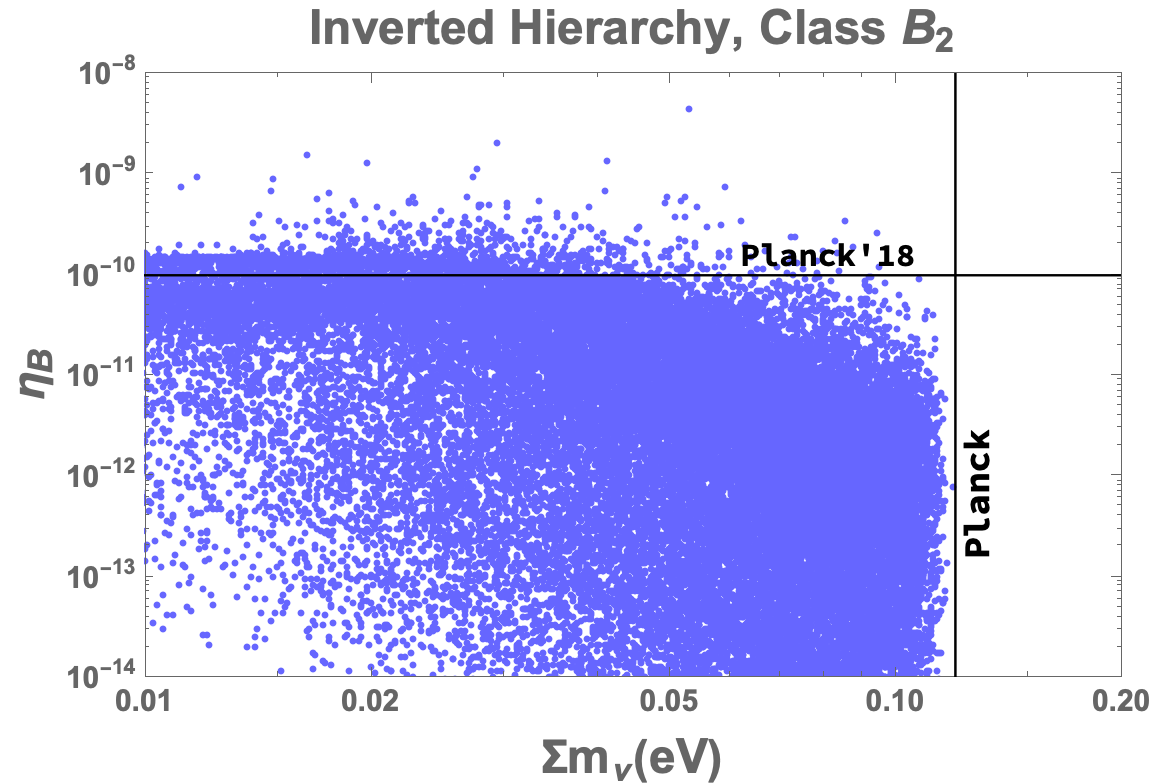}
 	\includegraphics[scale=0.25]{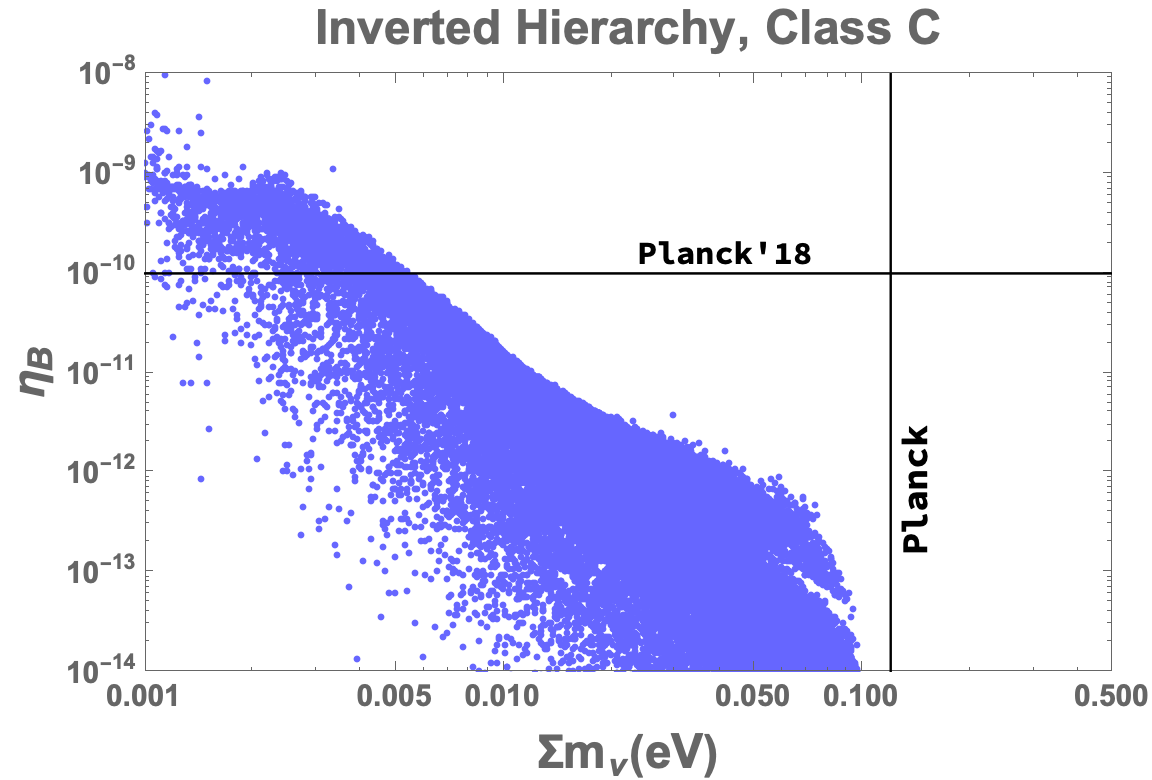}
 	\caption {\label{f8} Baryon asymmetry parameter with sum of neutrino masses, where the black horizontal line depict the observed value of BAU\cite{Planck:2018vyg} and the vertical line represents the Planck bound on sum of neutrino masses\cite{Planck:2018vyg} for Classes $B_{1}$,$B_{2}$ and $C$.}
 \end{figure}
 For the remaining classes as we observe the mass matrices, it has been found that RL cannot be mathematically achieved for the said textures.
 \subsection{Neutrinoless Double Beta Decay}
Because of the presence of several new heavy particles in LRSM, along with the standard light neutrino contribution to $0\nu\beta\beta$, several new physics contributions also come into the picture\cite{Mohapatra:1981pm,Awasthi:2013ff,Patra:2012ur,Chakrabortty:2012mh,Tello:2010am,Awasthi:2015ota,Huang:2013kma,Borah:2016iqd,Borah:2015ufa,Hirsch:1996qw,BhupalDev:2014qbx,Kakoti:2023isn,Kakoti:2023xkn}. There are total eight contributions of the phenomenon to LRSM, however, we will take into account four of the new physics contribution and then we take the total effective mass of the contributions, where in the heavy right-handed neutrino contribution the mediator particles are $W_L$ bosons, light neutrino contribution is mediated by $W_L$ and $W_R$, and the amplitude of the process depends upon the mixing between light and heavy neutrinos, leptonic mixing matrix elements, light neutrino masses and the mass of the gauge bosons, $W_L$ and $W_R$ ($\lambda$ contribution). The $\eta$ contribution is the one which is mediated by both $W_L$ and $W_R$, and the amplitude of the process depends upon the right handed leptonic mixing matrix elements, mixing between the light and heavy neutrinos, also the mass of the gauge bosons, $W_L$ and $W_R$ and the mass of the heavy right handed neutrino. Lastly, we also take into account the right-handed scalar triplet contribution in which the mediator particles are $W_R$ bosons, and the amplitude for the process depends upon the masses of the $W_R$ bosons, right-handed triplet Higgs, $\Delta_R$ as well as their coupling to leptons. The details of these contributions can be found in \cite{Chakrabortty:2012mh,Borgohain:2017akh}.
\begin{itemize}
	\item For $\lambda$ contribution, the dimensionless parameter $\eta_\lambda$ is given by,
	\begin{equation}
		\label{E:19}
		|{\eta_\lambda}| = \Biggl(\frac{M_{W_L}}{M_{W_R}}\Biggl)^{2}|\Sigma_{i}U_{ei}T_{ei}^*|
	\end{equation}
	The effective mass in this case is calculated using the formula,
	\begin{equation}
		m_{{eff}_{\lambda}}=m_{e}.|\eta_\lambda|
	\end{equation}
	\item For $\eta$ contribution, the dimensionless parameter describing $0\nu\beta\beta$ is given by,
	\begin{equation}
		\label{E:20}
		|\eta_{\eta}| = \tan \xi |\Sigma_{i}U_{ei}T_{ei}^*|
	\end{equation}
		The effective mass is given by,
	\begin{equation}
		m_{{eff}_{\eta}}=m_{e}.|\eta_\eta|
	\end{equation}
\end{itemize}
In the above equations, $U_{ei}$ represents the first row of the neutrino mixing matrix. $|\Sigma_{i}U_{ei}T_{ei}^*|$ can be simplified to the form $-[M_{D}M_{RR}^{-1}]_{ee}$ as described in \cite{Barry:2013xxa}. T is represented by the equation \eqref{E:28} and,
\begin{equation}
	\label{E:21}
	tan 2\xi = -\frac{2k_1k_2}{v_{R}^2 - v_{L}^2}
\end{equation}
The heavy right-handed (RH) neutrino contribution for $0\nu\beta\beta$ is given by,
\begin{equation}
	\label{E:22}
m_{{eff}_{N}} = p^2\Biggl(\frac{M_{W_{L}}^4}{M_{W_{R}}^4}\Biggl)\frac{U_{Rei}^2}{M_i}
\end{equation}
where, $<p^2> = m_e m_p \frac{M_N}{M_\nu}$ is the typical momentum exchange of the process. $m_e$ and $m_p$ are the masses of the electron and proton respectively and $M_N$ is the nuclear matrix element (NME) for the right-handed neutrino exchange. The allowed value of $p$ is in the range $(100-200)$ MeV. But for our analysis we have used $p = 180 MeV$ \cite{Chakrabortty:2012mh}. 
The effective mass for scalar triplet contribution is given by the following formula,
\begin{equation}
	m_{{eff}_{\Delta}} = |p^2 \frac{M_{W_L}^4}{M_{W_R}^4} \frac{2M_N}{M_{\Delta_R}}|
\end{equation}
where, $M_{\Delta_{R}}=3TeV$. Hence, the total contribution for neutrinoless double beta decay is given by,
\begin{equation}
	m_{{eff}_{(total)}}=m_{{eff}_{N}} +m_{{eff}_{\Delta}}+m_{{eff}_{\lambda}}+m_{{eff}_{\eta}}
\end{equation}
 \begin{figure}[H]
	\centering
	\includegraphics[scale=0.25]{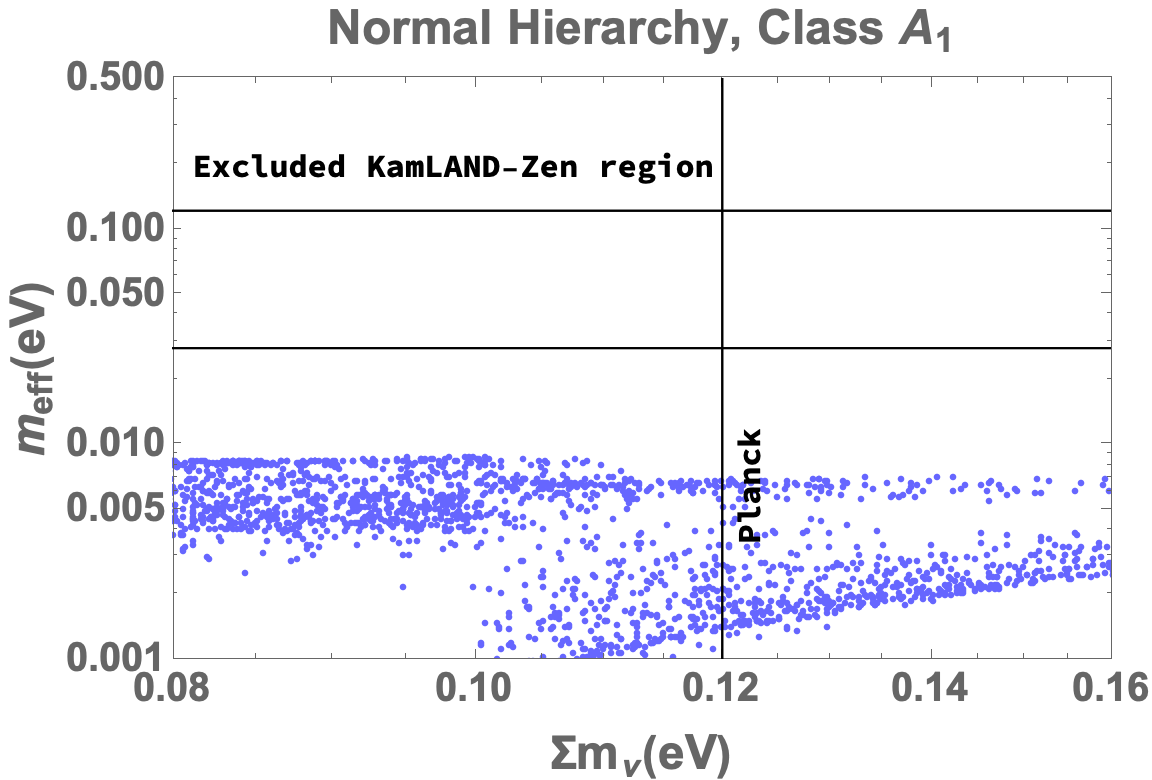}
	\includegraphics[scale=0.25]{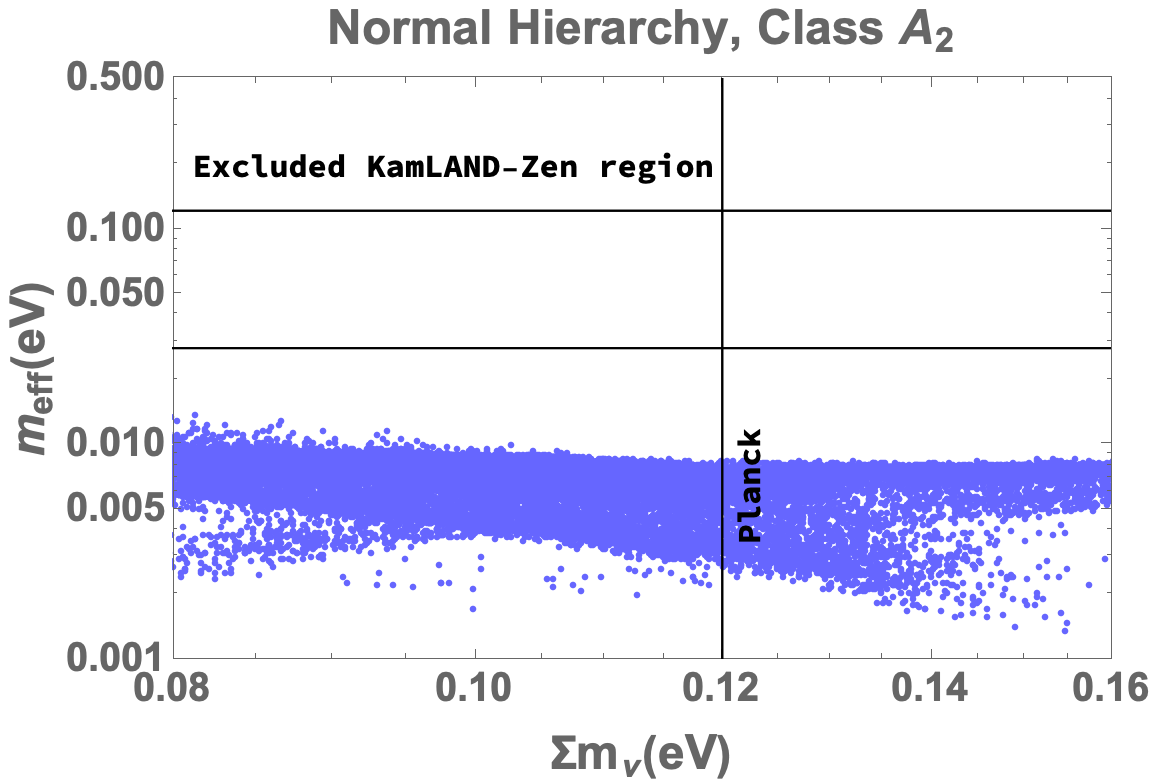}
	\includegraphics[scale=0.25]{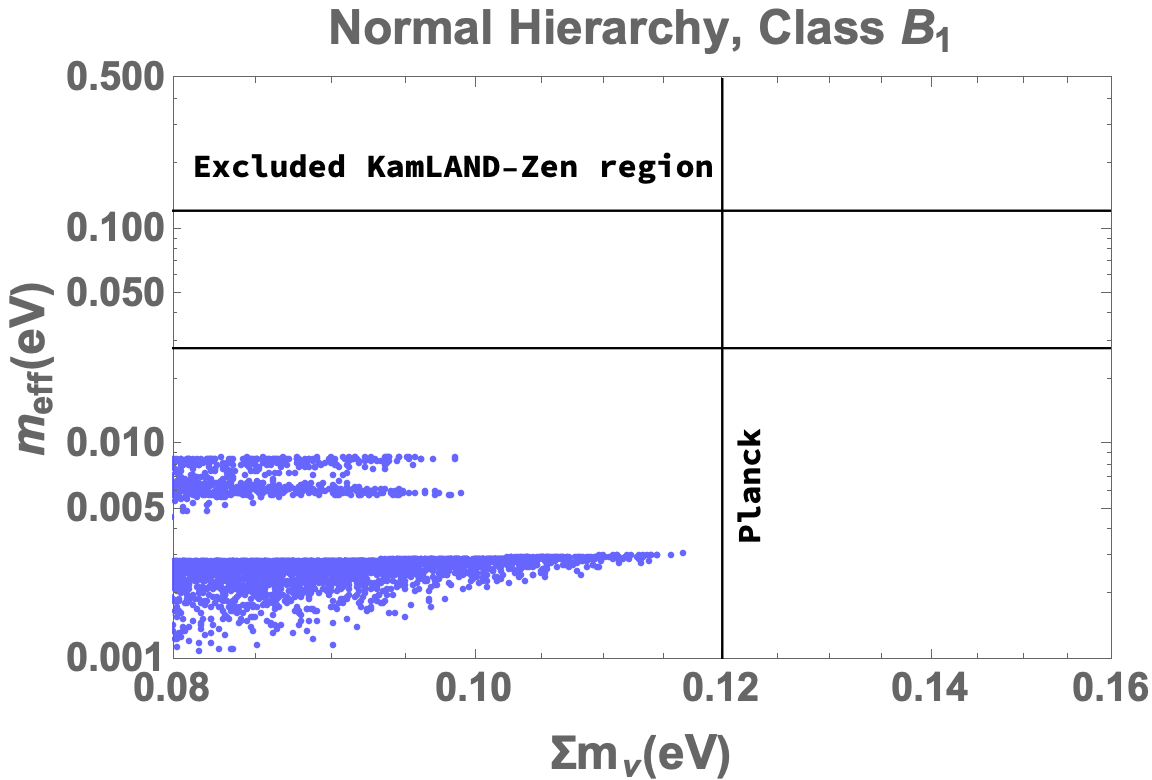}
	\includegraphics[scale=0.25]{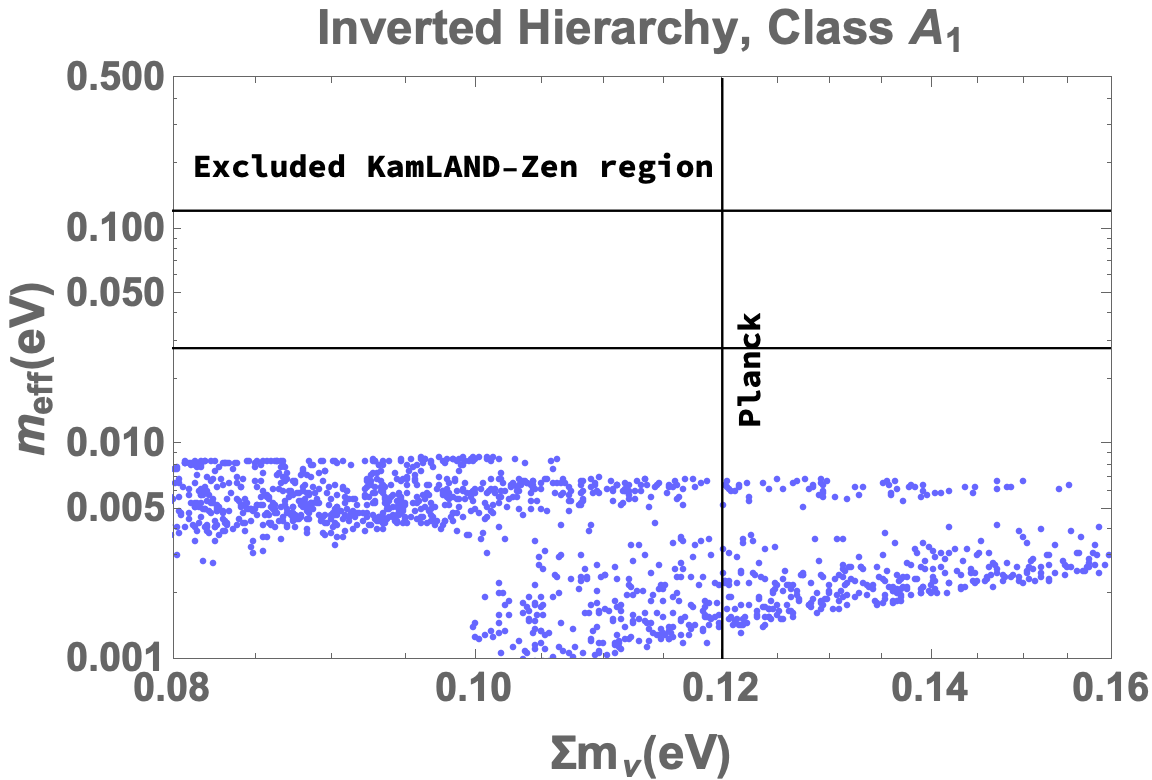}
	\includegraphics[scale=0.25]{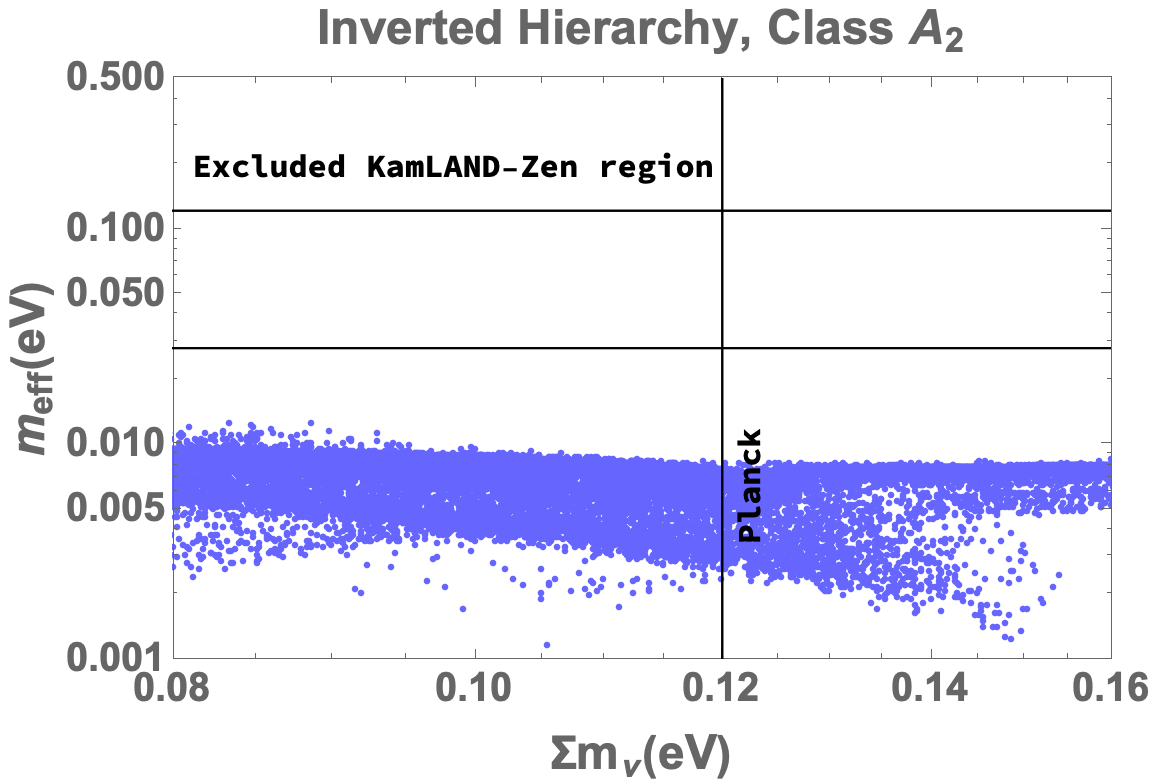}
	\includegraphics[scale=0.25]{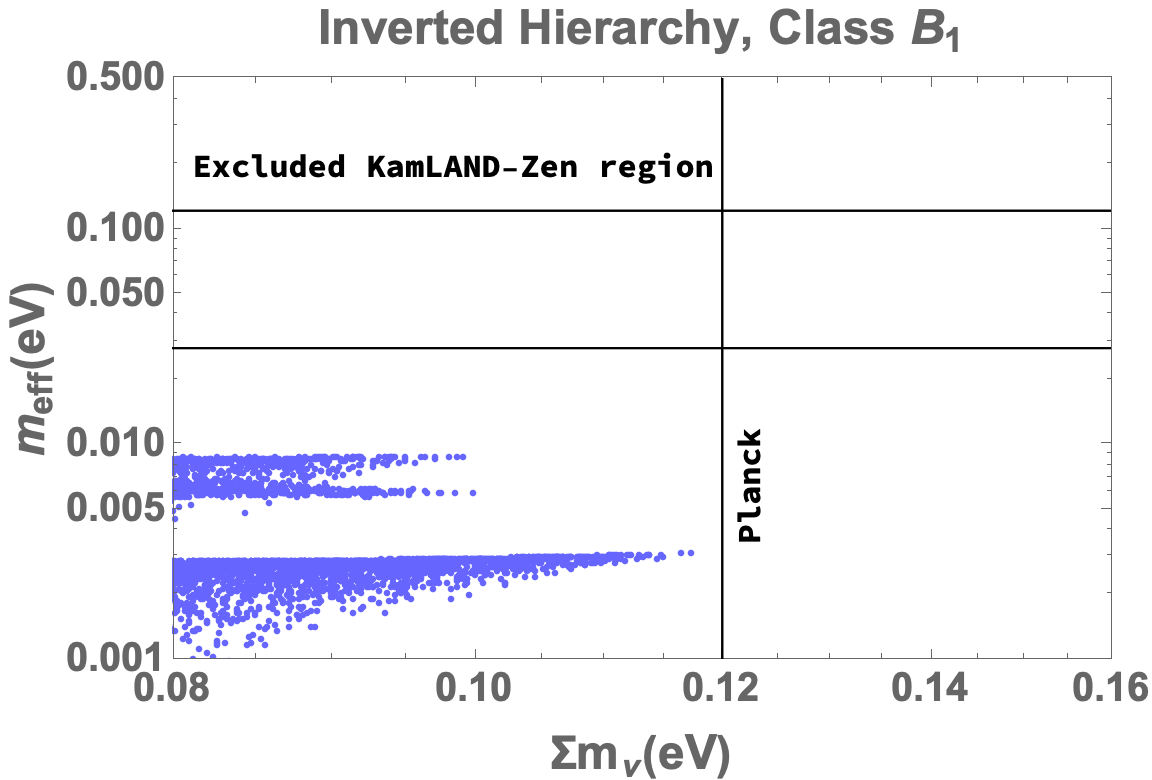}
	\caption {\label{f9} Effective mass with sum of neutrino masses, where region above the black horizontal lines depict the excluded data by KamLAND-Zen\cite{KamLAND-Zen:2024eml} and the vertical line represents the Planck bound on sum of neutrino masses\cite{Planck:2018vyg} for Classes $A_{1}$,$A_{2}$ and $B_{1}$.}
\end{figure}
\begin{figure}[H]
	\centering
	\includegraphics[scale=0.25]{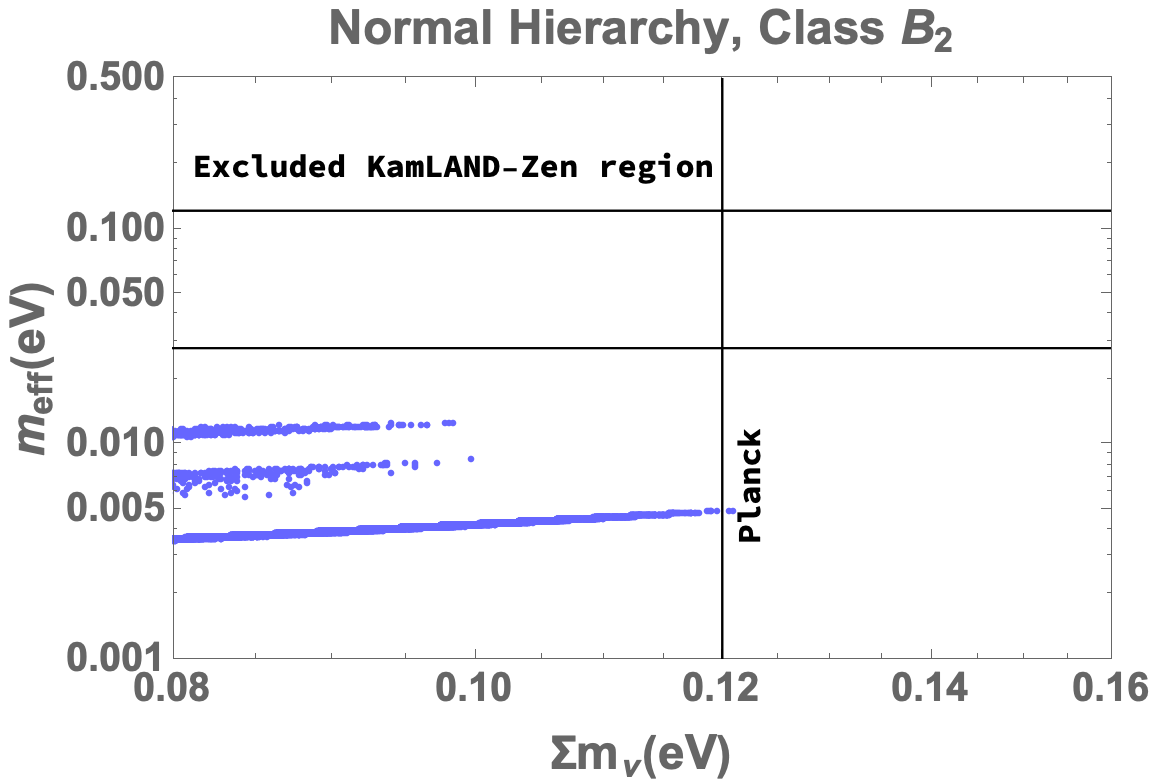}
	\includegraphics[scale=0.25]{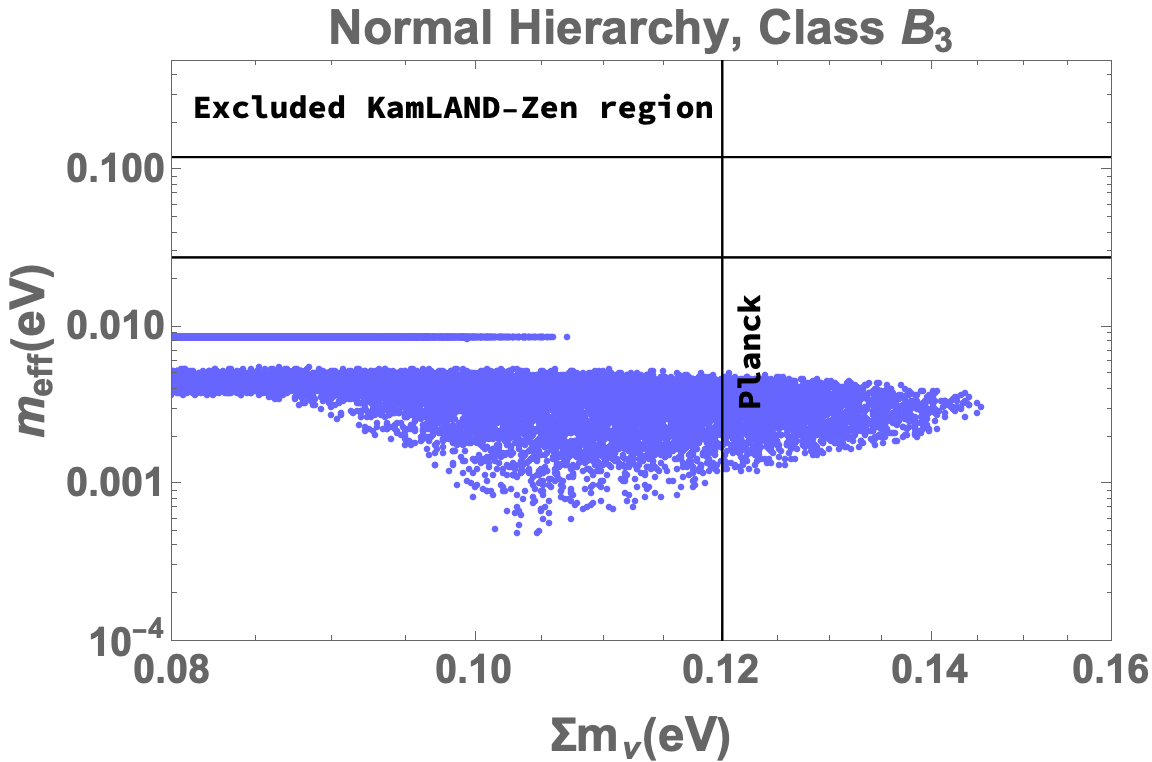}
	\includegraphics[scale=0.25]{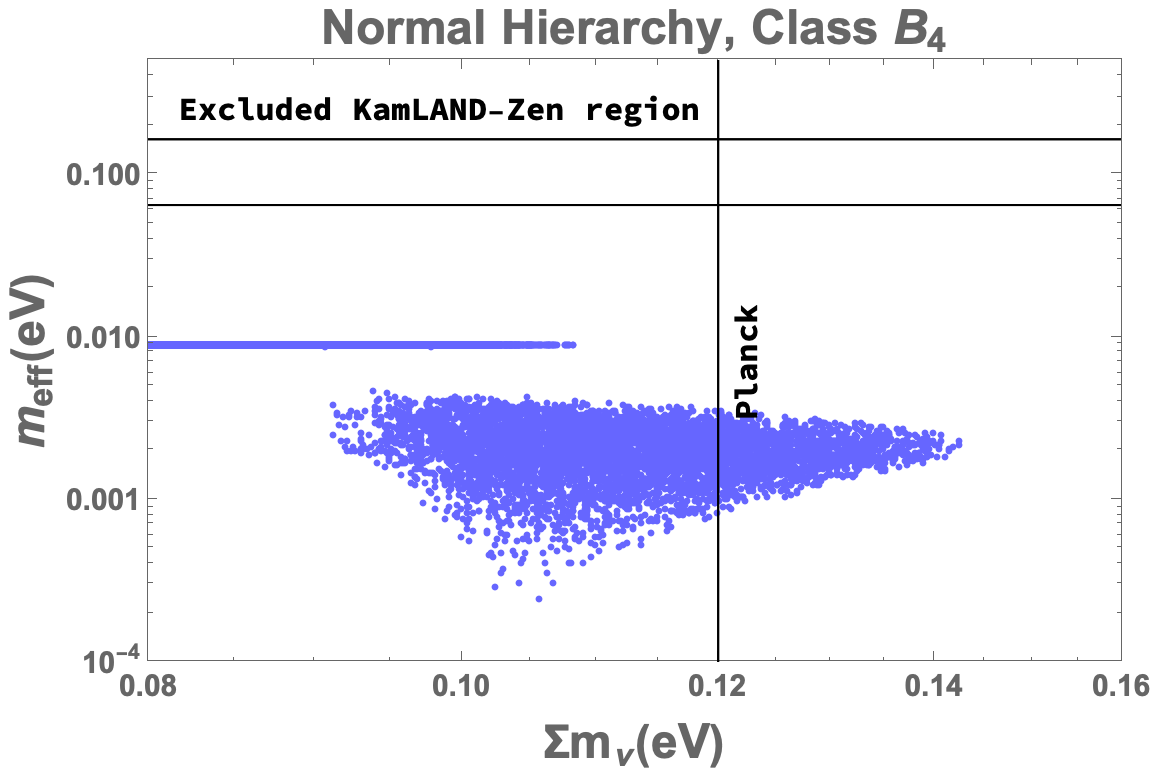}
\end{figure}
\begin{figure}[H]
\centering
	\includegraphics[scale=0.25]{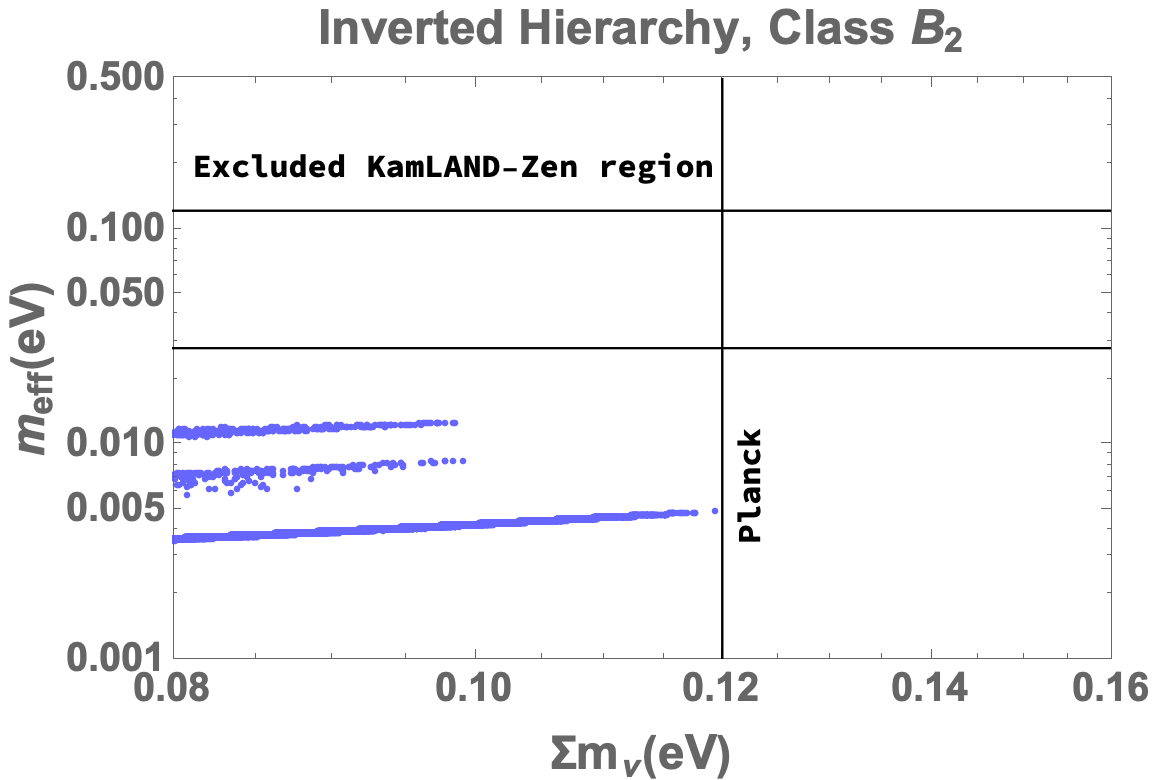}
	\includegraphics[scale=0.25]{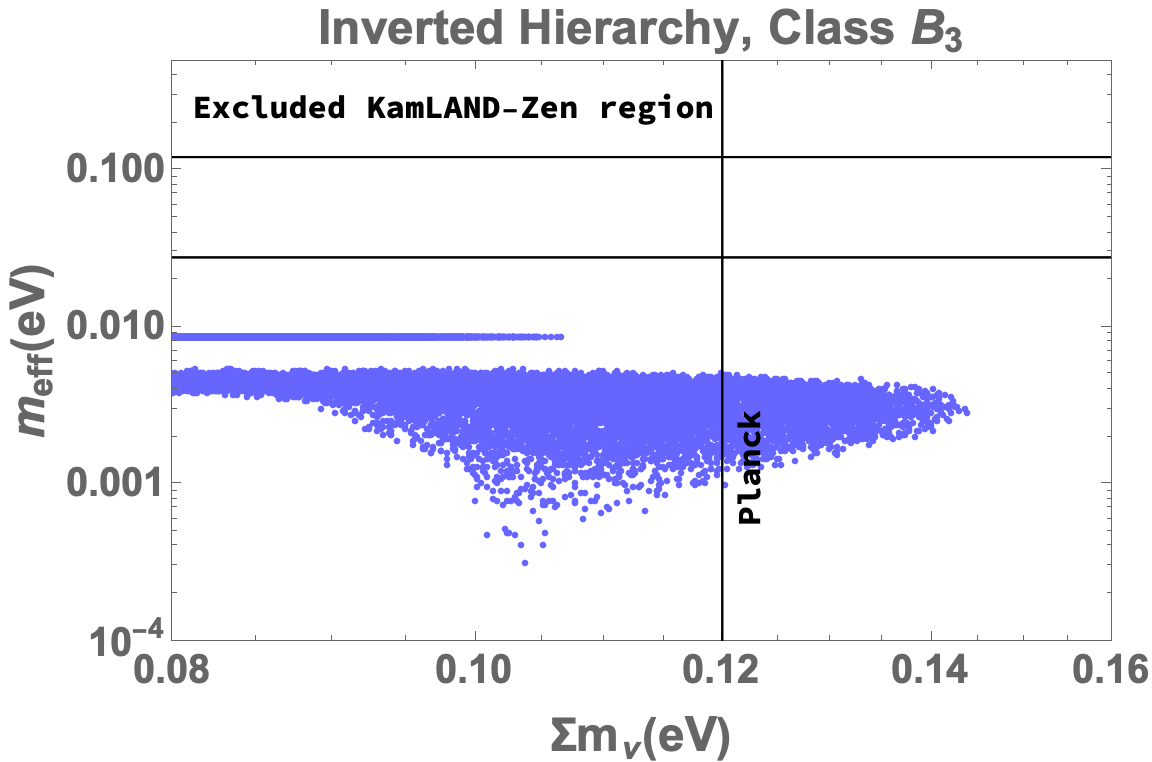}
	\includegraphics[scale=0.25]{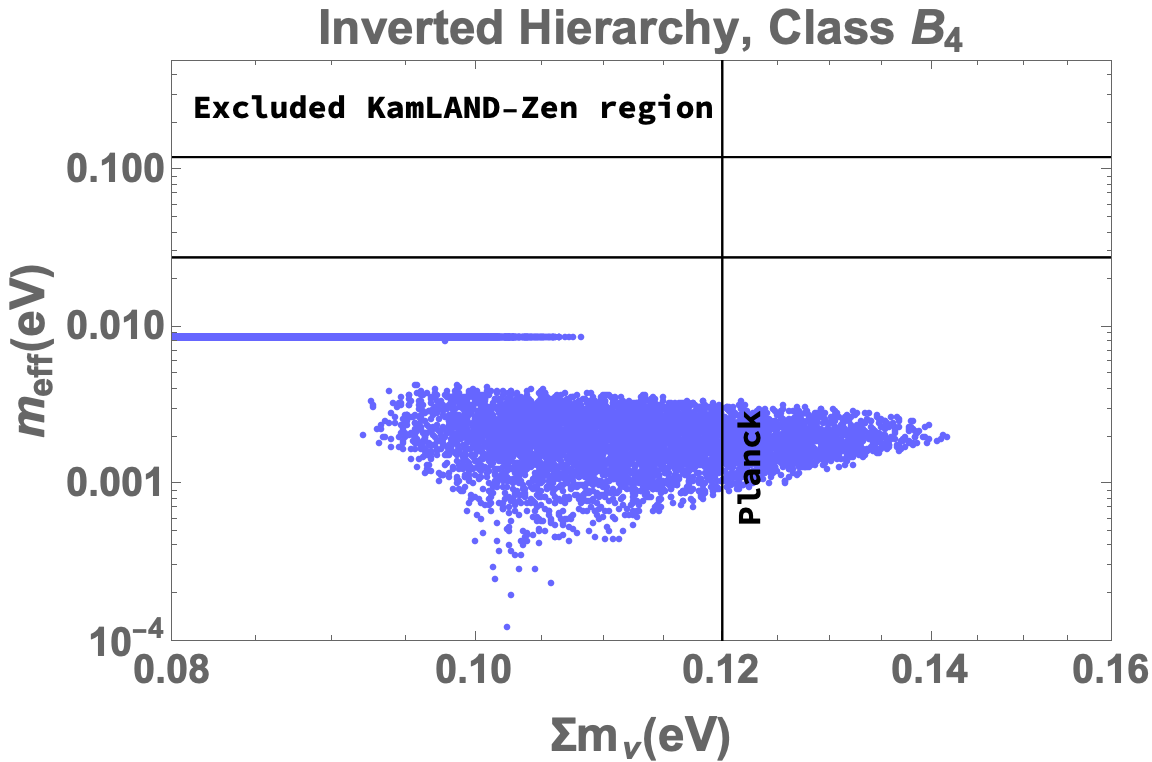}
	\caption {\label{f10} Effective mass with sum of neutrino masses, where region above the black horizontal lines depict the excluded data by KamLAND-Zen\cite{KamLAND-Zen:2024eml} and the vertical line represents the Planck bound on sum of neutrino masses\cite{Planck:2018vyg} for Classes $B_{2}$,$B_{3}$ and $B_{4}$.}
\end{figure}
\begin{figure}[H]
	\centering
	\includegraphics[scale=0.3]{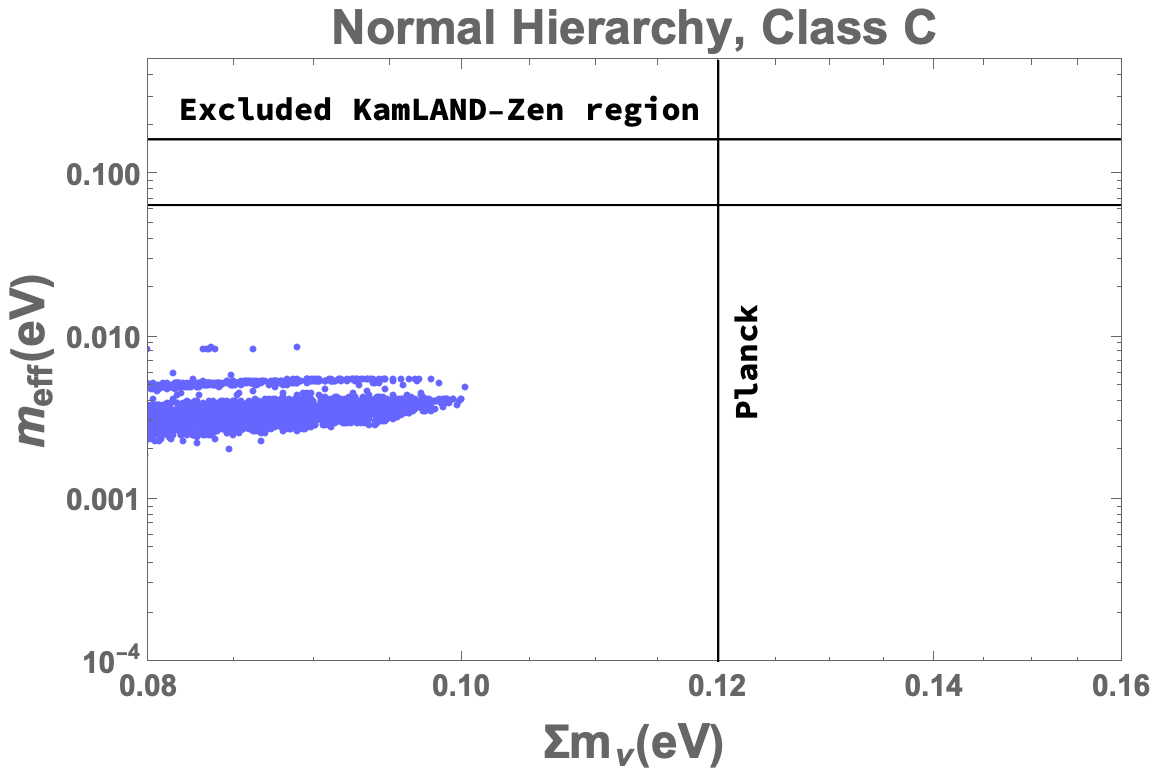}
	\includegraphics[scale=0.3]{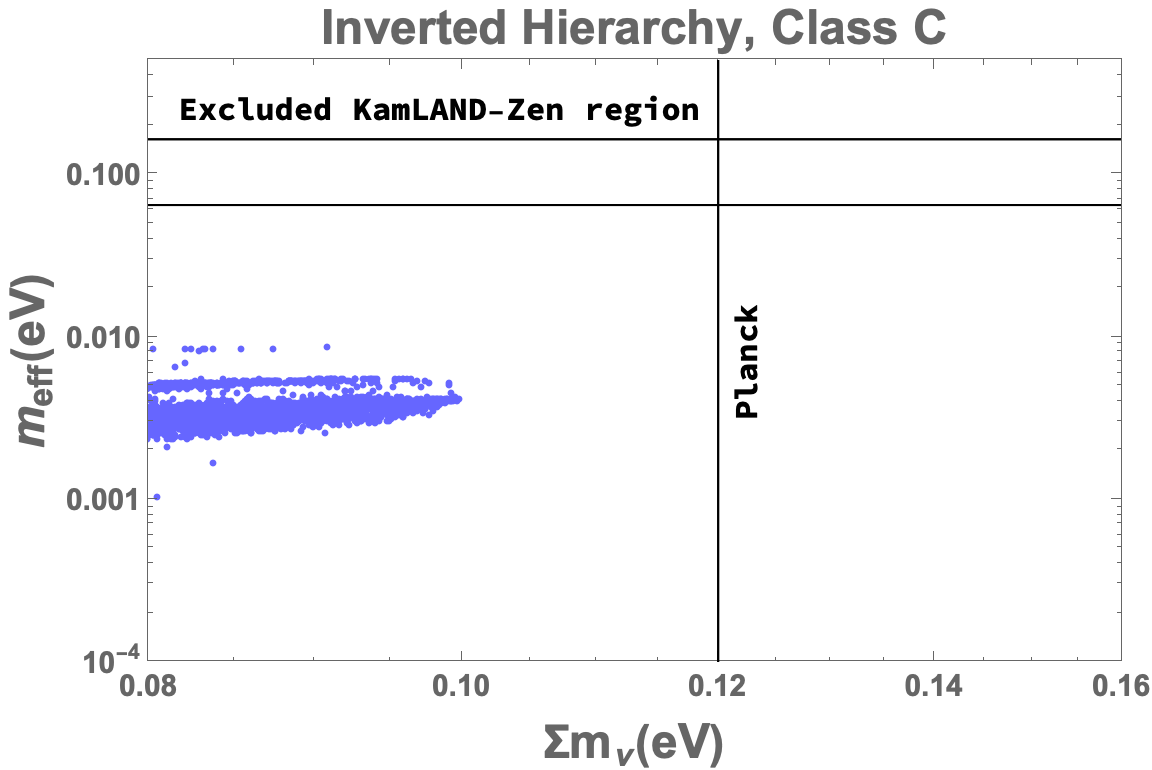}
	\caption {\label{f11} Effective mass with sum of neutrino masses, where region above the black horizontal lines depict the excluded data by KamLAND-Zen\cite{KamLAND-Zen:2024eml} and the vertical line represents the Planck bound on sum of neutrino masses\cite{Planck:2018vyg} for Class C.}
	\end{figure}
	\section{\label{lrsm15}Discussion and Conclusion}
	In this work, LRSM has been implemented using $A_4$ modular symmetry, which corresponds to the modular forms having distinct irreducible representations for various modular weights. The main goal of the work is to investigate the effects of giving various modular weights to the particle content of the model. This leads to potential two-zero neutrino mass matrix textures in the context of the current investigation. The findings of the phenomenomenological study of two-zero neutrino mass textures in the context of modular symmetry are explained in detail in the points as listed below.
	\begin{itemize}
		\item In LRSM, owing to the presence of RH neutrinos and scalar triplets, the resulting light neutrino mass arises as a summation of type-I and type-II seesaw masses. However, in some cases one of the seesaw terms may be dominant, but in the current work, we have considered equal contributions for both the seesaw mass terms. The use of modular symmetry is favorable as it does not require the use of extra particle fileds for obtaining the desired results. 
		\item In this work, we have used weights $k_{Y}=4,8$ and 10 for obtaining different two-zero neutrino mass textures. As it is known that out of the fifteen possible two-zero textures, only seven of them falls in the allowed regime, and as such we have obtained all the seven classes, namely $A_{1}$, $A_{2}$, $B_{1}$, $B_{2}$, $B_{3}$, $B_{4}$ and $C$ by virtue of present realization of the model. Generally in model building processes, we do not use modular forms with weights greater than 10, no matter whether they exist or not.
		\item In the first case, we have assigned weight $k_{Y}=4$ to the modular Yukawa form. As such, we assign modular weights to the particle content of the model, keeping in mind that sum of the weights in each term of the superpotential must be zero. This weight gives rise to Class $B_{1}$ and $B_{2}$ of two-zero neutrino mass texture. Figure \ref{f1} shows the variation of neutrino oscillation parameters with sum of neutrino masses for Class $B_{2}$, and it has been observed that for inverted hierarchy, no data points satisfy the 3$\sigma$ range for Dirac CP phase. For solar mixing angle ($sin^{2}\theta_{12}$), none of the orderings show satisfactory data which distinctly rules out class $B_2$, inverted ordering for the model under study.
		\item By interchanging the weight and charge assignments of the fields for $k_{Y}=4$ in the manner $L_{L_2} \leftrightarrow L_{L_3}$, and $L_{R_2} \leftrightarrow L_{R_3}$, we obtain a class $B_{1}$ of two-zero texture, and the results for the same are depicted in figure \ref{f2}. It has been observed that the atmospheric and solar mixing angles very well satisfy the 3$\sigma$ bound on oscillation parameters; however, no viable data points were obtained for the Dirac CP phase for this neutrino mass texture. From figure \ref{f2}, it can be seen that for $sin^{2}\theta_{23}$, dominant data falls in the regime of $0.5$ to $0.6$ radians, which is more prominent in the case of inverted hierarchy. As such, it can be stated that Class $B_{1}$ constrains the value of atmospheric mixing angle, more precisely in the case of inverted hierarchy.	
		\item The first case of $k_{Y}=8$ results in Class $B_{4}$ of two-zero texture, the results for which have been shown in figure \ref{f3}. Although a considerable amount of data falls within the range of NuFit-6.1, no satisfactory data points were obtained for $\delta_{CP}$. In case of $sin^{2}\theta_{12}$, this class has been successful in constraining the range of sum of neutrino masses from around 0.0 5eV to 0.12 eV.
		\item Interchanging the charge and weight assignments of the fields respectively as $L_{L_2}\leftrightarrow L_{L_3}$ and $L_{R_2}\leftrightarrow L_{R_3}$, $L_{L_3}\leftrightarrow L_{L_1}$ and $L_{R_3}\leftrightarrow L_{R_1}$ and $L_{L_1}\leftrightarrow L_{L_2}\leftrightarrow L_{L_3}\leftrightarrow L_{L_1}$ and $L_{R_1}\leftrightarrow L_{R_2}\leftrightarrow L_{R_3}\leftrightarrow L_{R_1}$ will result in Class $B_{3}$, $B_{4}$, $A_{2}$ and Class $A_{1}$ respectively. The results for these textures have been shown in figures \ref{f3},\ref{f4},\ref{f5} and \ref{f6} respectively.
		\item Figure \ref{f5}  shows that no data points for Class $A_{2}$ were found to satisfy the current 3$\sigma$ constraints on the parameter $sin^{2}\theta_{12}$. For Class $A_{1}$, all the three neutrino oscillation parameters have been observed to be satisfying NuFiT-6.1 bounds, however for Dirac CP phase, results are favorable for normal hierarchy only. For $sin^{2}\theta_{12}$, both normal and inverted hierarchies show satisfactory results, but the model best-fit value for normal hierarchy lies outside the 3$\sigma$ range, but the present results of JUNO have put stringent bound on its value and hence we would state that results for solar mixing angle has been found to be favorable for inverted hierarchy for Class $A_{1}$.
		\item For $k_{Y_{max}}=10$, we obtain Class $C$ of two-zero neutrino mass textures and results for the same have been shown in figure \ref{f7}. It has been observed that all three neutrino oscillation parameters satisfy the 3$\sigma$ bounds, and for $\delta_{CP}$, the results are found to be favorable for normal hierarchy.
		\item For phenomenological studies we have considered resonant leptogenesis and new physics contribution of neutrinoless double beta decay. For all the seven textures, satisfactory results were obtained for $0\nu\beta\beta$ as shown in figures \ref{f9} to \ref{f11}.
		\item  It has been noted that for RL, the mathematical expressions representing the baryon asymmetry parameter $\eta_B$ only yield data points that satisfy the Planck'18 bound on $\eta_B$ for Classes $B_{1}$, $B_{2}$ and Class $C$.  As such, figure \ref{f8} shows only the plot for baryon asymmetry paramater with sum of neutrino masses for these three classes.
	\end{itemize}
	The table provided below summarizes a clear picture of the work done and findings acquired for all the derived textures as a part of the current study.
	\begin{table}[H]
		\begin{center}
			\begin{tabular}{|c|c|c|c|c|c|}
				\hline
				Textures & $\delta_{CP}$(NH)[IH] & $sin^{2}\theta_{23}$(NH)[IH] & $sin^{2}\theta_{12}$(NH)[IH] & BAU (NH)[IH] & $0\nu\beta\beta$ (NH)[IH]\\
				\hline
				Class $A_{1}$ & $(\checkmark)[\times]$ & $(\checkmark)[\checkmark]$ & $(\checkmark)[\checkmark]$ & $(\times)[\times]$ & $(\checkmark)[\checkmark]$\\
				\hline
				Class $A_{2}$ & $(\checkmark)[\checkmark]$ & $(\checkmark)[\checkmark]$ & $(\times)[\times]$ & $(\times)[\times]$ & $(\checkmark)[\checkmark]$\\
				\hline
				Class $B_{1}$ & $(\times)[\times]$ & $(\checkmark)[\checkmark]$ & $(\checkmark)[\checkmark]$ & $(\checkmark)[\checkmark]$ & $(\checkmark)[\checkmark]$\\
				\hline
				Class $B_{2}$ & $(\checkmark)[\times]$ & $(\checkmark)[\checkmark]$ & $(\times)[\times]$ & $(\checkmark)[\checkmark]$ & $(\checkmark)[\checkmark]$\\
				\hline
				Class $B_{3}$ & $(\times)[\times]$ & $(\checkmark)[\checkmark]$ & $(\checkmark)[\checkmark]$ & $(\times)[\times]$ & $(\checkmark)[\checkmark]$\\
				\hline
				Class $B_{4}$ & $(\times)[\times]$ & $(\checkmark)[\checkmark]$ & $(\checkmark)[\checkmark]$ & $(\times)[\times]$ & $(\checkmark)[\checkmark]$\\
				\hline
				Class $C$ & $(\checkmark)[\times]$ & $(\checkmark)[\checkmark]$ & $(\checkmark)[\checkmark]$ & $(\checkmark)[\checkmark]$ & $(\checkmark)[\checkmark]$\\
				\hline
			\end{tabular}
			\caption{\label{t9} Summarization of the results obtained for each of the textures, where the $\checkmark$ denotes that observed value falls within the desired range and $\times$ denote that data points obtained do not satisfy the present bounds on the parameter.}
		\end{center}
	\end{table}
    Thus the present framework produces correlated predictions for some significant parameters like the sum of neutrino masses, $\delta_{CP}$, the absolute neutrino mass scale, and the effective Majorana mass governing $0\nu\beta\beta$ and also baryogenesis. This makes the model directly testable through future neutrino oscillation, cosmological, and $0\nu\beta\beta$ experiments. Embedding the current study into grand unified, or supersymmetric, frameworks is an obvious expansion that could offer even stronger phenomenological consequences and a much richer theoretical base. Furthermore, the prediction strength and robustness of modular flavor symmetries as a compelling foundation for comprehending neutrino masses and probing physics beyond the Standard Model could be further evaluated which we leave for our future studies.
    \newpage
\bibliographystyle{unsrt}
\bibliography{cite}
\end{document}